%% file: radio_stars_arxiv.tex
\documentclass[useAMS,usenatbib]{mn2e}
\usepackage[dvips]{graphicx}
\usepackage{hyperref}
\usepackage{amsmath}
\usepackage{amssymb}
\usepackage{caption}
\usepackage{xcolor}
\include{aas_macros}

\newcommand{\bz}{$\langle B_z \rangle$}
\newcommand{\nz}{$\langle N_z \rangle$}

\newcommand{\msun}{$M_{\odot}$}
\newcommand{\rsun}{$R_{\odot}$}

\newcommand{\kms}{km\,s$^{-1}$}

\newcommand{\vsini}{$v \sin i$}
\newcommand{\teff}{$T_{\rm eff}$}

\newcommand{\mdot}{$\dot{M}$}
\newcommand{\vinf}{v$_{\infty}$}

\newcommand{\rk}{$R_{\rm K}$}
\newcommand{\ra}{$R_{\rm A}$}
\newcommand{\rark}{$\log{(R_{\rm A}/R_{\rm K})}$}

\newcommand{\beq}{\begin{equation}}
\newcommand{\eeq}{\end{equation}}
\newcommand{\beqa}{\begin{eqnarray}}
\newcommand{\eeqa}{\end{eqnarray}}

\newcommand{\mnras}{$MNRAS$}
\newcommand{\apj}{$ApJ$}
\newcommand{\apjl}{$ApJ$}
\newcommand{\aj}{$AJ$}
\newcommand{\aap}{$A\&A$}

\newcommand{\aaps}{$A\&AS$}

\newcommand{\pasp}{$PASP$}
\newcommand{\apjs}{$ApJS$}

\newcommand{\nat}{$Nature$}

\newcommand{\an}{$Astron.\ Nachr.\ $}

\newcommand{\physscr}{Phys.\ Scr.}

\title[Rotation and radio magnetospheres]{MOBSTER -- VI. The crucial influence of rotation on the radio magnetospheres of hot stars}
\author[M.\ E.\ Shultz et al.]{M.\ E.\ Shultz$^{1}$\thanks{E-mail:
mshultz@udel.edu},
S.\ P.\ Owocki$^1$,
A.\ ud-Doula$^2$,
A.\ Biswas$^3$,
D.\ Bohlender$^4$,
P.\ Chandra$^3$,
\newauthor{
B.\ Das$^3$,
A.\ David-Uraz$^{5,6}$,
V.\ Khalack$^{7}$,
O.\ Kochukhov$^{8}$,
J.\ D.\ Landstreet$^{9,10}$,
}
\newauthor{
P.\ Leto$^{11}$,
D.\ Monin$^4$,
C.\ Neiner$^{12}$,
Th.\ Rivinius$^{13}$,
and G.\ A.\ Wade$^{14}$
}
\\
$^1$Department of Physics and Astronomy, University of Delaware, 217 Sharp Lab, Newark, Delaware, 19716, USA\\
$^2$Department of Physics, Penn State Scranton, Dunmore, PA 18512, USA\\
$^3$National Centre for Radio Astrophysics, Tata Institute of Fundamental Research, Pune University Campus, Pune-411007, India\\
$^4$National Research Council of Canada, Herzberg Astronomy and Astrophysics Research Centre, 5071 West Saanich Road, Victoria, BC V9E 2E7\\
$^{5}$Department of Physics and Astronomy, Howard University, Washington, DC 20059, USA\\
$^{6}$Center for Research and Exploration in Space Science and Technology, and X-ray Astrophysics Laboratory, NASA/GSFC, Greenbelt,\\
MD 20771, USA\\
$^{7}$D\'epartement de Physique et d’Astronomie, Universit\'e de Moncton, Moncton, N.B., Canada E1A 3E9\\
$^8$Department of Physics and Astronomy, Uppsala University, Box 516, Uppsala 75120, Sweden\\
$^9$Armagh Observatory and Planetarium, College Hill, Armagh BT61 9DG, UK\\
$^{10}$University of Western Ontario, London, Ontario, N6A 3K7, Canada\\
$^{11}$INAF - Osservatorio Astrofisico di Catania, Via S. Sofia 78, 95123 Catania, Italy\\
$^{12}$LESIA, Paris Observatory, PSL University, CNRS, Sorbonne Universit\'e, Universit\'e de Paris, 5 place Jules Janssen, 92195 Meudon\\
$^{13}$ESO - European Organisation for Astronomical Research in the Southern Hemisphere, Casilla 19001, Santiago 19, Chile\\
$^{14}$Department of Physics and Space Science, Royal Military College of Canada, Kingston, Ontario K7K 7B4, Canada\\
}
\begin{document}

\date{}

\pagerange{\pageref{firstpage}--\pageref{lastpage}} \pubyear{2002}

\maketitle

\label{firstpage}

\begin{abstract}
Numerous magnetic hot stars exhibit gyrosynchrotron radio emission. The source electrons were previously thought to be accelerated to relativistic velocities in the current sheet formed in the middle magnetosphere by the wind opening magnetic field lines. However, a lack of dependence of radio luminosity on the wind power, and a strong dependence on rotation, has recently challenged this paradigm. We have collected all radio measurements of magnetic early-type stars available in the literature. When constraints on the magnetic field and/or the rotational period are not available, we have determined these using previously unpublished spectropolarimetric and photometric data. The result is the largest sample of magnetic stars with radio observations that has yet been analyzed: 131 stars with rotational and magnetic constraints, of which 50 are radio-bright. We confirm an obvious dependence of gyrosynchrotron radiation on rotation, and furthermore find that accounting for rotation neatly separates stars with and without detected radio emission. There is a close correlation between H$\alpha$ emission strength and radio luminosity. These factors suggest that radio emission may be explained by the same mechanism responsible for H$\alpha$ emission from centrifugal magnetospheres, i.e.\ centrifugal breakout (CBO), however, whereas the H$\alpha$-emitting magnetosphere probes the cool plasma before breakout, radio emission is a consequence of electrons accelerated in centrifugally-driven magnetic reconnection. 
\end{abstract}


\section{Introduction}

Approximately 10\% of OBA stars possess magnetic fields \citep{2019MNRAS.483.2300S,2017MNRAS.465.2432G}, with properties that are remarkably consistent across the Hertzsprung-Russell diagram: they are strong  \citep[$10^2$ to $10^4$ G;][]{2019MNRAS.490..274S}; topologically simple \citep[i.e., with only a few exceptions, approximately dipolar;][]{2019A&A...621A..47K}; and, in all cases for which sufficient data is available for evaluation, stable over at least thousands of rotational cycles \citep[i.e.\ at least decades;][]{2018MNRAS.475.5144S}. Unlike stars with convective envelopes, for which surface magnetic field strength increases with rotation \citep{2014MNRAS.441.2361V,2016MNRAS.457..580F,2018MNRAS.474.4956F}, there is no such correlation with rotation for the magnetic fields of stars with radiative envelopes \citep{2019MNRAS.490..274S,2019MNRAS.483.3127S}. Instead, hot star magnetic fields decline in strength with age in a fashion consistent with conservation of magnetic flux in an expanding atmosphere \citep[for intermediate mass stars;][]{2019MNRAS.483.3127S} or gradual decay of magnetic flux \citep[for stars above about 5 \msun;][]{2007A&A...470..685L,2019MNRAS.490..274S,2016A&A...592A..84F}. These properties, together with the absence of a sustainable dynamo mechanism in radiative zones, has led to the interpretation of hot star magnetic fields as `fossils' left over from a previous epoch, a scenario supported by magnetohydrodynamic (MHD) calculations and simulations that have demonstrated the stability of fossil magnetic fields over evolutionary timescales as well as the ability of processes such as binary mergers to generate fossil fields \citep{2004Natur.431..819B,2009MNRAS.397..763B,2010ApJ...724L..34D,2019Natur.574..211S}.

Strong magnetic fields stabilize the atmospheres of hot stars, enabling various chemical elements to accumulate in long-lived surface patches via radiative diffusion \citep[e.g.][]{michaud1981,2015MNRAS.454.3143A,2019MNRAS.482.4519A}. This leads directly to modulation of the light curve on rotational timescales \citep[e.g.][]{2009A&A...499..567K,2012A&A...537A..14K,2015A&A...576A..82K}, making it straightforward to infer precise rotation periods from photometric time series \citep[e.g.][]{2001A&A...378..113R}. A key goal of the MOBSTER collaboration \citep[Magnetic OB(A) Stars with
TESS: probing their Evolutionary and Rotational properties;][]{2019MNRAS.487..304D} is to leverage space photometry from the {\em TESS} mission in order to dramatically expand the number of known rotational periods for magnetic chemically peculiar stars \citep[e.g.][]{2019MNRAS.487.4695S}, as a means of investigating the evolutionary and magnetospheric characteristics of this population.

The radiation-driven winds of hot stars serve as ion sources which feed their magnetospheres \citep{lb1978,bm1997,ud2002}. Hot star magnetospheres have a number of observable consequences. They were first detected by \cite{lb1978} via eclipsing of $\sigma$ Ori E by the dense plasma clouds of its magnetosphere. Ultraviolet observations demonstrated that the wind-sensitive resonance lines of magnetic hot stars exhibit clear rotational modulation indicating departures from spherical symmetry \citep[e.g.][]{2013A&A...555A..46H}. Optical and near-infrared H emission is also formed in the dense plasma of the magnetosphere \citep{petit2013,2015A&A...578A.112O}. Magnetically confined wind-shocks lead to X-ray emission \citep{2014ApJS..215...10N,ud2014}. Finally, a large fraction of magnetic hot stars show gyrosynchrotron radiation at high frequencies \citep[e.g.][]{1987ApJ...322..902D} and occasionally auroral radio emission at low frequencies \citep[e.g.][]{2000AA...362..281T,2018MNRAS.474L..61D}. 

With the exception of radio diagnostics, magnetospheric emission is believed to be formed within the inner magnetosphere, i.e.\ the magnetically dominated region within the Alfv\'en surface, in which the wind kinetic energy density is less than the magnetic energy density. By contrast, radio diagnostics are believed to be a consequence of activity in the middle magnetosphere, a region beyond the Alfv\'en radius in which magnetically enforced corotation of the plasma with the star breaks down, while the ram pressure of the winds opens the magnetic field lines, the combination of which leads to the formation of a current sheet. Inside the current sheet, electrons are accelerated to relativistic velocities, some of which then return to the star along magnetic fields lines, leading to gyrosynchrotron emission \citep{2004A&A...418..593T} and, for those that are caught in auroral circuits, electron-cyclotron maser emission \citep{2011ApJ...739L..10T,2016MNRAS.459.1159L,2020ApJ...900..156D}.

Rotation has emerged as a key parameter governing the structure of the inner magnetosphere. In the absence of rotation, inner magnetosphere plasma exists in dynamical equilibrium: flowing up along magnetic field lines, colliding at the magnetic equator, and then being pulled back to the star by gravity \citep{ud2002}. These dynamical magnetospheres are generally detectable in H$\alpha$ only for stars with high mass-loss rates \citep[i.e. O-type stars;][]{petit2013}. Due to corotation of the inner magnetosphere plasma, around rapid rotators centrifugal forces can prevent gravitational infall \citep{ud2008}. This leads to the formation of a centrifugal magnetosphere (CM) between the Kepler corotation radius (the equilibrium distance between the gravitational and centrifugal forces) and the Alfv\'en radius. Within the CM, plasma can accumulate to high enough densities for H$\alpha$ emission to be detectable even around stars with low mass-loss rates \citep[i.e.\ B-type stars;][]{petit2013,2019MNRAS.490..274S}. Rotational influence furthermore distorts the plasma distribution, such that (for a tilted dipole) it departs from a torus in the magnetic equator to two distinct clouds located at the intersections of the rotational and magnetic equatorial planes, as described by the Rigidly Rotating Magnetosphere \citep[RRM;][]{town2005c} model. In addition to the prototypical CM host star $\sigma$ Ori E \citep[e.g.][]{lb1978,2015MNRAS.451.2015O}, the variable H$\alpha$ profiles of a large number of CM host stars has been examined in detail and found to be phenomenologically consistent with the RRM model \citep[e.g.][]{2010MNRAS.401.2739L,2011AJ....141..169B,2012MNRAS.419.1610G,rivi2013,2015MNRAS.451.1928S,2016MNRAS.460.1811S,2021MNRAS.504.3203S}, with significant differences so far apparent only in the case of tidally locked binaries \citep{2018MNRAS.475..839S}.

The current understanding of radio magnetospheres assumes that the inner magnetosphere plasma makes no contribution to the current sheet \citep{2004A&A...418..593T}. Within this framework the only importance of the inner magnetosphere is absorption and diffraction of radio emission due to the denser plasma in this region, and the primary role of rotation is signal modulation due to the changing projection of a tilted dipole on the sky, and a reduced density in the inner magnetosphere due to centrifugal stress on the magnetic field. However, \cite{2020MNRAS.499.5379S} and \cite{2020MNRAS.499.5366O} have recently demonstrated that the H$\alpha$ emission properties of magnetic early B-type stars can only be explained if mass-balancing in the CM is accomplished by centrifugal breakout (CBO), rather than steady-state leakage mechanisms operating via a combination of diffusion and drift across magnetic field lines \citep{2018MNRAS.474.3090O}. This process, analogous to magnetotail reconnection in planetary magnetospheres, occurs when mass-loading by the wind drives the plasma density beyond the ability of the magnetic field to contain it, at which point the plasma is ejected outwards in a centrifugally driven reconnection process \citep{ud2006,ud2008}. In contrast to previous expectations that this should result in large-scale reorganization of the inner magnetosphere due to emptying of the plasma \citep[e.g.][]{town2013}, observations instead suggest that CBO events happen more or less continuously over small spatial scales, with the CM maintained at a constant state of near-breakout density \citep{2020MNRAS.499.5379S}.

Since plasma ejected by CBO must flow away from the star and, therefore, should pass through the middle magnetosphere, it is reasonable to ask whether there might be some connection between gyrosynchrotron emission and rotation. \cite{1992ApJ...393..341L} searched for just such a connection but were unable to find anything statistically significant. Since then the number of stars with precisely determined rotation periods has dramatically increased. A connection between rotation and gyrosynchrotron emission was suggested by \cite{2017MNRAS.465.2160K}, who did not detect radio emission from slow rotators; however, their small sample size prevented firm conclusions. \cite{2021MNRAS.507.1979L} have recently demonstrated a close connection between rotation and radio luminosity, suggesting that the wind-driven current sheet model advanced by \cite{2004A&A...418..593T} be abandoned in favour of a radiation belt model in which radio emission originates from a magnetic shell unrelated to the middle magnetospheric regions where the magnetic field lines are opened by the wind ram pressure. However, \cite{2021ApJ...921....9D} have recently reported the detection of correlated flux enhancements emanating via the electron cyclotron maser mechanism from auroral circuits above both magnetic poles of CU Vir, which they interpreted as a possible result of centrifugal breakout events in the inner magnetosphere injecting electrons into both magnetic hemispheres, suggesting that gyrosynchrotron emission may also be connected to CBO. 

In the current work we collect together all magnetic stars for which radio observations, magnetic data, and rotational periods have been obtained, both for radio-bright and radio-dim stars (i.e.\ stars from which radio emission respectively is and is not detected), in order to investigate the influence of rotation in gyrosynchrotron emission from hot star magnetospheres. Literature data are supplemented with unpublished magnetometry, photometry, and radio observations in order to provide the most comprehensive sample of radio emission from magnetic early-type stars that has been analyzed to date. In \S~\ref{sec:sample} the sample and observations are described, together with the determination of atmospheric, fundamental, rotational, and magnetic parameters. The parameter space distributions of radio-bright and -dim stars are examined in \S~\ref{sec:empirical}, together with comparison to H$\alpha$ emission, and analysis of correlations between radio luminosities and various parameters. The implications of these results are discussed in \S~\ref{sect:discussion}, and the conclusions are summarized in \S~\ref{sec:conclusions}. Stellar parameters are tabulated in Appendix \ref{append:summary_table}. The online appendices B, C, and D respectively provide the observation log of newly presented radio measurements, notes on individual stars for which new magnetic and rotational analyses are presented together with newly published magnetic data, and the tabulated radio flux density measurements for the individual stars.

\section{Sample}\label{sec:sample}

\begin{table}
\centering
\caption[]{Sources for radio observations.}
\label{radio_source_table}
\begin{tabular}{l r r}
\hline\hline
Source & Number of stars & Wavelength (cm) \\
\hline
\protect{\cite{1987ApJ...322..902D}} & 33 & 6 \\
\protect{\cite{1992ApJ...393..341L}} & 42 & 2, 3.6, 6, 20\\
\protect{\cite{1994A&A...283..908L}} & 40 & 6 \\
\protect{\cite{1996A&A...310..271L}} & 7 & 1.3, 2, 6, 20 \\
\protect{\cite{2004A&A...423.1095L}} & 11 & 0.3 \\
\protect{\cite{2006ESASP.604...73D}} & 19 & 6 \\
\protect{\cite{2015MNRAS.452.1245C}} & 9 & 20, 50 \\
\protect{\cite{2017ApJ...834..142K}} & 2 & 6 \\
\protect{\cite{2017MNRAS.465.2160K}} & 19 & 1, 3, 13 \\
\protect{\cite{2017MNRAS.467.2820L}} & 1 & 1, 2, 3 \\
\protect{\cite{2018MNRAS.476..562L}} & 1 & 1, 2, 3, 20 \\
\protect{\cite{2019ApJ...877..123D}} & 1 & 50 \\
\protect{\cite{2020MNRAS.493.4657L}} & 1 & 1, 2, 3, 6 \\
\protect{\cite{2020MNRAS.499L..72L}} & 1 & 2, 3, 6, 13, 20 \\
\protect{\cite{2021MNRAS.502.5438P}} & 5 & 20 \\
\protect{\cite{2021MNRAS.507.1979L}} & 1 & 3 \\
\protect{\cite{2021ApJ...921....9D}} & 1 & 50 \\
\protect\cite{2021arXiv210904043D} & 4 & 50 \\
Drake (priv. comm.) & 46 & 6 \\
This work & 19 & 20, 50 \\
\hline\hline
\end{tabular}
\end{table}

The sample started with all chemically peculiar or magnetic OBA stars which have been observed in at least one radio band. For Ap/Bp stars, we assume them to be magnetic even if magnetic data are not available, as chemical peculiarity of this type is invariably associated with strong surface magnetic fields. For magnetic OB stars (i.e.\ stars of spectral type B0 and hotter, in which strong winds inhibit the formation of surface chemical abundance spots), only those stars known to be magnetic via spectropolarimetric measurement of the Zeeman effect are included, as chemical peculiarity is not an indicator of magnetism at the top of the main sequence since stellar winds strip surface material faster than chemical abundance anomalies can accumulate. The sources consulted for radio data are summarized in Table \ref{radio_source_table}. In addition to literature measurements, we also include new radio measurements of 19 stars (see below). Note that there is considerable overlap in targets between the various surveys; across all papers, 192 unique targets were observed. 


Since some of the stars observed in the early surveys belong to non-magnetic classes (e.g., classical Be stars, HgMn stars), these stars (33 in total) were removed from the sample. After cross-referencing the catalogues and removing non-magnetic stars, 156 stars have at least one radio frequency observation, 50 of which are detected. These stars are listed in Table \ref{partab}, with the observed fluxes given online in Table D1. Where more than one observation is available at a given wavelength, the radio luminosity corresponds to the maximum observed flux density. 

\subsection{Stellar parameters}

We searched the literature for determinations of atmospheric parameters effective temperature \teff~and bolometric luminosity $\log{L_{\rm bol}}$, and projected rotational velocities \vsini. These are given together with references in Table \ref{partab}. When stellar parameters could not be found in existing compilations or single studies, they were determined photometrically. As a first step, the catalogue was cross-referenced with SIMBAD\footnote{\url{http://simbad.u-strasbg.fr/simbad/}}, in order to obtain spectral types and Johnson photometry. Distances were obtained from the Gaia early Data Release 3 Catalogue \citep{2021A&A...649A...6G}; in the few cases where these were not available, {\em Hipparcos} parallaxes \citep{vanleeuwen2007} generally were. Distances were calculated by inverting Gaussian parallax distributions, with the resulting asymmetric error bars propagated through to determinations of bolometric luminosity; however, in most cases the relative parallax errors are small enough (the median relative error is about 2\%) that the difference between positive and negative distance uncertainties is negligible. If Str\"omgren photometry is available \citep[using the catalogues provided by][]{1998A&AS..129..431H,2015A&A...580A..23P}, effective temperatures were determined with the {\sc idl} program {\sc uvbybeta}\footnote{\url{https://idlastro.gsfc.nasa.gov/ftp/pro/astro/uvbybeta.pro}} \citep[which uses the calibration determined by][]{1993A&A...268..653N}. If Str\"omgren photometry is not available, Johnson photometry was used to obtain \teff. All available de-reddened colours were compared to the empirical calibration provided by \cite{2011ApJS..193....1W}. Reddening was found using the {\em Stilism} three-dimensional tomographic dust map \citep{2014A&A...561A..91L,2017A&A...606A..65C,2018A&A...616A.132L} based on the positions and Gaia distances of the individual stars. While {\em Stilism} typically extends only out to around 1 kpc, the overwhelming majority of the sample stars are well within this distance; the few stars beyond this distance have stellar parameters available in the literature. Extinctions were determined with the usual reddening law ($A_V = 3.1E(B-V)$). For magnetic chemically peculiar (mCP) stars, the bolometric correction BC determined by \cite{2008AA...491..545N} for mCP stars was used to determine $L_{\rm bol}$. Since the \citeauthor{2008AA...491..545N} BC is only calibrated up to 19 kK, for mCP stars hotter than this limit a larger uncertainty was adopted following \cite{2019MNRAS.485.1508S}. For chemically normal stars, the \cite{nieva2013} BC was used.

We then searched the literature for determinations of rotational periods $P_{\rm rot}$ and magnetic oblique rotator model (ORM) parameters. In the simplest case of a tilted dipole (appropriate to first order for the vast majority of stars with fossil fields), an ORM consists of an inclination $i$ of the rotational axis from the line of sight, an obliquity angle $\beta$ of the magnetic axis from the rotational axis, and a polar surface strength $B_{\rm d}$ of the magnetic dipole at the stellar surface. In the simplest case of a tilted dipole, appropriate to the vast majority of stars \citep[e.g.][]{2019A&A...621A..47K}, the rotation of the star will lead to a sinusoidal variation in the longitudinal, or line-of-sight, magnetic field \bz~averaged over the stellar disk. If $P_{\rm rot}$ is known, the \bz~curve can then be used to obtain the ORM parameters \citep{preston1967}, however there is a degeneracy between the angular parameters $i$ and $\beta$. Breaking this degeneracy requires knowledge of \vsini~and the stellar radius $R_*$. 


Where ORM parameters were not already available, we searched for longitudinal magnetic field measurements \bz~with which to determine them. ORM parameters were determined simultaneously with fundamental, rotational, and magnetospheric parameters using the Monte Carlo Hertzsprung-Russell diagram (MCHRD) sampler described by \cite{2019MNRAS.490..274S}. The MCHRD sampler combines all available measurements with evolutionary models in order to infer self-consistent fundamental, ORM, wind, and magnetospheric parameters, automatically accounting for correlated error bars. In this case we utilized the rotating or non-rotating Geneva evolutionary models calculated by \cite{ekstrom2012}, as appropriate for a given stellar rotational period (non-rotating models were used if $P_{\rm rot} > 10$~d). In some cases, ORM parameters have been revised to those obtained from the MCHRD sampler, in order to ensure methodological consistency across the full sample; it is these values which are reported in Table \ref{append:summary_table}.

\subsection{Radio observations}

\subsubsection{VLA}

We report previously unpublished 6 cm observations of 46 stars acquired at the Very Large Array (VLA). The data were acquired in 1992 and 1994 in the context of the survey presented by \cite{1987ApJ...322..902D,2006ESASP.604...73D} and \cite{1992ApJ...393..341L}, and were reduced and analyzed following the procedures described in those works. They were provided by Drake (priv. comm.). All 46 observations are non-detections. One of the stars in this sample, HD\,118022, was reanalyzed by \cite{2021MNRAS.507.1979L} and found to be a detection. 

\subsubsection{uGMRT}

We report new 20 and 50 cm radio observations of 19 magnetic hot stars, including 4 new detections (HD\,11503, HD\,64740, HD\,189775, and HD\,200775). These data were acquired with the upgraded Giant Metrewave Radio Telescope (uGMRT), located at Pune, India. The uGMRT is a radio interferometer consisting of 30 antennae, and operates over the frequency range of 120--1450 MHz divided into four bands. Our observation frequency corresponds to bands 4 (550--900 MHz) and 5 (1050--1450 MHz). For each observation, we observed a set of calibrators in order to calibrate the absolute flux density scale and the bandpass (flux calibrator), and the time-dependent antenna gains (phase calibrator). The details of these observations, including the calibrators used, are provided online in Table B1. The data were analyzed using the Common Astronomy Software Applications \citep[{\sc casa},][]{mcmullin2007} following the procedure described in \citet{2019ApJ...877..123D,2019MNRAS.489L.102D}. 

Nine stars were observed in the context of the GMRT legacy survey. Ten stars, indicated in Table D1, were acquired in the context of an ongoing uGMRT survey aiming to detect and characterize auroral radio pulses emitted via the electron cyclotron maser mechanism \citep[ECM;][]{2018MNRAS.474L..61D,2019MNRAS.489L.102D,2019ApJ...877..123D,2021ApJ...921....9D,2021arXiv210904043D}. These pulses occur at or near \bz~nulls (i.e. at phases corresponding to the magnetic equator bisecting the stellar disk) since they are emitted tangent to the auroral circuits above the magnetic poles \citep[e.g.][]{2000A&A...362..281T,2011ApJ...739L..10T,2016MNRAS.459.1159L,2020ApJ...900..156D}. For this reason, observations were acquired close to magnetic nulls, and care is required to ensure that the adopted flux density reflects basal gyrosynchrotron emission rather than the much stronger ECM pulse. For 5 additional stars for which phase coverage was insufficient to cover the basal flux density level, uGMRT data were not included. It should be noted that, since gyrosynchrotron emission is rotationally modulated and, unlike ECM pulses, is at a minimum rather than a maximum at magnetic nulls \citep[e.g.][]{2017MNRAS.467.2820L,2018MNRAS.476..562L}, there is the possibility that these data systematically under-estimate the peak 50 cm flux densities of these targets. However, in most cases when observations at other wavelengths are available, the measurements are comparable, consistent with expectations that the radio spectrum is approximately flat and that rotational modulation of the flux density is generally only a factor of a few \citep[e.g.][]{2004A&A...418..593T,leto2012,2017MNRAS.467.2820L,2018MNRAS.476..562L,2020MNRAS.493.4657L}.

\subsection{Spectropolarimetric and photometric observations}

When neither ORM parameters nor published \bz~measurements were available, or when rotation periods were unknown, we utilized both public and private archives of spectropolarimetric and space photometric data with which to constrain magnetic and rotational properties. These were then used in conjunction with stellar parameters and the MCHRD sampler to infer ORM models as described above. The data used for this analysis are described in detail in Appendix C. In total, we provide new magnetic data for 30 stars, of which magnetic fields were detected in 16, and utilized magnetic and/or photometric data to evaluate rotational periods for 59 stars, of which we refined the published periods of 14 stars and determined new periods for 16 stars. In some cases (HD\,36629, HD\,37041, HD\,49606, and HD\,89822), these analyses also led to the rejection of published rotational periods and magnetic data as spurious results arising from noisy data; these stars were removed from the sample.

\subsubsection{Dominion Astrophysical Observatory spectropolarimetry}

The dimaPol spectropolarimeter is a medium-resolution ($\lambda/\Delta\lambda \sim 10,000$) instrument covering the 25 nm region centred on the laboratory wavelength of the H$\beta$ line. It is mounted on the 1.8 m Dominion Astrophysical Observatory (DAO) Plaskett Telescope. The instrument and reduction pipeline are described in detail by \cite{2012PASP..124..329M}. Magnetic measurements are obtained primarily using the wings of H$\beta$ and are therefore fairly insensitive to either \vsini~or surface chemical abundance patches \citep[e.g.][]{1977ApJ...212..141B}.

Unpublished DAO measurements are available for 20 stars in the sample, although in some cases no magnetic field can be detected at the available precision (generally hundreds of G). Of the 12 stars for which a magnetic field can certainly be detected and good constraints do not already exist, 217 individual measurements are available, with a median of 18 measurements per star. These data are analyzed in detail in Appendix C, and the measurements are available as supplementary material through Vizier.

\subsubsection{ESPaDOnS and Narval spectropolarimetry}

ESPaDOnS (\'Echelle SPectropolarimetric Device for the Observations of Stars) and Narval are identical high-resolution ($\lambda/\Delta\lambda$) spectropolarimeters respectively mounted at the 3.6 m Canada-France-Hawaii Telescope (CFHT) and the 2 m Bernard Lyot Telescope (TBL). They cover a wavelength range of approximately 370 nm to 1050 nm across 40 overlapping spectral orders. Each observation consists of 4 differently polarized subexposures, yielding four unpolarized (Stokes $I$) spectra, one circularly polarized (Stokes $V$) spectrum, and two diagnostic null ($N$) spectra with which to check for normal instrument operation and determination of noise. The characteristics of the instruments and data reduction were described in detail by \cite{2016MNRAS.456....2W}. 

We queried the PolarBase database of Narval and ESPaDOnS spectropolarimetry for unpublished spectropolarimetric measurements \citep{2014PASP..126..469P}. These were found for 20 stars (overlapping with the DAO dataset). Magnetic fields were detected via the multiline least-squares deconvolution \citep[LSD;][]{d1997,koch2010} method in 6 stars. The magnetic analysis of these measurements is described in Appendix C.

\subsubsection{Space photometry}

The surface abundance spots of magnetic chemically peculiar stars lead to photometric variability that can be used to infer their rotational periods. We searched public archives (the {\em Hipparcos} archive and MAST, the Mikulsi Archive for Space Telscopes) for the light curves from the High precision parallax collecting satellite ({\em Hipparcos}), {\em Kepler}, and Transiting Exoplanet Survey Satellite ({\em TESS}) space telescopes. These light curves are provided in Appendix~C. Period analysis was performed using the Lomb-Scargle program {\sc period04} \citep{2005CoAst.146...53L}. This was accomplished by identifying the lowest-frequency term in a harmonic series, fixing higher harmonics to whole number multiples of the rotational harmonic, and then optimizing the phases and amplitudes of the terms to minimize residuals, as is standard practice for the strictly periodic rotational variability of mCP stars \citep[e.g.][]{2019MNRAS.487..304D,2019MNRAS.487.4695S}.

{\em Hipparcos} was an astrometric space telescope, whose mission lasted from 1989 to 1993. While the primary aim was to obtain high-precision trigonometric parallaxes, it also obtained time series photometry for a large number of stars \citep{perry1997,vanleeuwen2007}, which is available for 12 stars without published rotation periods. 

The NASA {\em Kepler} satellite was a $\mu$mag-precision space photometer with a 110 square degree field of view operating in the 400 to 850 nm bandpass, intended for high-cadence, long-duration observations with the goal of detecting transiting exoplanets \citep{2010Sci...327..977B}. The {\em K2} mission was an extension of the original {\em Kepler} mission, following the failure of two of the satellite's reaction wheels; by utilizing pressure from the solar wind, the satellite could be stabilized on a given field of view for about 3 months, enabling it to observe fields along the ecliptic \citep{2014PASP..126..398H}. A {\em K2} light curve is available for 1 star. 

{\em TESS} uses four cameras with a total field of view of $24^\circ \times 96^\circ$, with a bandpass covering 600 to 1050 nm \citep{2015JATIS...1a4003R}. The initial two-year TESS mission began in 2018, during which it completed coverage of almost the entire sky. During each year, 13 sectors were observed for 27 days each, with a nominal precision of 60 ppm hr$^{-1}$ (although this varies between fields and targets). High-priority targets are observed with a two-minute cadence, and the processed light curves made available on the MAST archive immediately following reduction. Two-minute cadence TESS data are available for 9 stars. In other cases, we used the 30-minute cadence data extracted from Full Frame Images, obtained from MAST when available or, for 9 stars for which this was not the case, extracted ourselves. In total we utilized {\em TESS} data for 46 stars.

\subsection{Final sample}

In the end, magnetic data are available for 142 stars, rotational periods for 138 stars, and both for 131 stars, of which 50 have detected radio emission (note that these numbers do not include the magnetic O-type stars, which are dropped from the analysis for reasons explained below in Sect. \ref{subsec:hrd}.). Dipolar magnetic field strengths and rotation periods are given together with references in Table \ref{partab}, along with all quantities necessary to calculate the various parameters examined in the subsequent analysis. In the cases in which ORM parameters were determined here using published \bz~measurements, the references to the measurements are also included. 

   \begin{figure}
   \centering
   \includegraphics[width=9.cm]{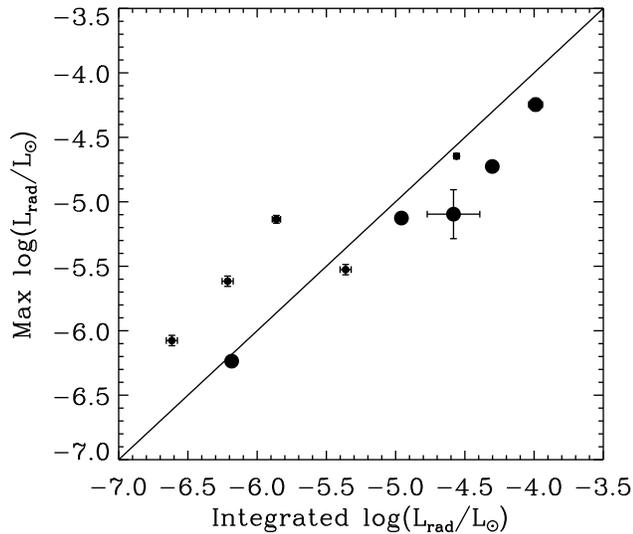}
      \caption[]{Radio luminosities inferred from the maximum flux density vs.\ radio luminosities obtained by integrating flux density across the full frequency range, for those stars with observations sampling at least 4 frequencies. Symbol size is proportional to number of observations (either 4, small, or 5, large).}
         \label{lradmax_lradint}
   \end{figure}

   \begin{figure}
   \centering
   \includegraphics[width=9.cm]{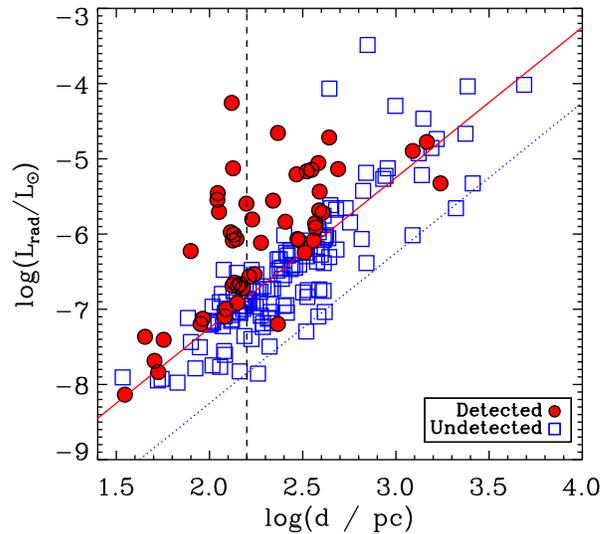}
      \caption[]{Radio luminosity as a function of a distance. Solid and dotted lines indicate distance-dependent detection limits as defined by the lower bounds of detected and non-detected stars, respectively. The vertical dashed line indicates the distance beyond which the observed lower detection limit begins to rise with increasing distance. Non-detections are upper limits.}
         \label{radio_distance_flux_luminosity}
   \end{figure}

   \begin{figure*}
   \centering
   \includegraphics[width=\textwidth]{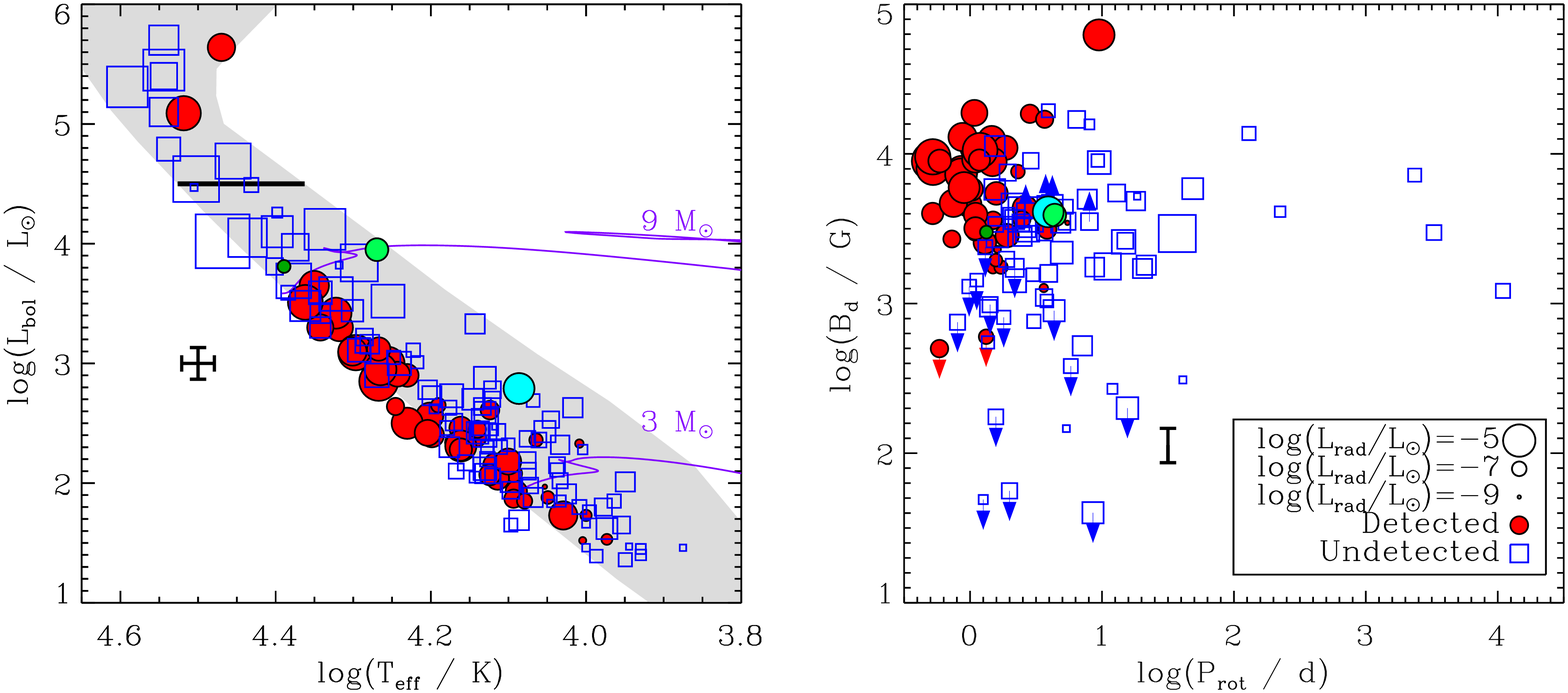}
   \includegraphics[width=\textwidth]{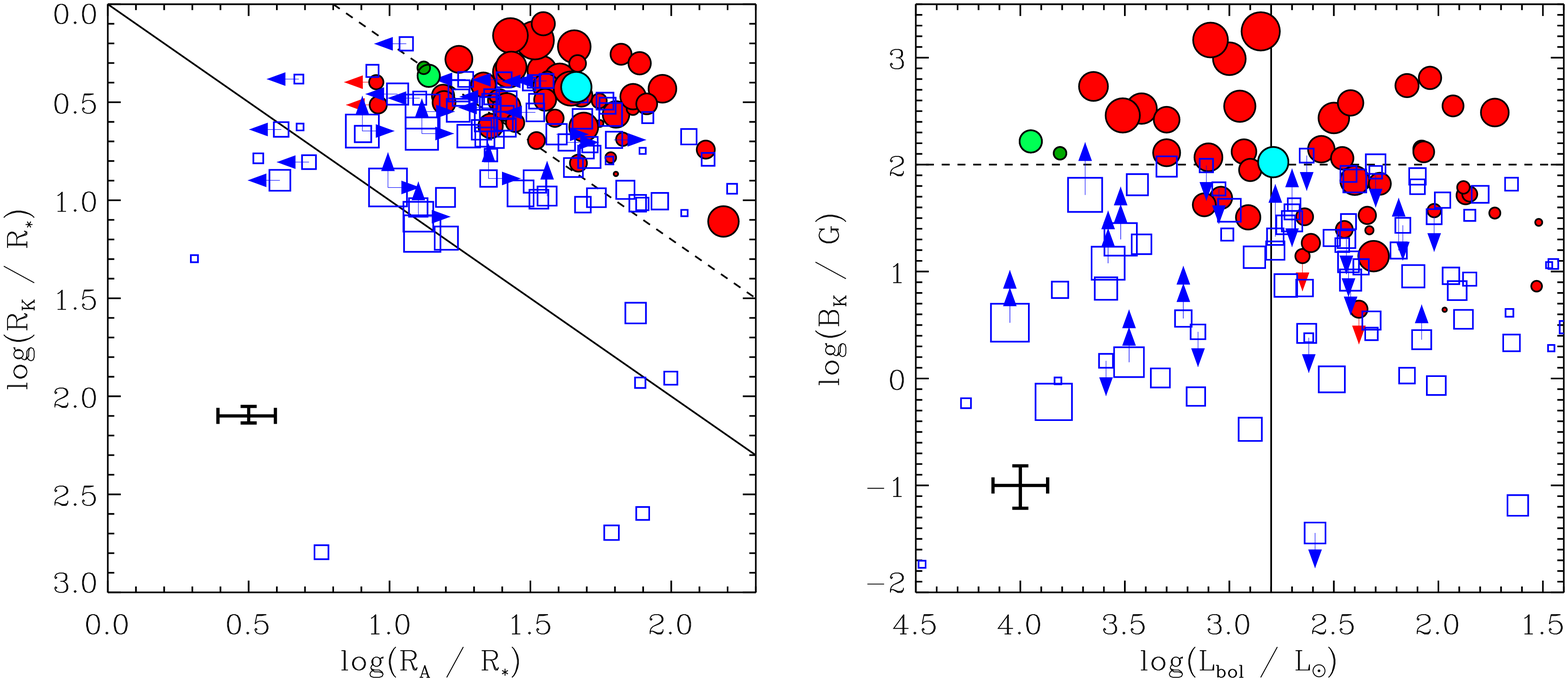}
      \caption[]{Parameter space distribution of the sample. Filled red circles indicate radio-bright stars, open blue squares radio-dim stars, and symbol size is proportional to radio luminosity (or its upper limit). Mean uncertainties are indicated by error bars. Filled dark green, light blue, and light green circles highlight HD\,64740, HD\,171247, and HD\,200775 respectively (discussed in the text). {\em Top Left}: Hertzsprung-Russell diagram showing all magnetic stars with radio observations. The grey shaded region indicates the main sequence. The thick line indicates the empirical bolometric luminosity cutoff applied to the subsequent analysis. {\em Top right}: the sample on the $\log{B_{\rm d}} - \log{P_{\rm rot}}$ plane. {\em Bottom left}: the sample on the rotation-magnetic wind confinement diagram. The solid line indicates $R_{\rm A} = R_{\rm K}$: points below have dynamical magnetospheres only, points above possess centrifugal magnetospheres. The dashed line shows $\log{R_{\rm A}/R_{\rm K}} = 0.8$, the approximate minimum threshold for H$\alpha$ emission. {\em Bottom right}: the $\log{B_{\rm K}} - \log{L_{\rm bol}}$ plane. The minimum value of $\log{B_{\rm K}}$ extends to about $-6$; none of the stars not shown are detected in radio. The dashed line indicates the approximate $B_{\rm K}$ threshold for H$\alpha$ emission, while the solid line indicates the lower luminosity limit for H$\alpha$.}
         \label{radio_hrd_prot_bd}
   \end{figure*}

Radio luminosities were determined using parallax distances. When multiple radio measurements are available, the highest flux density measurement was chosen as representative of the radio luminosity of the star. When they have been measured, the spectral indices of radio emission from magnetic hot stars are approximately flat between 1 and 100 GHz \citep[as has been shown by][for the largest sample to date of stars with a sufficient number of multifrequency observations to perform this analysis]{2021MNRAS.507.1979L}, and the difference between measurements at different frequencies for a given star in the present sample is in general small. It is therefore likely that radio luminosities can be estimated with reasonable accuracy from single observations at a single frequency (which are all that are available for much of the sample). Following this, radio luminosity was determined by integrating a trapezoidal function between between 600 MHz (50 cm) and 100 GHz (0.3 cm), with values of unity between 1.5 GHz (20 cm) and 30 GHz (1 cm), and zero at the extrema. This was then scaled by the peak specific intensity measured across all observations (when more than one observation is available). Integrating with values at unity at all wavelengths, or only integrating between 1.5 GHz and 30 GHz, were also tried; however, the trapezoidal approximation gives the closest agreement with radio luminosities acquired for stars with observations at 4 or more wavelengths. In the end, 0.3 cm measurements were discarded as likely outliers due to significant discrepancies between these and observations at other wavelengths for the same stars; only 2 stars are detected at 0.3 cm, and in both cases the stars were also detected at other wavelengths, therefore this does not affect the detection statistics. While this is a less-than-perfect approximation of the actual spectral energy distributions (SEDs) of the sample stars, in the absence of multi-wavelength measurements constraining the variation of SEDs across stellar parameters it is not yet possible to adopt a more sophisticated approach. Furthermore, rotational modulation of the signal and the reliance on snapshot observations makes it likely that the maximum flux density is under-estimated for much of the sample, for which this trapezoidal function approach may partially compensate given that it may over-estimate the radio luminosity by failing to account for departures from perfectly flat spectral indices. As a check on this approximation, Fig.\ \ref{lradmax_lradint} shows the radio luminosity approximated from the maximum flux density, vs.\ the radio luminosity measured via integration of measured flux densities across the same frequency range, for those stars with observations sampling at least 4 frequencies. While there are outliers by up to about 1 dex, there is generally a good correlation between the two quantities, suggesting this approach is a reasonable approximation of the actual radio luminosities of the sample.


As can be seen in Fig.\ \ref{radio_distance_flux_luminosity}, radio luminosity varies over about 4 orders of magnitude. While radio emission is rotationally modulated, the amplitude of this modulation is a factor of a few \citep[e.g.][]{2004A&A...418..593T,leto2012,2017MNRAS.467.2820L,2018MNRAS.476..562L}, i.e.\ much smaller than the differences between individual stars in the sample. That radio observations sampling the entire phase curve are in general unavailable, and that the true peak luminosity is therefore unknown, is unimportant at the level of the full population. 

Another consideration that is apparent from Fig.\ \ref{radio_distance_flux_luminosity} is that the detection limit is a function of distance. However, below a distance of $\log{(d/{\rm pc})} = 2.2$, the lower detection limit is fairly constant, with radio non-detections being comparable in luminosity to the weakest radio detections. Beyond this distance, it is more likely that radio-dim stars would have been detected if they were closer; below it, this scenario is less likely. This nearby sub-sample is therefore in a sense more complete than the full sample, and can be used to test conclusions derived from the full sample of stars.

\section{Parameter study}\label{sec:empirical}

We begin our analysis by examining the distributions of radio-bright and radio-dim stars in atmospheric, magnetic, rotational, and magnetospheric parameter space, examining the effectiveness of each parameter in separating the two populations, as well as the strength of the correlation between radio luminosity and a given parameter. 

\subsection{Hertzsprung-Russell diagram}\label{subsec:hrd}

The top left panel of Fig.\ \ref{radio_hrd_prot_bd} shows all magnetic stars with radio observations on the Hertzsprung-Russell diagram (HRD), where we have shown the non-rotating evolutionary models calculated with the solar metallicity Geneva evolutionary code by \cite{ekstrom2012}. Most radio-bright stars are between about 3 and 9 \msun, and are generally close to the Zero-Age Main Sequence (ZAMS). They are relatively evenly distributed within this mass range, with no obvious tendency to cluster at high luminosities, consistent with the finding from \cite{2021MNRAS.507.1979L} that gyrosynchrotron emission is more or less independent of the wind power. There are two stars which are very obviously not near the ZAMS, highlighted in Fig.\ \ref{radio_hrd_prot_bd}. These are HD\,200775, which is a magnetic Herbig Be star \citep{2008MNRAS.385..391A}, and HD\,171247, which is examined in further detail below.

As discussd by \cite{2015MNRAS.452.1245C}, the strong winds of O-type stars lead to radio photospheres that are, in general, much larger than their Alfv\'en radii, and swallow any gyrosynchrotron emission that might be produced. Thermal radio emission from O-type stars can be produced by their winds \citep[e.g.][]{1989ApJ...340..518B,1993ApJ...412..771L}, and while this can in principle be rotationally modulated due to symmetry-breaking in the presence of a magnetic field \citep{2019MNRAS.489.3251D}, this is unrelated to the gyrosynchrotron emission of interest here. Furthermore, non-thermal synchrotron emission can be produced in the colliding wind shocks of close binaries \citep[e.g.][]{2006A&A...446.1001P,2010A&A...519A.111B}. Only two O-type stars are detected in the sample \citep{2017MNRAS.465.2160K}, these being $\zeta$ Ori A\footnote{This system is actually a spectroscopic binary, in which the Aa component is magnetic \citep{2013A&A...554A..52H,2015A&A...582A.110B}. However, given the long 7.3~yr orbit, the Aa and Ab components are not interacting, and the radio emission is dominated by the effectively single wind of the Aa component.} (which has a thermal radio spectrum) and Plaskett's Star \citep[a spectroscopic colliding wind binary;][]{2008A&A...489..713L}. O-type stars were therefore excluded from the sample, as indicated by the horizontal thick bar in Fig.\ \ref{radio_hrd_prot_bd}. This removed 11 stars from the sample. 

\subsection{Rotation and magnetic field strength}

The top right panel of Fig.\ \ref{radio_hrd_prot_bd} shows the sample on the $\log{B_{\rm d}} - \log{P_{\rm rot}}$ plane. The period axis is truncated for clarity, omitting three stars with periods on the order of several years, none of which are detected in the radio. Notably, all radio-bright stars are both strongly magnetic (as expected) and rapidly rotating ($P_{\rm rot} \lesssim 5$~d, with Babcock's Star, HD\,215441, the only exception -- a `slow' rotator with a period of about 10~d). There is some indication in Fig.\ \ref{radio_hrd_prot_bd} that the stronger the magnetic field, the slower the rotation can be while still producing detectable radio emission.

Comparing radio-dim and radio-bright stars, their rotational and magnetic properties are clearly different. The mean rotational period and surface magnetic dipole strengths of the radio emitters are $\log{(P_{\rm rot / {\rm d})} = 0.14^{+0.23}_{-0.14}}$ and $\log{(B_{\rm d} / {\rm G})} = 3.70^{+0.23}_{-0.25}$, while the corresponding means for the radio-dim stars are $\log{(P_{\rm rot / {\rm d})} = 0.81^{+0.87}_{-0.22}}$ and $\log{(B_{\rm d} / {\rm G})} = 3.31^{+0.24}_{-0.49}$, where the error bars correspond to standard deviations above and below the mean value. Notably, radio emission is not detected in any star with $\log{P_{\rm rot} > 1}$, regardless of magnetic field strength.

\subsection{The rotation-magnetic wind confinement diagram}

The bottom left panel of Fig.\ \ref{radio_hrd_prot_bd} shows the sample on the rotation-magnetic wind confinement diagram \citep[introduced as a fundamental plane of magnetospheres by][]{petit2013}. The vertical axis shows the Kepler corotation radius $R_{\rm K}/R_* = W^{-2/3}$, where the critical rotation parameter $W$ is given by the ratio of the equatorial velocity $v_{\rm eq}$ to the orbital velocity $v_{\rm orb}$ necessary to maintain a Keplerian orbit at the stellar equator \citep{ud2008}:

\begin{equation}\label{wrot}
W = \frac{v_{\rm rot}}{v_{\rm orb}} = \frac{2\pi R_*}{P_{\rm rot}}\left(\frac{GM_*}{R_*}\right)^{-1/2},
\end{equation}

\noindent where $R_*$ and $M_*$ are the stellar radius and mass, and $G$ is the gravitational constant. The Kepler radius corresponds to the distance from the star at which gravity and the centrifugal force due to magnetically enforced corotation are in balance, and therefore decreases with increasing rotational velocity. 

The horizontal axis of the bottom left panel of Fig.\ \ref{radio_hrd_prot_bd} shows the Alfv\'en radius \ra, i.e.\ the distance from the star at which the wind ram pressure and magnetic pressure equalize. The Alfv\'en radius was determined from the wind magnetic confinement parameter $\eta_*$ as $R_{\rm A}/R_* = 0.3 + (\eta_* + 0.25)^{1/4}$, where $\eta_*$ is the ratio of the magnetic energy to the wind kinetic energy given by \cite{ud2002}:

\begin{equation}\label{etastar}
\eta_* = \frac{B_{\rm eq}^2R_*^2}{\dot{M}v_\infty},
\end{equation}

\noindent with $B_{\rm eq} = B_{\rm d}/2$ the surface magnetic field strength at the magnetic equator, \mdot~the mass-loss rate in the absence of a magnetic field (i.e., the surface mass flux), and \vinf~the wind terminal velocity. For mass-loss we adopted the usual \citet*[][CAK]{cak1975} scaling formula,

\begin{equation}
{\dot M} = \frac{\alpha}{1-\alpha} \, \frac{L_{\rm bol}}{c^2} \, \left (\frac{{\bar Q} \Gamma_e}{1-\Gamma_e} \right  )^{-1+1/\alpha} \sim L_{\rm bol}^{1/\alpha}
\, ,
\label{eq:Mdcak}
\end{equation}

\noindent where ${\bar Q} \approx 1000$ \citep{gayley1995}, $c$ is the speed of light, and the electron Eddington parameter scales as $\Gamma_e = \kappa_e L_{\rm bol}/(4 \pi G M_\ast c)$ for electron opacity $\kappa_e$. The effective CAK exponent can range from $\alpha \approx 1/2$ to $\alpha \approx 2/3$, with $\alpha \approx 0.55$ applicable for the magnetic B-stars considered here \citep[see e.g.][]{petit2013}. We used CAK mass-loss in preference to the B-star mass-loss rates developed by \cite{krticka2014} because the latter are effectively zero for stars below about 14 kK for the default solar metallicity\footnote{While essentially all of these stars are chemically peculiar, detailed mean surface abundances are not generally available.}. Wind terminal velocities were scaled with the escape speed $v_{\rm esc}$ 

\begin{equation}\label{vesc}
v_{\rm esc} = \left ( \frac{2GM_*(1 - \Gamma_e)}{R*} \right )^{1/2},
\end{equation}

\noindent where we adopted a scaling factor $f$, such that $v_\infty = fv_{\rm esc}$, where $f=1.3$ and 2.6 on either side of the bistability jump at 25 kK \citep{vink2000,vink2001}. We did not, however, adopt an abrupt change in \mdot~across the bistability jump as, in contrast to the change in \vinf~which is observationally motivated \citep{lamers1995}, the predicted change in \mdot~has not been confirmed \citep{mark2008a}.


If $R_{\rm K} > R_{\rm A}$ the inner magnetosphere is purely dynamical, meaning that rotation plays no role; no stars in this regime show radio emission. When $R_{\rm K} < R_{\rm A}$ the inner magnetosphere forms a centrifugal component. The dashed line indicates $\log{(R_{\rm A}/R_{\rm K})} = 0.8$, the approximate threshold for H$\alpha$ emission \citep{petit2013,2019MNRAS.490..274S}. Essentially all of the radio-bright stars are above this threshold, once again indicating that rotation plays a crucial role. It is also noteworthy that radio and H$\alpha$ emission occur in the same part of the rotation-magnetic confinement diagram. Furthermore, while there are relatively few stars in the DM-only regime with radio observations, there are numerous stars in the small-CM regime ($0 < \log{(R_{\rm A}/R_{\rm K})} < 0.8$), almost all of which are undetected in the radio (with the two detected stars having limiting values of $B_{\rm d}$). Since it seems to be necessary for a star to have a large CM for it to display gyrosynchrotron emission, it also seems unlikely that additional observations will detect DM stars with non-thermal radio (although this should naturally be verified in the future).


\subsection{Magnetic field at the Kepler radius}

\cite{2020MNRAS.499.5379S} showed that H$\alpha$ emission is regulated directly by the strength of the magnetic field at the Kepler radius in the magnetic equatorial plane, which for a dipole is $B_{\rm K} = B_{\rm eq} / R_{\rm K}^3$, for \rk~in units of stellar radii. H$\alpha$ emission appears only in stars with $B_{\rm K} \sim 100$~G. As demonstrated by \cite{2020MNRAS.499.5366O}, this is the magnetic field strength necessary for the plasma density at $R_{\rm K}$ to reach an H$\alpha$ optical depth of unity, under the assumption that mass balancing is governed by centrifugal breakout. 

The bottom right panel of Fig.\ \ref{radio_hrd_prot_bd} shows the sample on the $\log{B_{\rm K}} - \log{L_{\rm bol}}$ plane \citep[compare to the right panel of Fig.\ 3 in][]{2020MNRAS.499.5379S}. The dashed line indicates the H$\alpha$ emission threshold; essentially all stars above this threshold are radio-bright. The vertical line indicates the low-luminosity cutoff for H$\alpha$ emission; notably, radio emission extends to lower luminosities, including essentially the entire B-type spectral sequence. Gyrosynchrotron emission is also seen at lower values of $B_{\rm K}$ than those at which H$\alpha$ can be detected, down to about 10 G. 

\subsection{Evolution of radio luminosity}

   \begin{figure}
   \centering
   \includegraphics[width=0.5\textwidth]{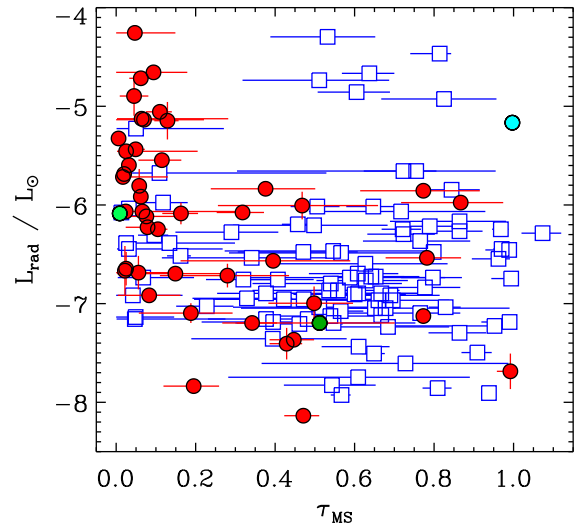}
      \caption[]{Radio luminosity as a function of fractional main sequence age $\tau_{\rm MS}$. HD\,64740, HD\,171247, and HD\,200775 are highlighted as in Fig.\ \ref{radio_hrd_prot_bd}.}
         \label{radio_ttams}
   \end{figure}

As is apparent from the HRD in Fig.\ \ref{radio_hrd_prot_bd}, the majority of radio-bright stars are found close to the ZAMS. Fig.\ \ref{radio_ttams} shows radio luminosity as a function of fractional main sequence age $\tau_{\rm MS}$, and demonstrates that radio luminosity drops precipitously by about 2 dex beyond a fractional main-sequence age of $\tau_{\rm MS} \sim 0.2$. The stars with the weakest radio emission are furthermore found in the second half of the main sequence. This is just as would be expected if radio emission is tied to rotation, since magnetic braking rapidly removes angular momentum \citep{ud2009,2019MNRAS.485.5843K,2020MNRAS.493..518K}. A similar phenomenon has been seen in the H$\alpha$ magnetospheres of early B-type stars: emission is found only in young stars \citep{2019MNRAS.490..274S}, and drops in strength steeply with age \citep{2020MNRAS.499.5379S}.

The one exception to this trend is HD\,171247, highlighted in Figs.\ \ref{radio_hrd_prot_bd} and \ref{radio_ttams} with a filled light blue circle. This is a somewhat curious object as its radio luminosity is relatively high ($\log{L_{\rm rad}} = -5.16 \pm 0.02$) despite being a relatively slow rotator ($P_{\rm rot} = 3.9$~d) with a surface magnetic field of average strength ($B_{\rm d} \sim 4.1$~kG). Furthermore, in contrast to the general case in which radio-bright stars are found close to the ZAMS, HD\,171247 is apparently a fairly evolved object very near to the terminal age main sequence. As described in Appendix C, there is considerable uncertainty regarding HD\,171247's rotational period, as strikingly different values (about 1 d vs.\ 4 d) are found from \bz~and photometry. 

It is possible that HD\,171247 is affected by some other factor. For example, an undetected binary companion might lead to an overestimated bolometric luminosity or, in the case of an interacting system, enhance its radio luminosity; however, there is nothing particularly strange about the measurements from the well-studied binary systems HD\,36485 or HD\,37017 \citep{2010MNRAS.401.2739L,1998AA...337..183B}, and there is furthermore no indication of asymmetry or radial velocity variability in the available DAO spectra. The star does, however, have a substellar companion of approximately 46 Jupiter masses at a separation of about 2 AU, detected via {\em Gaia} astrometry \citep{2019A&A...623A..72K}; if the companion is also magnetic, it may be an additional source of radio emission. Alternatively, its reported radio flux density measurement might have been obtained at a rotational phase corresponding to an auroral radio emission pulse, which can result in substantial enhancements over the basal flux \citep[while its 6 cm observations are not in the usual wavelength regime for this phenomenon, which is predominantly seen at longer wavelengths, ECM was detected at this wavelength by][in the case of HD\,124224]{2021ApJ...921....9D}. Given HD\,171247's anomalous position on the HRD, and the uncertainty in its rotational period, this object was removed from the subsequent analysis as likely suffering from one or more systematic errors.




\subsection{Wind absorption}\label{subsec:wind}

   \begin{figure}
   \centering
   \includegraphics[width=0.45\textwidth]{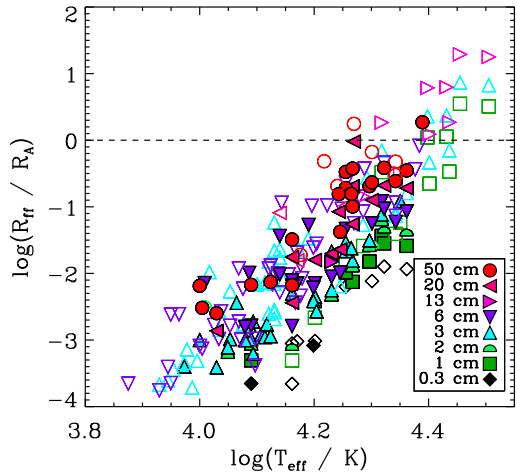}
      \caption[]{Ratio between the radius of the free-free emission photosphere $R_{\rm ff}$ and the Alfv\'en radius \ra~as a function of \teff. Symbol colour and type indicates wavelength; filled symbols correspond to radio-bright stars. Above the dashed line, the radio photosphere is larger than \ra~for the given frequency. Only one radio-bright star, HD\,64740, is above the dashed line.}
         \label{rff_teff}
   \end{figure}

To determine to what degree the remaining sample might still be affected by wind absorption, following \cite{2015MNRAS.452.1245C} we calculated the ratio between the radius of free-free emission $R_{\rm ff}$ and \ra, where $R_{\rm ff}$ gives the extent of the radio photosphere at a given frequency. If $R_{\rm ff} > R_{\rm A}$, it is likely that gyrosynchrotron emission will be absorbed by the wind, and any radio emission detected from the source will arise from free-free emission in the wind. Fig.\ \ref{rff_teff} shows $R_{\rm ff}/R_{\rm A}$ as a function of \teff. Since $R_{\rm ff}$ is a strong function of wavelength, this analysis was done for observations at specific wavelengths rather than integrated values. As can be seen in Fig.\ \ref{rff_teff}, for all but one radio-bright star $R_{\rm ff} \ll R_{\rm A}$. The sole exception is HD\,64740, which is the hottest and most luminous of the radio-bright stars remaining in the sample after removing the O-type stars, and the only radio-bright B-type star with a mass above 9~\msun. This star is highlighted in Fig.\ \ref{radio_hrd_prot_bd} by a small green circle. HD\,64740 has a relatively low radio luminosity, $\log{L_{\rm rad}/L_\odot} = -7.16 \pm 0.06$, and was subsequently found to be under-luminous in comparison to stars with similar rotational, magnetic, and stellar parameters. Following \cite{2017MNRAS.465.2160K}'s Eqn.\ 1, the minimum mass-loss rate that could explain the star's radio luminosity via free-free emission is $\sim 2\times 10^{-7}~{\rm M_\odot~yr^{-1}}$, which is about $100$ times higher than the star's CAK mass-loss rate, indicating that the detected radio emission cannot be due to free-free emission from the wind. While HD\,64740's radio emission is therefore almost certainly gyrosynchrotron, it seems probable that the sole 50 cm observation of this star is strongly attenuated by self-absorption in the wind, and it was therefore removed from the subsequent analysis.

\subsection{Comparison to H$\alpha$ emission}\label{subsec:halpha}

   \begin{figure}
   \centering
   \includegraphics[width=0.45\textwidth]{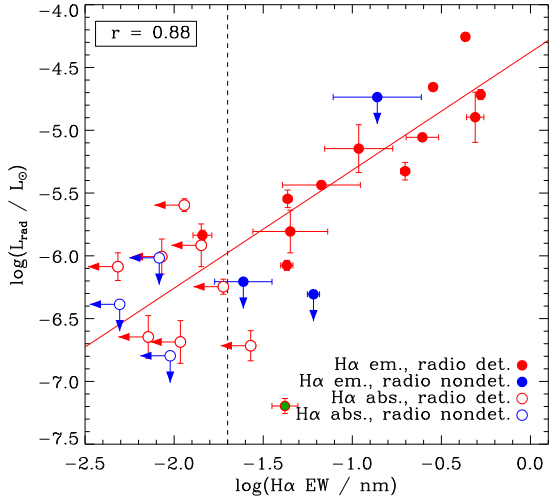}
      \caption[]{Radio luminosity as a function of H$\alpha$ emission equivalent width. The solid diagonal line shows a fit to the measurements of stars detected in both H$\alpha$ and radio. The vertical dashed line indicates the approximate noise floor identified by Shultz et al.\ (2020). Red and blue points are stars detected and not detected in the radio; filled and open symbols are stars with H$\alpha$ in emission and absorption respectively. HD\,64740 is highlighted with dark green.}
         \label{lrad_vs_halpha}
   \end{figure}

The co-occurence of radio-bright and H$\alpha$-bright stars in the same part of the rotation-magnetic confinement diagram (see Fig.\ \ref{radio_hrd_prot_bd}) is suggestive of a relationship between the two forms of magnetospheric emission. Fig.\ \ref{lrad_vs_halpha} demonstrates that the two forms of emission do in fact correlate. H$\alpha$ emission {equivalent widths (EWs)} were taken from the measurements of \cite{2020MNRAS.499.5379S}, with the addition of measurements of HD\,156424 \citep{2021MNRAS.504.4850S}, ALS\,9522 \citep{2021MNRAS.504.3203S}, and HD\,147932 (Shultz et al., in prep.). For stars in which both gyrosynchrotron emission and H$\alpha$ emission are detected, Pearson's correlation coefficient is $r=0.88$. 

The only outlier to the trend is HD\,64740 (highlighted in Fig.\ \ref{lrad_vs_halpha}), for which its radio luminosity is underluminous compared to its H$\alpha$ emission EW. This is consistent with its gyrosynchrotron emission being partially absorbed by its large free-free radio photosphere, as described is \S~\ref{subsec:wind}. HD\,64740 was therefore not included in the fit in Fig.\ \ref{lrad_vs_halpha}.

Stars without H$\alpha$ emission (open symbols) are of course all at or below the noise level (dashed line) inferred from the median EW error bar. Furthermore, the radio luminosities of these stars are systematically lower than those of stars with H$\alpha$ emission, consistent with magnetic confinement in their CMs being too weak to contain plasma that is optically thick in H$\alpha$. Only two stars have H$\alpha$ emission but are not detected in radio; in these cases, the upper limits on their radio luminosities lie very close to the regression line. 

\subsection{Regression analysis}\label{subsec:regression}

   \begin{figure*}
   \centering
   \includegraphics[width=\textwidth]{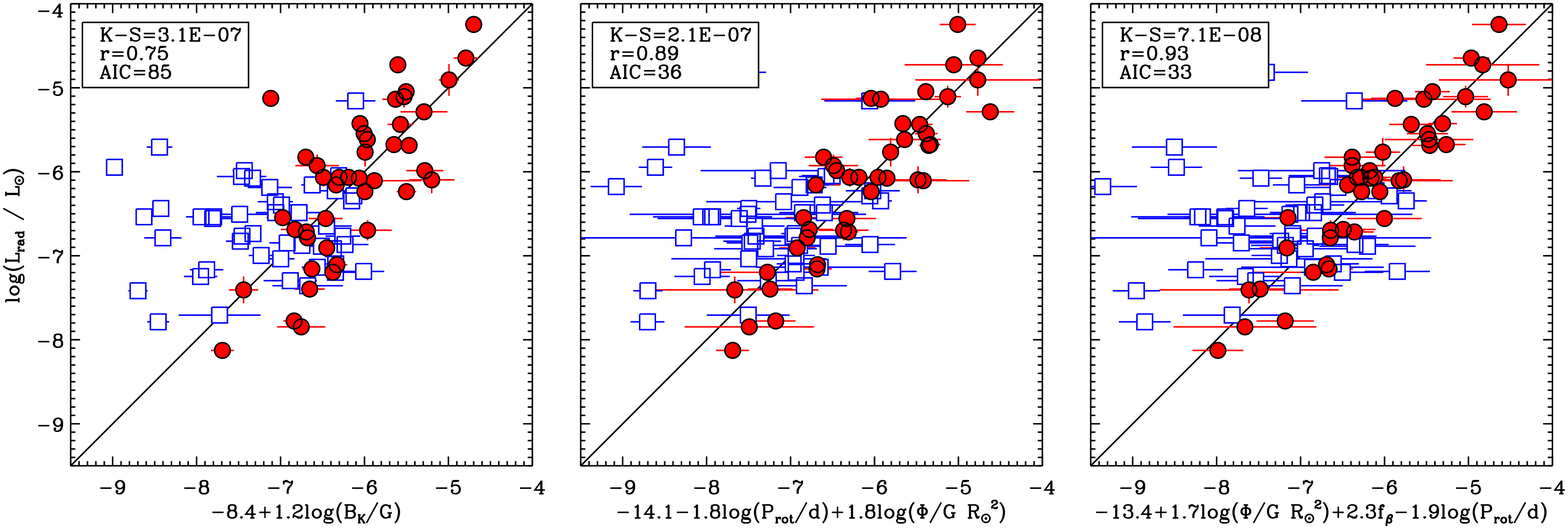}
      \caption[]{Best results for ({\em left -- right}) single-, double-, and triple-variable regressions of various parameters vs.\ radio luminosity. Red circles show radio-bright stars; open squares are upper limits for radio-dim stars. Legends give the K-S probability for separating radio-bright and -dim stars into different populations; Pearson's $r$; and the Akaike Information Criterion (AIC).}
         \label{multiregress}
   \end{figure*}

\begin{table}
\centering
\caption[]{Regression parameters for maximum radio luminosity. From left to right, the columns give: the tested variable; the two-sample K-S test probability that the variable separates radio-bright and -dim stars into separate populations; Pearson's correlation coefficient $r$; the reduced $\chi^2$ of the regression; the Akaike Information Criterion (AIC) for the regression; and the slope of the regression.}
\label{regression_table}
\begin{tabular}{l r r r r r}
\hline\hline
                            Variable      &  K-S                &  $r$  & $\chi^2/\nu$ &  AIC & Slope \\
\hline
\multicolumn{6}{c}{{\em One-Variable Regressions}} \\
$\log({L_{\rm bol}/{\rm L}_\odot})$       & 0.87                &  0.64  &  5.1         & 103  &  1.0$\pm$0.2 \\
    $\log{(T_{\rm eff} / {\rm K})}$       & 0.53                &  0.78  &  4.1         & 84   &  6.6$\pm$0.8 \\
      $\log{(R_* / {\rm R}_\odot)}$       & 0.06                &  0.18  &  6.3         & 124  &  1.3$\pm$1.1 \\
      $\log{(M_* / {\rm M}_\odot)}$       & 0.66                &  0.70  &  4.2         & 88   &  4.0$\pm$0.6 \\
    $\log{(P_{\rm rot} / {\rm d})}$       & $10^{-7}$           & $-0.50$  &  6.4       & 126  & $-1.5\pm$0.4 \\
      $\log{(B_{\rm d} / {\rm G})}$       & $10^{-2}$           &  0.60  &  3.5         & 75   &  1.6$\pm$0.3 \\
$\log{(\dot{M} / {\rm M_\odot~yr^{-1}})}$ & 0.83                &  0.50  &  5.2         & 105  &  0.5$\pm$0.1 \\
          $\log{(R_{\rm A} / R_*)}$       & 0.08                & $-0.17$  & 6.2        & 121  & $-0.7\pm$0.6 \\
          $\log{(R_{\rm K} / R_*)}$       & $10^{-5}$           & $-0.45$  &  6.1       & 120  & $-2.1\pm$0.6 \\
    $\log{(R_{\rm A} / R_{\rm K})}$       & 0.35                & $-0.28$  &  6.0       & 120  & $-0.7\pm$0.4 \\
            $\log{(B_{\rm K} / G)}$       & $10^{-7}$           &  0.75  &  4.1         & 86   &  1.2$\pm$0.2 \\
 $\log{(\Phi / {\rm G~R_\odot^2})}$       & 0.19                &  0.67  &  4.0         & 84   &  1.6$\pm$0.3 \\
$\log{f_\beta}$                           & 0.29                &  0.33  &  7.1         & 137  &  2.9$\pm$1.3 \\
\hline
\multicolumn{6}{c}{{\em Best Two-Variable Regression}} \\
 $\log{(\Phi / {\rm G~R_\odot^2})}$       & $10^{-7}$           &  0.89  &  1.2         & 36   &  1.8$\pm$0.2 \\
    $\log{(P_{\rm rot} / {\rm d})}$       &                     &        &              &      & $-1.8\pm$0.2 \\
\hline
\multicolumn{6}{c}{{\em Best Three-Variable Regression}} \\
 $\log{(\Phi / {\rm G~R_\odot^2})}$       & $10^{-8}$           &  0.93  &  0.8          & 33   &  1.7$\pm$0.2 \\
    $\log{(P_{\rm rot} / {\rm d})}$       &                     &        &              &      & $-1.9\pm$0.2 \\
$f_\beta$                                 &                     &        &              &      &  2.3$\pm$0.5 \\
\hline\hline
\end{tabular}
\end{table}

In order to identify the primary parameters affecting radio emission in a relatively hypothesis-independent fashion, we compared radio luminosities to a variety of stellar, magnetic, and rotational parameters, using one-, two-, and three-variable regressions (regressions with four variables yielded no statistical improvement). The results of these tests are summarized in Table \ref{regression_table}. The best regressions are shown in Fig.\ \ref{multiregress}. 

The particular quantities chosen for regression analysis are: $L_{\rm bol}$; \teff; the stellar radius $R_*$; the stellar mass $M_*$; the rotation period $P_{\rm rot}$; the surface magnetic dipole strength $B_{\rm d}$; the mass-loss rate \mdot; the Alfv\'en radius \ra; the Kepler radius \rk; the dimensionless size of the CM \rark; the strength of the equatorial magnetic field at the Kepler radius $B_{\rm K}$; the unsigned magnetic flux $\Phi = B_{\rm d}R_*^2$; and as a test of the dependence on the geometry of the magnetic field, $f_\beta = (1 + \cos{\beta})/2$, where $\beta$ is the obliquity angle of the magnetic axis from the rotational axis. The inclusion of the geometric parameter $f_\beta$ is motivated by the RRM model, since at higher $\beta$ the amount of plasma retained in the CM is reduced \citep{town2005c}.

Each parameter was tested in several ways. First, the two-sample Kolmogorov-Smirnov test was used to compare the distributions of stars with and without detected radio emission, in order to determine if the parameter or combinination of parameters effectively separates the two populations. Second, for the radio-bright stars, Pearson's correlation coefficient $r$ was calculated for each parameter or set of parameters, where $r$ values close to $\pm$1 indicate a strong correlation, and values close to 0 no correlation. Third, the reduced $\chi^2/\nu$ (where $\nu$ is the number of degrees of freedom) was calculated, in order to estimate the quality of the fit of the linear regression to the data. Finally, the Akaike Information Criterion (AIC) was calculated, which provides a relative estimator of the quality of a given model based upon the fit and the number of variables (a lower value indicating a superior fit despite additional model parameters). Since adding additional parameters to a regression will naturally improve the fit to the data, $\chi^2/\nu$ and AIC help to determine whether the improvement is a meaningfully better fit, or simply a consequence of the additional degrees of freedom. In calculating $\chi^2/\nu$ and the AIC, we used the uncertainties in the radio luminosities, rather than also including the uncertainties in the tested parameters, since the latter are widely variable between parameters (e.g., on the order of 10\% or higher in $B_{\rm d}$, as compared to around 0.0001\% in $P_{\rm rot}$), and including these uncertainties results in the goodness-of-fit tests simply reflecting the parameter uncertainties, making meaningful comparison difficult. 

For one-variable regressions, stellar parameters (\teff, $\log{L_{\rm bol}}$, $R_*$, $M_*$, \mdot) have large K-S probabilities, indicating that they do not separate the radio-bright and -dim populations. However, $r$ is relatively high for \teff, $\log{L_{\rm bol}}$, and $M_*$, indicating that stellar parameters play some role. By contrast, parameters associated with the magnetic field or rotation achieve K-S probabilities close to 0, indicating that they do a good job of distinguishing between radio-bright and -dim stars, with parameters involving rotation ($P_{\rm rot}$, \rk, $B_{\rm K}$) achieving the smallest K-S probabilities. Interestingly, the correlation coefficients associated with $B_{\rm d}$ and $P_{\rm rot}$ are lower than those achieved for some stellar parameters. {Of the magnetic and rotational parameters, the highest $r$ is achieved for $B_{\rm K}$, while $B_{\rm d}$ gives the smallest AIC.

The one-variable results indicate that radio emission is primarily an effect of magnetic field strength and rotation, however they also point to at least some role for stellar parameters. With the addition of a second variable, the best $r = 0.89$ and ${\rm AIC} = 36$ is provided by $L_{\rm rad} \propto \Phi^{1.8 \pm 0.2}P_{\rm rot}^{-1.8 \pm 0.2}$, which also yields a very small K-S probability. Adding a third variable yields the best $r = 0.93$ for $L_{\rm rad} \propto \Phi^{1.7 \pm 0.2}P_{\rm rot}^{-1.9 \pm 0.2}f_\beta^{2.3 \pm 0.5}$, with with a smaller AIC from the best two-variable result. {Both the two-and three-variable regressions yield $\chi^2/nu$ close to 1, indicating a good fit. While the three-variable result is slightly less than 1, suggesting a possible over-fit to the data, the lower AIC indicates that the improvement in the fit achieved by adding a third variable is real.





The overall results favour a strong dependence of radio luminosity on surface magnetic field strength, rotational period, and the size of the star, with a possible residual dependence on the magnetic geometry. The overall basic best-fit regression seems to go as $L_{\rm rad} \propto (\Phi/P_{\rm rot})^2 = (B_{\rm d}R_*^2/P_{\rm rot})^2$. This confirms the basic result found by \cite{2021MNRAS.507.1979L}. 

As demonstrated by Fig.\ \ref{radio_distance_flux_luminosity}, beyond a distance of $\log{(d/{\rm pc})} = 2.2$ the lower limit on $L_{\rm rad}$ is a strong function of distance. If the above analysis is repeated only using those stars closer than this distance, the basic results are qualitatively unchanged. The best single-variable regression (K-S~$= 0.01$, $r = 0.85$, AIC~$=42$) is given by $L_{\rm rad} \propto B_{\rm K}^{1.4 \pm 0.2}$. Two variables yield a best fit (K-S~$= 0.02$, $r = 0.93$, AIC~$=28$) for $L_{\rm rad} \propto B_{\rm K}^{1.1 \pm 0.1}T_{\rm eff}^{4.2 \pm 0.9}$. Adding a third variable provids the overall best model (K-S~$= 0.02$, $r = 0.97$, AIC~$=21$) for $L_{\rm rad} \propto \Phi^{1.7 \pm 0.2}P_{\rm rot}^{-2.4 \pm 0.2}f_\beta^{2.6 \pm 0.6}$. Once again, the results favour a dependence of radio luminosity on magnetic field strength, and an inverse dependence on rotation period. The best one- and three-variable regressions are identical to those obtained with the full dataset. 


In order to test for robustness against individual outliers, the analysis was repeated removing individual stars. Results were qualitatively unchanged in all cases. Results were also qualitatively unchanged if HD\,171247 and HD\,64740 were reintroduced to the analysis (see \S~\ref{sec:empirical}), although $r$ was reduced and the AIC increased (further suggesting them to be outliers).

\subsection{Summary}

Radio gyrosynchrotron emission is found in the same parameter space in which H$\alpha$ emission from centrifugal magnetospheres is seen -- i.e., in young, strongly magnetic, and rapidly rotating stars \citep{2019MNRAS.490..274S}. Indeed, radio-bright stars occupy essentially the same part of the rotation-magnetic wind confinement diagram as that occupied by H$\alpha$-bright stars. Radio luminosity drops rapidly with age, declining by about 2 orders of magnitude over the first 10\% of a star's main sequence lifetime. This is consistent with the abrupt spindown that is an expected and observed consequence of hot star magnetic fields \citep{2019MNRAS.490..274S,2020MNRAS.493..518K}, and is similar to the precipitious decline in H$\alpha$ emission strength observed in magnetic early B-type stars \citep{2020MNRAS.499.5379S}. Amongst those stars with both radio and H$\alpha$ emission, there is a strong correlation between the two. Finally, 91\% of the variance in radio luminosity is explained by the total unsigned magnetic flux and the rotational period, with a residual dependence on the obliquity angle of the magnetic field explaining a further 3\% of the variance.

\section{Discussion}\label{sect:discussion}

\subsection{Comparison to previous results}

\cite{1992ApJ...393..341L} found an empirical relationship of $L_{\rm rad} \propto \dot{M}^{0.38}B_{\rm rms}^{1.06}P_{\rm rot}^{-0.32}$, where $B_{\rm rms}$ is the root-mean-square \bz~(as $B_{\rm d}$ was not available for most of the stars). The improvement of this relationship over a two-parameter scaling relationship involving only $\dot{M}$ and $B_{\rm rms}$ was only marginal. The much stronger dependence on magnetic field strength and rotation period is due to our much larger sample, as well as the fact that magnetic field strengths and rotational periods have now been derived for a much larger number of stars. 

\cite{2021MNRAS.507.1979L} found an essentially identical scaling relationship to that found here, i.e.\ a dependence of radio luminosity on the ratio $\Phi/P_{\rm rot}$. Our results therefore confirm those of \citeauthor{2021MNRAS.507.1979L}, albeit with a signficantly larger sample. Further, since our sample includes stars that are not detected in radio, we have been able to demonstrate that this magneto-rotational empirical scaling relationship efficiently separates stars with radio emission from those in which such emission has not yet been detected. The principle difference between our results, and those given by \citeauthor{2021MNRAS.507.1979L}, is the weak dependence on $\beta$, a factor which they did not consider. 

Scaling relationships for the luminosity of auroral radio emission were also explored by \cite{2021arXiv210904043D} in their analysis of the largest sample to date of `main sequence radio pulse emitters' (MRPs), i.e.\ early type stars exhibiting pulsed electron-cyclotron maser emission (ECME). \citeauthor{2021arXiv210904043D} found most of the variarance of their sample to be explained by the relationship $L_{\rm ECME} \propto B_{0,{\rm max}}/(T_{\rm eff} - 16.5~{\rm kK})^2$, i.e.\ a linear dependence on the maximum surface magnetic field strength, and a dependence on the inverse square of the difference between the effective temperature and a reference value of 16.5 kK. They interpreted this as indicating that for stars with \teff~below 16.5 kK, the increasingly weak winds lead to less populated magnetospheres and therefore weaker emission, while above this temperature the increasing circumstellar density acts to attenuate the beamed emission via self-absorption. While no strong \teff~dependence was found in the present work, it is notable that HD\,64740 is under-luminous compared to expectations, which seems to be a consequence of self-absorption. Notably, \citeauthor{2021arXiv210904043D} found that ECME luminosity is independent of rotation; however, since all but one of their stars were rapid rotators (periods between 0.7 and 2~d), the small variance in $P_{\rm rot}$ may have hidden any such dependence. Given the similarity between the ECME luminosity scaling relationship found by \cite{2021arXiv210904043D}, and the initial gyrosynchrotron scaling relationship found by \cite{1992ApJ...393..341L} -- both exhibiting linear dependences on the magnetic field strength, with the remaining dependence explained by the strength of the wind -- it will be instructive to revisit the relationship when a larger sample spanning a wider range of rotational properties is available. 

\subsection{Intepretation of the results}

   \begin{figure*}
   \centering
   \includegraphics[width=0.95\textwidth]{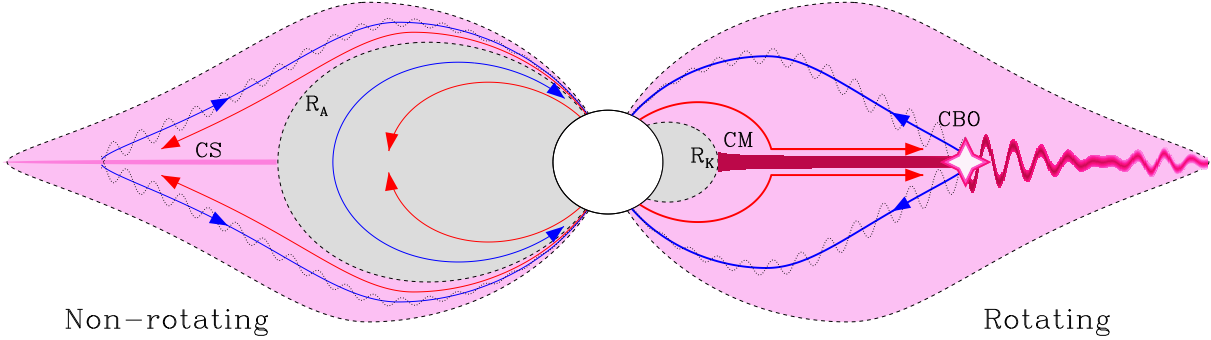}
      \caption[]{Schematic of the proposed interpretation. Pink shaded regions indicate magnetic field lines contributing plasma to the electron acceleration region; grey shaded regions indicate magnetospheric regions isolated from the locus of electron acceleration. In the non-rotating case (meaning that rotation is dynamically unimportant), plasma in the inner magnetosphere is in dynamical equilibrium, with upflow (red arrows) and downflow (blue arrows) occuring at the same rate. Beyond the Alfv\'en radius \ra, corotation ceases, and ram pressure from the wind stretches magnetic field lines, leading to the formation of a current sheet (CS) which accelerates electrons to relativistic velocities. The electrons return to the star along magnetic field lines, leading to the emission of gyrosynchrotron radiation. The inner magnetosphere is entirely isolated from the current sheet, as all plasma flow is internal. By contrast, in the rotating case, centrifugal support of plasma above the Kepler corotation radius \rk~leads to the formation of a centrifugal magnetosphere (CM), in which plasma accumulates to high density (below \rk~the magnetosphere remains dynamical, as in the non-rotating case). When gas pressure overloads the ability of the magnetic field to confine the plasma, plasma is ejected outwards by a centrifugal breakout (CBO) event. Magnetic reconnection during CBO leads to flaring, which accelerates electrons to high energies (indicated by the starburst), thereby providing the source electrons to populate the radio magnetosphere. Note also that the fraction of the wind plasma captured by the CM is much higher than that captured by the CS in the non-rotating case.}
         \label{cartoon}
   \end{figure*}

Until now the prevailing paradigm explaining gyrosynchrotron emission from hot stars has been the wind-powered current sheet model described by \cite{2004A&A...418..593T}. This model is illustrated in the left half of Fig.\ \ref{cartoon}. In this interpretation, a current sheet forms in the `middle magnetosphere' just beyond the Alfv\'en surface, where the wind's ram pressure opens the magnetic field lines, forming helmet streamers in which the opposite polarities of the magnetic field reconnect in the magnetic equatorial plane. Electrons injected into the current sheet by the wind are accelerated to relativistic velocities, following which they return to the star along the magnetic field lines, emitting gyrosynchrotron radiation as they go. This model is now challenged on two fronts. First, it makes absolutely no reference to the rotational properties of the star, since the power source is provided directly by the wind; yet, as shown by \cite{2021MNRAS.507.1979L}, and as verified here, rotation is absolutely crucial. Second, and more fundamentally, \cite{2021MNRAS.507.1979L} demonstrated via detailed modelling that the wind does not actually contain enough power to explain the observed radio luminosities. For the coolest stars examined by \citeauthor{2021MNRAS.507.1979L}, the difference between the required mass-loss rates and those predicted by the theoretical prescription given by \cite{krticka2014} are up to 4 orders of magnitude. The higher \cite{vink2001} or CAK mass-loss rates do not qualitatively change this picture. 

The close correlation with H$\alpha$ emission EWs is suggestive of a resolution. \cite{2020MNRAS.499.5379S} and \cite{2020MNRAS.499.5366O} demonstrated via a combined empirical and theoretical analysis that H$\alpha$ emission from CMs is fully explained by a centrifugal breakout (CBO) process in which the plasma density in the CM is set by the ability of the magnetic field to confine the plasma. The lack of secular variation, demonstrated by both \cite{2020MNRAS.499.5379S} and \cite{town2013}, indicates that the magnetosphere must be constantly maintained at the breakout density. This means that the large-scale emptying and reorganization of the CM observed in the 2D MHD simulations conducted by \cite{ud2008} does not in practice happen in three dimensions; instead, breakout events must be small in azimuthal extent and effectively continuous. 

The dependence of radio on rotation, and the close correlation with H$\alpha$, suggest that CBO may also be the explanation for radio emission. There are two, not necessarily mutually exclusive, mechanisms by which this might take place, both illustrated in the right half of Fig. \ref{cartoon}. 

First, CBO involves an outward ejection of material, necessarily establing a flow of plasma from the inner to the middle magnetosphere. In the absence of a CM, no such flow takes place, and the middle and inner magnetospheres should be effectively isolated from one another. This means that the total fraction of the wind captured by that part of the magnetosphere capable of contributing to gyrosynchrotron is greatly increased by the presence of a CM. 

Second, and perhaps most importantly, CBO is by its very nature a magnetic reconnection process. This means that each CBO event should be accompanied by an explosive release of energy, in which electrons will naturally be accelerated to high energies. In this case, the electron acceleration region is no longer the current sheet, but directly within the CM. 

Notably, following from their detection of a `giant pulse' from the MRP CU Vir, apparently originating simultaneously from both magnetic hemispheres, \cite{2021ApJ...921....9D} speculated that CBO in the inner magnetosphere might have led to an enhanced injection of electrons into the auroral current systems around both magnetic poles. If so, CBO might also play a role in auroral radio emission. 

The dependence of radio luminosity on the tilt angle $\beta$ of the magnetic field is consistent with the CBO hypothesis. In the RRM model, the amount of plasma trapped in the CM is a function of $\beta$. When $\beta = 0$ (i.e.\ a magnetic axis aligned with the rotational axis), the CM is an azimuthally symmetric torus in the common magnetic and rotational equatorial planes, with the inner edge coinciding with $R_{\rm K}$. Since plasma is most strongly confined at the intersections of the two planes, as $\beta$ increases the disk becomes increasingly warped, ultimately separating into two clouds concentrated at the intersections. The $\beta$ dependence found here is consistent with the radio luminosity of aligned rotators being intrinsically stronger, and dropping by a factor of about 4 as $\beta$ increases to $90^\circ$. This is exactly as would be expected if the plasma trapped in the CM is the ultimate source population for the high-energy electrons in the radio magnetosphere. 

\cite{2021MNRAS.507.1979L} noted that the scaling relationship has the physical dimension of an electromotive force. However, dimensional analysis of the correlation with $P_{\rm rot}$, $B_{\rm d}$, and $R_*$ is suggestive of an alternative interpretation. $B^2$ is the magnetic energy density, while $R_*^3$ suggests the volume of the star; combined, this yields the magnetic energy of the system. The relationship $B_{\rm d}^2R_*^3/P_{\rm rot}$ then directly yields a luminosity: the magnetic energy of the star being tapped on a rotational timescale. Indeed, if $B_{\rm d}$, $R_*$, and $P_{\rm rot}$ are allowed to vary independently, the best-fit relationship is $L_{\rm rad} \propto B_{\rm d}^{2.0 \pm 0.2}P_{\rm rot}^{-1.8 \pm 0.2}R_*^{3.4 \pm 0.6}$, i.e.\ a somewhat weaker $R_*$ dependence is favoured than in the case of the magnetic flux. The dependence on the inverse {\em square} of the rotation period introduces an extra dimension of time, which must somehow be accounted for, as must the possible additional $R_*$ term. It is suggestive, however, that the scaling relationship contains within it the natural units of luminosity. In Paper II by Owocki et al., we show that this scaling relationship is a natural consequence of CBO, and demonstrate the origin of the extra dependence on $P_{\rm rot}$. 

\subsection{Indirect magnetometry}\label{subsect:indirect}

\begin{table}
\centering
\caption[]{Parameters for radio-bright stars with known rotational periods but without detected magnetic fields. References: $a$, \protect\cite{2012MNRAS.420..757W}; $b$ \protect\cite{1998A&AS..127..421C}; $c$ \protect\cite{1992ApJ...393..341L}; $d$ \protect\cite{2021MNRAS.502.5438P}.}
\label{bd_pred_tab}
\begin{tabular}{l c c c}
\hline\hline
Parameter & HD\,143699 & HD\,146001 & HD\,77653 \\
\hline
$P_{\rm rot} / {\rm d}$ & 0.894 & 0.586$^a$ & 1.488$^b$ \\
$R_* / {\rm R_\odot}$ & $3.0 \pm 0.3$ & $2.5 \pm 0.3$ & $2.6 \pm 0.1$ \\
$M_* / {\rm M_\odot}$ & $5.3 \pm 0.3$ & $4.3 \pm 0.2$ & $3.4 \pm 0.1$ \\
$\log{(L_{\rm rad}/L_\odot)}$ & $-7.08 \pm 0.17^c$ & $-6.69 \pm 0.17^c$ & $-5.71 \pm 0.08^d$ \\
$B_{\rm d} / {\rm G}$~(obs.) & $<$600 & $<$500 & -- \\
$B_{\rm d} / {\rm G}$~(pred.) & 300 & 430 & 2900 \\
\hline\hline
\end{tabular}
\end{table}

The magneto-rotational scaling law discovered by \cite{2021MNRAS.507.1979L}, confirmed in the present work, and explained in the companion paper by Owocki et al. as a consequence of CBO reconnection shows the potential, as pointed out by \citeauthor{2021MNRAS.507.1979L}, to be utilized as a reliable form of `indirect magnetometry', enabling measurement of stellar magnetic fields in objects beyond the reach of contemporary spectropolarimeters. Three stars in the present sample are radio-bright and have known rotational periods, but do not have detected magnetic fields. While these stars could not be used to constrain the scaling law, they can serve as test cases for the predictive ability of the scaling law. Table \ref{bd_pred_tab} summarizes their key parameters. 

For HD\,143699 and HD\,146001, the `observed' values of $B_{\rm d}$ correspond to the $1\sigma$ upper limits derived via modelling their \bz~error bars (both around 70~G; see Appendix C) using the MCHRD sampler, where in both cases high-resolution spectropolarimetry was used. The predicted $B_{\rm d}$ was found via solving the scaling relationship from Paper II in this series by Owocki et al. for $B_{\rm d}$, using an efficiency factor of $\epsilon = 10^{-8}$ (i.e.\ ignoring the correction for $\beta$, which is unknowable). For the two stars with available \bz~measurements, the predicted $B_{\rm d}$ -- a few hundred G in both cases -- is in both cases just below the upper limits. For HD\,77653, for which spectropolarimetry is not available, the scaling relationship predicts $B_{\rm d} \sim 3$~kG, which should be easily detectable. Followup magnetometry of these stars will provide a useful test of this scaling relationship.

\subsection{Radio emission from stars with ultra-weak magnetic fields}

The nearby A7\,V star Altair was recently discovered by \cite{2021ApJ...912L...5W} to emit non-thermal radio at cm wavelengths, with a brightness temperature around $10^4$ K and a luminosity of around $\log{L_{\rm rad}/L_\odot} \sim -10.5$. \citeauthor{2021ApJ...912L...5W} interpreted this as chromospheric emission, possibly related to the equatorial convection zone formed due to the star's extremely rapid rotation. \cite{2009A&A...497..511R} furthermore detected X-ray emission from Altair, which they interpreted as magnetic activity. 

Altair was observed with Narval as part of the BRIght Target Explorer \citep[BRITE;][]{2014PASP..126..573W} spectropolarimetric survey \citep[BRITEpol;][]{neiner2017ppas}. No magnetic field was detected, with an uncertainty in \bz~of about 10 G, implying that a surface magnetic dipole of around 100 G could well have gone undetected. Using the fundamental parameters (equatorial radius $R_{\rm eq} = 2.008 \pm 0.006$~\rsun, $M_* = 1.86 \pm 0.03$~\msun) and rotation period $P_{\rm rot} = 0.323$~d determined via careful interferometric modelling performed by \cite{2020A&A...633A..78B} yields a critical rotation parameter $W = 0.75$ and a Kepler corotation radius $R_{\rm K} = 1.2~R_*$. The star's CAK mass-loss rate is $\dot{M} = 10^{-13}~{\rm M_\odot~yr^{-1}}$; assuming a terminal velocity of 3000 \kms, $R_{\rm K}$ will be inside the Alfv\'en surface so long as $B_{\rm d} > 0.1$~G, well within the upper limits on Altair's surface magnetic field and consistent with the range of ultra-weak fields detected in other main sequence A-type stars \citep{2010AA...523A..41P,2020MNRAS.492.5794B}. 

To see if the star's non-thermal radio emission might be consistent with a magnetospheric origin given the limits on the surface magnetic field, we follow the same method as above in \S~\ref{subsect:indirect}. for $B_{\rm d}$. We again assumed the efficiency $\epsilon \sim 10^{-8}$. This yields a predicted surface magnetic field of $B_{\rm d} \sim 10$~G. Altair's radio emission may therefore be consistent with a magnetospheric origin, although actually detecting such a weak field (which would require uncertainties on the order of 1 G) is a challenging prospect given the star's broad spectral lines.

Unlike Altair, magnetic fields have actually been detected in Vega and Sirius \citep{2010AA...523A..41P,2011A&A...532L..13P}, with both stars having sub-gauss \bz. Radio observations at mm and sub-mm wavelengths of both stars are consistent with thermal emission \citep{2012ApJ...750...82H,2019ApJ...875...55W}. Sirius is a slow rotator and therefore unlikely to produce gyrosynchrotron emission. Taking Vega's stellar parameters \citep{2010ApJ...708...71Y} and 0.732~d rotation period \citep{2010A&A...523A..41P} yields $R_{\rm K} = 1.5$~\rsun. With the CAK mass-loss rate $\dot{M} = 10^{-11.9}~{\rm M_\odot~yr^{-1}}$ and the same assumption of a 3000 \kms~wind terminal velocity, the miniumum surface dipole strength capable of confining the wind out to \rk~is 2.3 G, $4\times$ higher than the dipolar component of about 0.5 G recovered via Zeeman Doppler Imaging \citep{2010A&A...523A..41P}. The expected radio luminosity from the breakout scaling is then $\log{L_{\rm rad}/L_\odot} = -12$, translating at 1 cm to 0.15 $\mu$Jy at Vega's 7.67 pc distance: certainly undetectable, since this is much less than the expected 1 cm photospheric flux of about 0.5 mJy. Radio observations of other stars with ultra-weak fields do not seem to be available, although at least in the case of Alhena the relatively long $\sim 9$~d period and $\sim$30~G surface field makes it unlikely the star would produce detectable emission \citep{2016MNRAS.459L..81B,2020MNRAS.492.5794B}, while in the cases of $\beta$~UMa and $\theta$~Leo \citep{2016A&A...586A..97B} the rotational periods are not known, making their radio luminosities impossible to estimate.

\section{Conclusions}\label{sec:conclusions}

By combining both published and unpublished radio observations, published rotational and magnetic data, and new determinations of magnetic models and rotational periods via space photometry and previously unpublished high- and low-resolution spectropolarimetry, we have conducted the largest analysis of the gyrosynchrotron emission properties of magnetic early-type stars undertaken to date. 

We find that radio-bright stars occur in the same part of the rotation-magnetic confinement diagram as stars with H$\alpha$ emission originating from centrifugal magnetospheres: that is to say, gyrosynchrotron emission requires rapid rotation as well as a strong magnetic field. This confirms the central result of \cite{2021MNRAS.507.1979L}. Radio-bright stars are additionally generally young, with a steep drop in radio luminosity with age consistent with magnetospheric braking rapidly removing the angular momentum necessary to power the radio magnetosphere, similar to the drop observed by \cite{2020MNRAS.499.5379S} for H$\alpha$ emission. Furthermore, there is a close correlation between the H$\alpha$ emission equivalent width and radio luminosity, which is strongly suggestive of a unifying mechanism. 

Multivariable regression analysis of radio luminosity yields a relation of the form $L_{\rm rad} \propto B^2R_*^4/P_{\rm rot}^2 = (\Phi/P_{\rm rot})^2$, further confirming the results of \cite{2021MNRAS.507.1979L}, although we add the refinement of an additional dependence on the geometry of the magnetic dipole such that radio luminosty declines with increasing tilt angle $\beta$. 

We propose that the close correlation between H$\alpha$ and radio emission strengths, and their cohabitation in parameter space, imply a unifying mechanism for the two phenomena, i.e.\ centrifugal breakout, which has already been shown to explain H$\alpha$ emission properties. Paper II by Owocki et al.\ provides a preliminary theoretical exploration of this concept. 

The empirical relationship found by \cite{2021MNRAS.507.1979L} and confirmed here suggests that radio observations may have utility as a form of indirect magnetometry. If the distance, stellar radius, and rotational period are known, a single radio observation may be sufficient to infer the global surface magnetic field strength of the star. In the era of {\em Gaia} and {\em TESS}, in which distances and rotational periods can be determined for a much larger number of stars than can be easily observed with optical spectropolarimetry, this may prove to be an important means of dramatically increasing the number of stars for which the surface magnetic field strength is known. 


There is a pressing need for more radio observations of magnetic early-type stars to be acquired. SEDs have been measured for only a small number of magnetic hot stars, and it is not known how these vary with fundamental, magnetic, or rotational parameters. Rotational phase coverage is likewise available in only a small number of cases; the geometrical dependence found here for the radio luminosity suggests that comparable effects might be seen in phase curves, which may be important in reconstructing plasma distributions out of the magnetic-equatorial plane probed by visible data. More sensitive observations might seek to discover if gyrosynchrotron emission disappears entirely in stars without centrifugal magnetospheres, or if slowly rotating stars in fact emit ultra-weak radio driven by the classical middle magnetosphere current sheet mechanism. Indeed, while gyrosynchroton emission has not yet been detected from slow rotators, there are very few stars in the dynamical magnetosphere regime with radio data. Finally, as pointed out by \cite{2021MNRAS.507.1979L}, the close correlation between radio luminosity and magnetic field strength suggests that radio data might become an important form of indirect magnetometry for stars that are too dim for their surface magnetic fields to be measured using Zeeman effect spectropolarimetry, but for which rotational periods are known via e.g.\ {\em TESS} space photometry.

\section*{Acknowledgments}
The authors thank Dr.\ Stephen Drake for providing his unpublished radio measurements, which they hope have been put to good use. This work is based on observations obtained at the Canada-France-Hawaii Telescope (CFHT) which is operated by the National Research Council of Canada, the Institut National des Sciences de l'Univers (INSU) of the Centre National de la Recherche Scientifique (CNRS) of France, and the University of Hawaii, and at the Observatoire du Pic du Midi (France), operated by the INSU; and based on observations obtained at the Dominion Astrophysical Observatory, National Research Council Herzberg Astronomy and Astrophysics Research Centre, National Research Council of Canada. This paper includes data collected by the TESS mission, which are publicly available from the Mikulski Archive for Space Telescopes (MAST). Funding for the TESS mission is provided by NASA’s Science Mission directorate. We thank the staff of the GMRT that made the uGMRT observations possible. The GMRT is run by the National Centre for Radio Astrophysics of the Tata Institute of Fundamental Research. This work has made use of the VALD database, operated at Uppsala University, the Institute of Astronomy RAS in Moscow, and the University of Vienna. This research has made use of the SIMBAD database, operated at CDS, Strasbourg, France. MES acknowledges the financial support provided by the Annie Jump Cannon Fellowship, supported by the University of Delaware and endowed by the Mount Cuba Astronomical Observatory. AuD acknowledges support by NASA through Chandra Award 26 number TM1-22001B issued by the Chandra X-ray Observatory 27 Center, which is operated by the Smithsonian Astrophysical Observatory for and behalf of NASA under contract NAS8-03060. VK acknowledges support by Natural Sciences and Engineering Research Council of Canada (NSERC). AB, BD, and PC acknowledge support of the Department of Atomic Energy, Government of India, under project no. 12-R\&D-TFR-5.02-0700. GAW acknowledges Discovery Grant support from NSERC. JDL acknowledges support from NSERC, funding reference number 6377--2016. OK acknowledges support by the Swedish Research Council and the Swedish National Space Board. The material is based upon work supported by NASA under award number 80GSFC21M0002.

\section*{Data Availability Statement}
Reduced ESPaDOnS spectra are available at the CFHT archive maintained by the CADC\footnote{\url{https://www.cadc-ccda.hia-iha.nrc-cnrc.gc.ca/en/}}, where they can be found via standard stellar designations. ESPaDOnS and Narval data can also be obtained at the PolarBase archive\footnote{\url{http://polarbase.irap.omp.eu/}}. {\em Kepler-2} and {\em TESS} data are available at the MAST archive\footnote{\url{https://mast.stsci.edu/portal/Mashup/Clients/Mast/Portal.html}}. {\em Hipparcos} data are available through SIMBAD and Vizier\footnote{\url{https://vizier.u-strasbg.fr/viz-bin/VizieR}}. DAO and uGMRT observations are available from the authors at request. 

\bibliography{bib_dat.bib}{}
\label{lastpage}

\appendix

\section{Sample summary table}\label{append:summary_table}

\pagebreak

\clearpage

\input radio_partab.tex

\pagebreak


\begin{center}

\onecolumn
\thispagestyle{plain} 
\vspace*{\fill}
\textbf{\Huge Online Material}
\vspace*{\fill}
\end{center}

\pagebreak
\clearpage
%

%

\twocolumn

\section{New radio observations}\label{append:new_radio}

\begin{table*}
    \centering
    \caption{New radio observations acquired with the uGMRT. Superscripts $e$ next to the star name indicate stars observed in the context of the ECME survey; the remainder were observed in the context of the GMRT legacy survey. HJD stands for Heliocentric Julian Date. `Eff. band' is the effective frequency range of observation after the edges of the original frequency band were removed due to low gain. For the stars that were not detected, we report the $3\sigma$ upper limit. For the detected stars (in bold), the quoted uncertainty includes the fitting error, the map rms, and the uncertainty associated with the absolute flux density calibration (assumed to be 10\% of the flux density).
    \label{tab:ugmrt_obs}}
    \begin{tabular}{lccllr}
    \hline
        Star  & HJD range & Eff. band & Flux calibrator & Phase calibrator & Flux density \\
              & $-2450000$ & (MHz) & & &(mJy)\\
        \hline\hline
	HD\,3360            & 6982.34$\pm$0.07 & 1374--1401 & 3C48 & 	J2355+498  &  $<0.097$\\
        {\bf HD\,11503}$^e$ & 8439.26$\pm$0.18 & 570--804 &  3C48, 3C147 & J0204+152 & $0.25\pm0.01$\\ 
        {\bf HD\,35502}$^e$ & 8595.03$\pm$0.06 & 570--716 & 3C147, 3C286 & J0607--085 & $0.64\pm0.16$\\
        HD\,36526$^e$ & 8231.00$\pm$0.04, 8242.86$\pm$0.05 & 560--726 & 3C48, 3C147 & J0503+020, J0607--085 & $<0.1$\\
        {\bf HD\,37061}$^e$ & 8608.90$\pm$0.10 & 570--804 & 3C48, 3C147 & J0607--085 & $0.54\pm0.14$\\
	{\bf HD\,37742}     & 7280.48$\pm$0.05 & 594--621 & 3C48 & J0607--085 & $0.948 \pm 0.105$\\ 
                            & 6984.34$\pm$0.21 & 1374--1401 & 3C48, 3C286 & J0503--020 & $0.281 \pm 0.053$\\ 
        HD\,37776$^e$ & 8711.65$\pm$0.16 & 570--804 & 3C48, 3C286 & J0607--085 &  $<0.1$\\
	HD\,55522     & 7280.57$\pm$0.04 &  594--621 & 3C48 & 		J0837--192 & $<0.223$\\
        {\bf HD\,61556}$^e$ & 8273.92$\pm$0.06 & 570--726 & 3C48 & J0735--175 & 1.21$\pm$0.18\\
        {\bf HD\,64740}$^e$ & 8602.02$\pm$0.07 & 570--794 & 3C48, 3C286 & J0828--375 & $0.06\pm0.01$\\
        HD\,66765$^e$ & 8342.74$\pm$0.05, 8391.63$\pm$0.01 & 560--726 & 3C48, 3C286 & J0828--375 & $<0.1$ \\
        {\bf HD\,175362}    & 7228.32$\pm$0.05 & 594--621 & 3C48 & 		J1830--360 &  $0.540 \pm 0.076$\\
                            & 6951.07$\pm$0.07 & 1374--1401 & 3C48, 3C286 & J1911--201 &  $0.271 \pm 0.040$\\
        {\bf HD\,182180}$^e$ & 8348.30$\pm$0.02 & 560--726 & 3C48 & J1924--292 & $4.17\pm0.07$\\
	HD\,184927          & 7277.09$\pm$0.06 &  594--621 & 3C286 & 	J2052+365  & $<0.187$ \\
                            & 7032.97$\pm$0.08 &  1374--1401 & 3C48 & 	J1924+334  & $<0.114$\\
	{\bf HD\,189775}    & 7231.32$\pm$0.04 & 594--621 & 3C48 &  J2022+616 &  $1.094 \pm 0.125$\\
                            & 6951.22$\pm$0.07 &  1374--1401 & 3C48 & 	J2022+616  &  $0.41 \pm 0.07$\\
	HD\,191612          & 7231.41$\pm$0.04 &  594--621 & 3C48 & 		J2052+365  & $<0.342$\\
                            & 6952.14$\pm$0.08 &  1374--1401 & 3C48 & 	J1924+334 & $<0.236$\\
	{\bf HD\,200775}    & 7230.52$\pm$0.04 & 594--621 & 3C48 & 		J2344+824  &  $<0.369$ \\
                            & 6982.09$\pm$0.07 & 1374--1401 & 3C48 & 	J2022+616  & $0.297 \pm 0.058$\\
	HD\,205021          & 7231.58$\pm$0.04 & 594--621 & 3C48, 3C147 & 	J2344+824  & $<0.252$\\
                            & 6982.22$\pm$0.06 & 1374--1401 & 3C48 & 	J2344+824  &  $<0.149$\\
        HD\,208057$^e$ & 8239.47$\pm$0.04 & 560--726 & 3C286 & J2202+422 & $<0.05$\\
        \hline
    \end{tabular}
    \label{tab:my_label}
\end{table*}



\section{Stars with new rotational and magnetic constraints}\label{append:indstars}

\subsection{Least-squares deconvolution}

\begin{table*}
\centering
\caption[]{Summary of Least-squares deconvolution analysis. The columns give the stellar designation, the number $N_{\rm ESP}$ of ESPaDOnS observations, $N_{\rm Nar}$ of Narval observations, the \teff~of the line mask, the number $N_{\rm lines}$ of lines in the mask used for analysis as a fraction of the total number of lines, the number of definite, marginal, and non-detections, the root-mean-square longitudinal magnetic field $B_{\rm rms}$, the root-mean-square null field $N_{\rm rms}$, the mean uncertainty $<\sigma_B>$, and the maximum longitudinal magnetic field \bz$_{\rm max}$.}
\label{lsd_table}
\begin{tabular}{l r r r r r r r r r}
\hline\hline
Star & $N_{\rm ESP}$ & $N_{\rm Nar}$ & \teff & $N_{\rm lines}$ & DD/MD/ND & $B_{\rm rms}$ & $N_{\rm rms}$ & $<\sigma_B>$ & \bz$_{\rm max}$ \\
     &               &               & (kK)  &                 &          & (G)           & (G)           & (G)          & (G)             \\
\hline
HD\,21699 & 26 & -- & 16 & 946/1203 & 26/0/0 & 431 & 11 & 19 & $697 \pm 33$ \\
HD\,22920 & 4  & 4  & 14 & 1284/2059 & 8/0/0 & 354 & 16 & 16 & $487 \pm 23$ \\
HD\,23408 & 2  & 1  & 13 & 1499/2379 & 0/0/3 & 5   & 12 & 13 & $-13 \pm 17$ \\
HD\,28843 & -- & 11 & 15 & 1099/1799 & 0/2/9 & 162 & 77 & 95 & $-264 \pm 119$ \\
HD\,36429 & 1  & --  & 14 & 309/561   & 0/0/1 & 120 & 59 & 67 & $120 \pm 67$ \\
HD\,36540 & 1  & --  & 13 & 1498/2328 & 1/0/0 & 430 & 94 & 59 & $430 \pm 59$ \\
HD\,36629 & 2  & --  & 22 & 302/502   & 0/0/2 & 4   & 3 & 4 & $-6 \pm 4$ \\
HD\,36960 & 1  & --  & 28 & 313/558   & 0/0/1 & 5   & 4 & 10 & $-5 \pm 10$ \\
HD\,37140 & 1  & --  & 14 & 1315/2059 & 1/0/0 & 186 & 17 & 22 & $-186 \pm 22$ \\
HD\,49606 & 2  & --  & 13 & 1513/2328 & 0/0/2 & 0   & 9  & 7  & $0 \pm 7$ \\
HD\,89822 & 4  & 1   & 10 & 616/1138 & 0/0/5  & 2   & 1  & 2  & $2 \pm 3$ \\
HD\,131120 & 1 & --  & 19 & 247/389  & 0/0/1  & 77  & 19 & 54 & $77 \pm 54$ \\
HD\,142884 & -- & 1 & 14   & 330/461  & 0/0/1  & 295 & 137 & 147 & $295 \pm 147$ \\
HD\,143699 & 1 & -- & 18 & 250/462 & 0/0/1    & 78 & 57 & 77 & $78 \pm 77$ \\
HD\,144844 & 1 & -- & 15 & 1288/1799 & 0/0/1  & 3 & 9 & 9 & $-3 \pm 9$ \\
HD\,146001 & 2 & -- & 14 & 1287/2059 & 0/0/2  & 67 & 76 & 66 & $97 \pm 66$ \\
HD\,147084 & 2 & -- & 10 & 623/1138 & 0/0/2 & 2 & 1 & 2 & $-2 \pm 1$ \\
HD\,162374 & 1 & -- & 16 & 1014/1504 & 0/0/1 & 12 & 25 & 27 & $12 \pm 27$ \\
HD\,170973 & 3 & -- & 10 & 623/1138 & 3/0/0 & 297 & 2 & 7 & $-431 \pm 7$ \\
HD\,207840 & 2 & -- & 13 & 1663/2328 & 0/0/2 & 2 & 5 & 4 & $-3 \pm 5$ \\
\hline\hline
\end{tabular}
\end{table*}

   \begin{figure*}
   \centering
   \includegraphics[width=18.5cm]{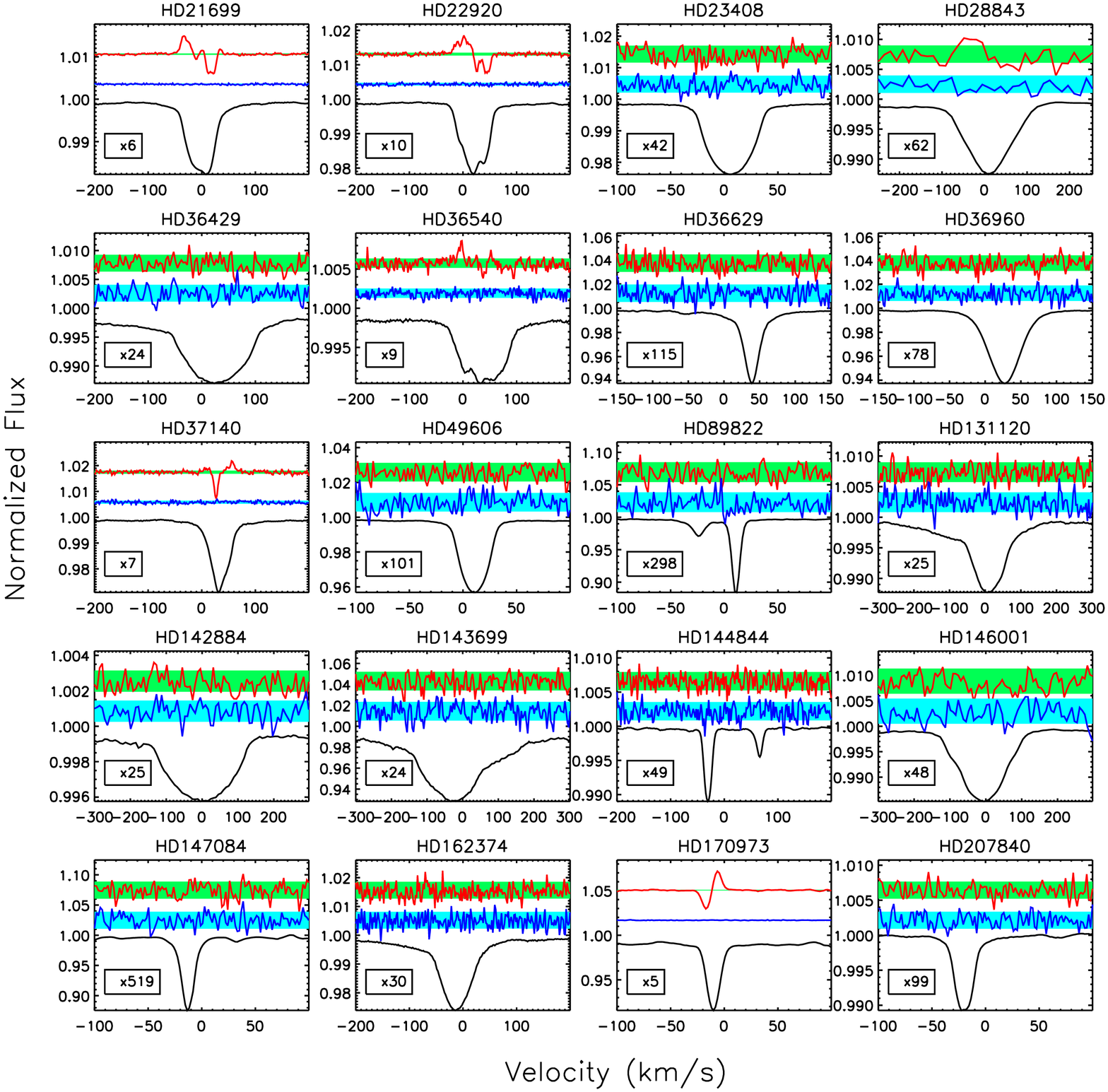}
      \caption[]{LSD profiles extracted from ESPaDOnS and Narval observations. Each panel shows Stokes $I$ (black, bottom), $N$ (blue, middle), Stokes $V$ (red, top). Blue and green shaded regions show the mean error bars for $N$ and Stokes $V$, respectively. Box gives the amplification factor for $N$ and Stokes $V$.}
         \label{lsd_all}
   \end{figure*}

\begin{table}
\centering
\caption[]{Summary of DAO \bz~measurements}
\label{dao_table}
\begin{tabular}{l r r r r}
\hline\hline
Star & $N$ & $B_{\rm rms}$ & $<\sigma_B>$ & \bz$_{\rm max}$ \\
     &     & (G)              & (G)          &                 \\
\hline
HD\,21699  & 29 & 453   & 157 & $-955 \pm 106$ \\
HD\,23408  & 5  & 105   & 93  & $316 \pm 108$  \\
HD\,36313  & 1  & --    & --  & $1275 \pm 491$ \\
HD\,36668  & 7  & 1726  & 538 & $3363 \pm 1048$ \\
HD\,37150  & 3  & 514   & 781 & $639 \pm 1413$ \\
HD\,37642  & 13 & 2782  & 595 & $-5162 \pm 307$ \\
HD\,41269  & 3  & 367   & 185 & $-408 \pm 178$ \\
HD\,51418  & 2  & 350   & 125 & $-550 \pm 157$ \\
HD\,133029 & 25 & 2302  & 178 & $4056 \pm 311$ \\
HD\,135679 & 22 & 792   & 190 & $1546 \pm 262$ \\
HD\,164429 & 24 & 614   & 288 & $-1467 \pm 409$ \\
HD\,165474 & 2  & 59    & 487 & $-82 \pm 285$ \\
HD\,170973 & 15 & 412   & 115 & $768 \pm 126$ \\
HD\,171247 & 27 & 348   & 260 & $1179 \pm 188$ \\
HD\,175744 & 4  & 171   & 146 & $-250 \pm 100$ \\
HD\,177410 & 3  & 196   & 251 & $-333 \pm 213$ \\
HD\,184961 & 18 & 646   & 175 & $1212 \pm 375$ \\
HD\,188041 & 11 & 793   & 71  & $1292 \pm 27$ \\
HD\,196178 & 31 & 920   & 182 & $-1470 \pm 422$ \\
HD\,209339 & 6  & 385   & 316 & $-1268 \pm 315$ \\
\hline\hline
\end{tabular}
\end{table}

\begin{table*}
\centering
\caption[]{New and revised rotation periods and \bz~fitting parameters. Periods with an asterisk may not be related to rotation. Superscripts $b$ and $p$ respectively indicate that $T_0$ is defined by \bz$_{\rm max}$ and maximum light.}
\label{bz_table}
\begin{tabular}{l r r r r r r r r r}
\hline\hline
Star & $P_{\rm rot}$ & $T_0$      & $B_0$ & $B_1$ & $\Phi_1$ \\
     & (d)           & $-2400000$ & (kG)  & (kG)  & (rad)    \\
\hline
HD\,21699 & 2.49187(7) & 55170.8(1)$^b$ & $0.158 \pm 0.005$ & $0.614 \pm 0.006$ & $1.91 \pm 0.01$ \\
HD\,22920 & 3.9472(1) & 56556.7(1)$^p$  & $0.312 \pm 0.009$ & $0.19 \pm 0.01$ & $1.47 \pm 0.06$ \\
HD\,28843 & 1.37382(6) & 47908.54(4)$^b$ & $0.01 \pm 0.04$ & $0.19 \pm 0.06$ & $2.0 \pm 0.3$ \\
HD\,35575 & 0.9841(2) & 58467.55(1)$^p$ & $0.08 \pm 0.08$ & $0.15 \pm 0.06$ & $0.23 \pm 0.21$ \\
HD\,36313 & 0.58884(2) & 58468.608(2)$^p$ & $-0.1 \pm 0.4$ & $1.7 \pm 0.6$ &  $1.5 \pm 0.3$ \\
HD\,36429 & 15.6(7)$^*$ & 58471.2(4)$^p$ & -- & -- & -- \\
HD\,36540 & 2.1732(1) &	58469.833(2)$^p$ & $0.09 \pm 0.07$ & $-0.3 \pm 0.1$ & $1.6 \pm 0.4$ \\
HD\,36668 & 2.1192(2) & 58466.50(2)$^p$ & $0.0 \pm  0.1$ & $1.2 \pm 0.1$ & $1.7 \pm 0.2$ \\
HD\,36916 & 1.56548(8) & 58467.186(7)$^p$ & $-0.53 \pm 0.09$ & $-0.49 \pm 0.04$ & $1.5 \pm 0.3$ \\
HD\,37140 & 2.7606(9) &	55557.1(2)$^b$ & $-0.18 \pm 0.05$ & $-0.8 \pm 0.1$ &  $1.52 \pm 0.06$ \\
HD\,37210 & 11.043(8) &	52682.8(7)$^b$ & $0.07 \pm 0.02$ & $0.45 \pm 0.04$ & $1.55 \pm 0.06$ \\
HD\,37642 & 1.07876(1) & 55582.75(7)$^p$ & $-1.2 \pm 0.1$ & $-3.9 \pm 0.2$ & $1.58 \pm 0.06$ \\
HD\,37808 & 1.098535(5) & 58469.585(9)$^p$ & $1.2 \pm  0.1$ & $0.6 \pm 0.2$ & $2.9 \pm 0.1$ \\
HD\,51418 & 5.431(3) & 58843.62(2)$^p$ & $-0.01 \pm 0.03$ & $-0.56 \pm 0.03$ & $1.53 \pm 0.07$ \\
HD\,89897 & 1.11037(5) & 58545.434(3)$^p$ & -- & -- & --    \\
HD\,131120 & 1.56873(1) & 58598.552(4)$^p$ & $0.06 \pm 0.03$ & $0.04 \pm 0.05$ & $3 \pm 1$ \\
HD\,133029 & 2.88767(5) & 54480.9(7)$^b$ & $2.23 \pm 0.05$ & $0.48 \pm 0.07$ & $1.26 \pm 0.14$ \\
HD\,135679 & 5.321(1) &	55259(1)$^b$ & $0.78 \pm 0.04$ & $0.55 \pm 0.05$ & $1.72 \pm 0.09$ \\
HD\,137193 & 4.867(3) &	59330(3)$^p$ & $0.6 \pm 0.2$ & $0.2 \pm 0.4$ & $0.4 \pm 1.2$ \\
HD\,143699 & 0.89421(5)$^*$ & 58627.368(5)$^p$ & -- & -- & --    \\
HD\,145482 & 5.804(2) &	58628.09(1)$^p$ & -- & -- & --    \\
HD\,149822 & 1.9661(1) & 8982.72(8)$^p$ & $-0.09 \pm 0.06$ & $-0.69 \pm 0.07$ & $1.6 \pm 0.1$\\
HD\,157779 & 3.2566(8) & 58981.29(3)$^p$ & -- & -- & --    \\
HD\,164429 & 1.081751(1) & 54714.5(1)$^b$ & $-0.06 \pm 0.06$ & $-0.84 \pm 0.07$ & $1.5 \pm 0.1$ \\
HD\,168856 & 2.4277(1) & 47992.2(5)$^b$ & $-0.30 \pm 0.05$ & $0.75 \pm 0.09$ & $0.8 \pm 0.1$ \\
HD\,170973 & 18.064(5) & 46547(2)$^b$ & $0.22 \pm 0.02$ & $0.66 \pm 0.02$ & $1.62 \pm 0.03$ \\
HD\,171247 & 3.9098(8) & 55077(1)$^p$ & $0.11 \pm 0.08$ & $0.3 \pm 0.1$ & $1.7 \pm 0.3$ \\
HD\,175132 & 8.0295(2) & 58106.5(2)$^p$ & -- & -- & --    \\ 
HD\,175744 & 2.799(1)  & 47958.9(5)$^p$ & $0.03 \pm 0.07$ & $0.12 \pm 0.08$ & $2.4 \pm 0.6$ \\
HD\,183339 & 4.2040(5)$^*$ & 58682.10(6)$^p$ & -- & -- & --    \\ 
HD\,184961 & 6.335(5) & 55720(1)$^b$ & $0.65 \pm 0.03$ & $0.24 \pm 0.05$ & $1.4 \pm 0.12$ \\
HD\,188041 & 224.0(2) &	46319.5(5)$^b$ & $0.82 \pm 0.01$ & $0.51 \pm 0.02$ & $1.68 \pm 0.04$ \\
HD\,196178 & 1.10111(1) & 42592.6(2)$^b$ & $-0.94 \pm 0.04$ & $-0.20 \pm 0.05$ & $1.6 \pm 0.3$ \\
\hline\hline
\end{tabular}
\end{table*}

In order to maximize the signal-to-noise ($S/N$) of the Stokes $V$ spectrum from the ESPaDOnS and Narval observations we used the standard multiline analyis technique least-squares deconvolution \citep[LSD;][]{d1997}, specifically the {\sc iLSD} package \citep{koch2010}. LSD applies a line mask composed of the rest wavelengths of spectral lines expected for a given \teff~and $\log{g}$, their line depths, and their Land\'e factors, to the observed spectrum in order to extract a mean line profile. The line masks were obtained from the Vienna Atomic Line Database  \citep[VALD3;][]{piskunov1995, ryabchikova1997, kupka1999, kupka2000,2015PhyS...90e4005R} using `extract stellar' requests over the wavelength interval 370 nm to 1050 nm, for solar metallicity, effective temperatures from 10 kK to 25 kK in 1 kK intervals (with the closest \teff~to the measured value used for a given star), and $\log{g} = 4$. Since LSD relies on the assumption that all lines can be reproduced with a common line profile, the line masks were cleaned by removing H lines, strong He lines, metallic lines formed in the wings of these lines, as well as regions contaminated by telluric bands, interstellar lines as appropriate for a given star, and parts of the spectrum affected by instrumental ripples. The line masks were then `tweaked' by hand such that the depths matched the observed depths of each star. 

The results of the LSD profile analysis are summarized for each star in Table \ref{lsd_table}, and representative LSD profiles are each star are shown in Fig.\ \ref{lsd_all}. Observations were classified as magnetic definite detections (DD), marginal detections (MD), or non-detections (ND) based upon the false alarm probabilities (FAPs) in Stokes $V$ inside the line profile according to the usual criteria, i.e.\ an ND if FAP$>10^{-3}$, MD if $<10^{-5}$~FAP~$<10^{-3}$, and a DD if FAP~$<10^{-5}$ \citep{dsr1992,d1997}. The number of DDs, MDs, and NDs are given in Table \ref{lsd_table}. Out of the 20 stars analyzed, 5 yielded at least 1 DD, 1 yielded 2 MDs, and the remainder were NDs. All null $N$ profiles yielded NDs, verifying normal instrument performance. 

To quantify the strength of the magnetic field the line-of-sight magnetic field averaged over the stellar disk \bz~was measured using the moment method \citep{mat1989}. The same measurement was performed on the $N$ profiles. Table \ref{lsd_table} gives the root-mean-square of \bz~and \nz, the mean \bz~error bar, and the maximum \bz~value measured. Note that for stars for which only one observation was obtained, there is no distinction between the root-mean-square and maximum \bz. 

\subsection{Oblique rotator models}

In order to determine the surface strength and geometry of the surface magnetic field, \bz~measurements were folded with the rotation periods determined using photometric or magnetic data, and harmonic functions in the same method as that described by \cite{2018MNRAS.475.5144S}. For this purpose \bz~measurements found in the literature, DAO \bz~measurements, and the LSD measurements described above were utilized as available. DAO measurements are summarized in Table \ref{dao_table}. The fitting parameters $B_0$, $B_1$, and $\Phi_1$, corresponding to the mean value of \bz, the amplitude of the \bz~sine wave, and the phase offset, are given in Table \ref{bz_table}. In all but one case a pure sine wave gives an adequate description of \bz, as evaluated by the reduced $\chi^2$; the exception being HD\,21699, which requires two additional harmonics. Table \ref{bz_table} also gives the rotational periods $P_{\rm rot}$ and zero points $T_0$ for the rotatational ephemerides, where the last number in brackets gives the uncertainty in the last significant digit. In most cases $T_0$ corresponds to magnetic maximum; when an insufficient number of \bz~measurements were available to constrain this, maximum light was used instead.

The \bz~fitting parameters were then used to determine the obliquity $\beta$ and surface strength of the magnetic dipole $B_{\rm d}$ from the geometric relations developed by \cite{preston1967}, following the same procedure as that described by \cite{2019MNRAS.490..274S}. In a few cases there were not enough \bz~measurements to constrain a fit; in those cases $B_0$ and $B_1$ were approximated by the mean and standard deviation of \bz. 

\subsection{Notes on individual stars}

In the following, the characteristics of the individual datasets for each star, and the particular considerations affecting its analysis, are briefly described. 


\noindent {\bf ALS\,8988}: This is a magnetic Herbig Be star, with 4 \bz~measurements published by \cite{2008A&A...481L..99A}, all negative and of approximately the same magnitude. The star is in a crowded field, and the {\em TESS} light curve (obtained in sectors 6 and 33) is certainly contaminated by multiple objects. The light curves in the two sectors do not resemble one another, and yield different sets of significant periods: in order of decreasing signal-to-noise, about 33~d, 3~d, and 8~d in the sector 6 light curve, and 12.5~d, 6.3~d, 27.8~d, 4.3~d, ad 3.2~d in the sector 33 light curve. The longer periods likely reflect systematics, while th shorter periods are not harmonically related. The rotation period cannot be determined.

   \begin{figure}
   \centering
   \includegraphics[width=0.45\textwidth]{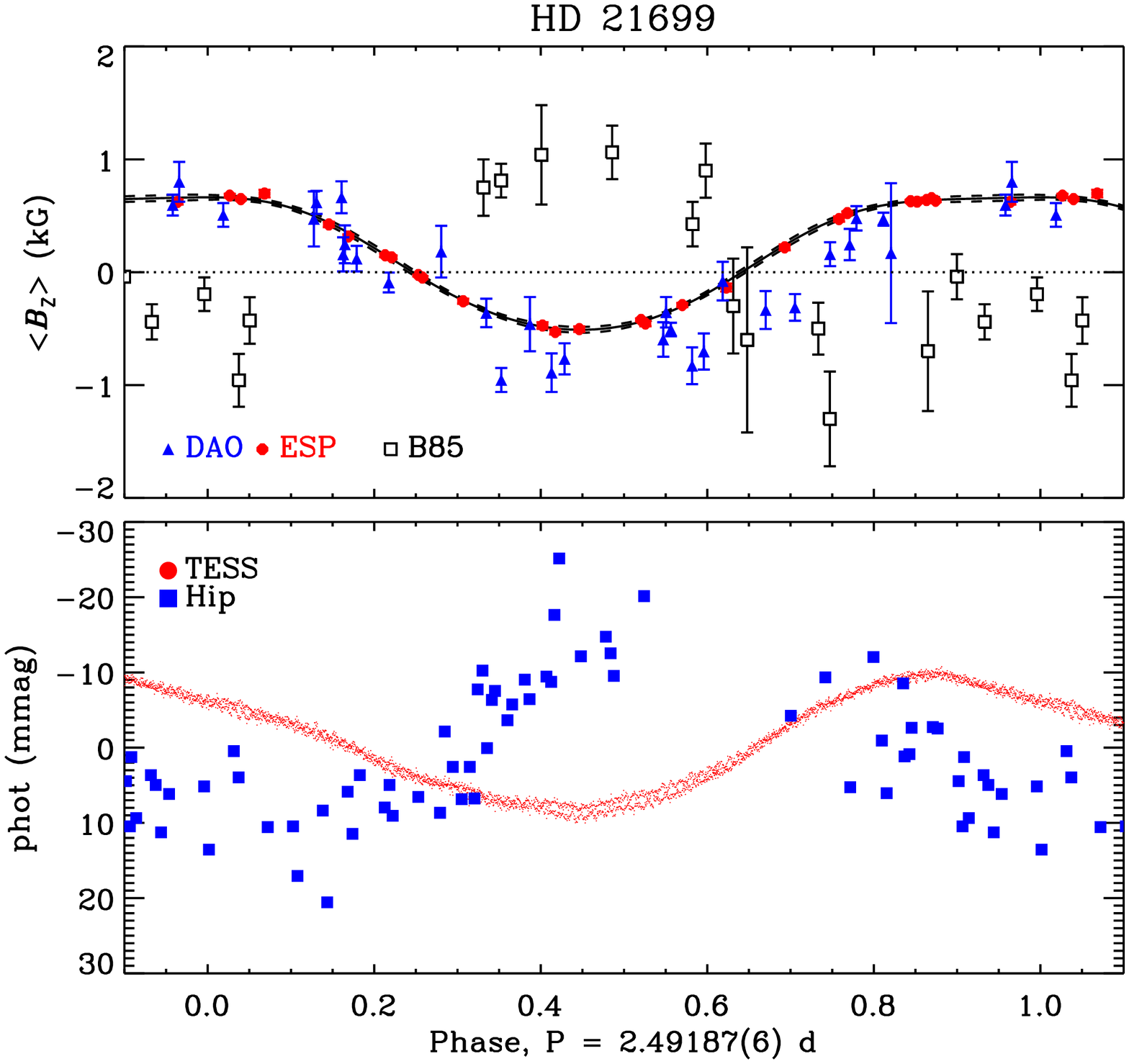}
      \caption[]{{\em Top}: \bz~measurements for HD\,21699, obtained from \protect\cite{1985AJ.....90.1354B}, DAO, and ESPaDOnS data, folded with the rotation period. Solid/dashed curves show the third-order harmonic fit to the ESPaDOnS data and the fit uncertainty. {\em Bottom}: {\em TESS} and {\em Hipparcos} light curves folded with the rotation period.}
         \label{hd21699_bz}
   \end{figure}

\noindent {\bf HD\,21699}: This star is listed in the \cite{2009A&A...498..961R} catalogue as a B8 He-weak Si star. \cite{2020MNRAS.493.3293B} give a photometric period of 2.4928(2)~d based on ASAS-3, KELT, and MASCARA data. There is a large ESPaDOnS dataset, with 26xDD (Fig.\ \ref{lsd_all}). The period from \cite{2020MNRAS.493.3293B} coherently phases the ESPaDOnS measurements, but not the full magnetic dataset. Period analysis of the ESPaDOnS \bz~measurments yields 2.49187(7)~d, however this period does not coherently phase the ESPaDOnS measurements with the \bz~measurements published by \cite{1985AJ.....90.1354B}, from which a period of 2.4928(9)~d is obtained, in formal agreement with the \citeauthor{2020MNRAS.493.3293B} period. A period obtained from Lomb-Scargle analysis of the full magnetic dataset can approximately phase the data, but the phasing of the ESPaDOnS measurements is noticeably worse. The {\em Hipparcos} and TESS light curves respectively yield 2.493(4)~d and 2.487(3)~d; combining them gives 2.49208(2)~d, however this considerably degrades the phasing of the ESPaDOnS data. We therefore revert to the ESPaDOnS period in Fig.\ \ref{hd21699_bz}. It is possible that the rotational period of this star may be variable. The ESPaDOnS \bz~curve, which is much more precise than the \cite{1985AJ.....90.1354B} measurements, is anharmonic, indicating that the surface field of this star is not a pure dipole.

   \begin{figure}
   \centering
   \includegraphics[width=0.45\textwidth]{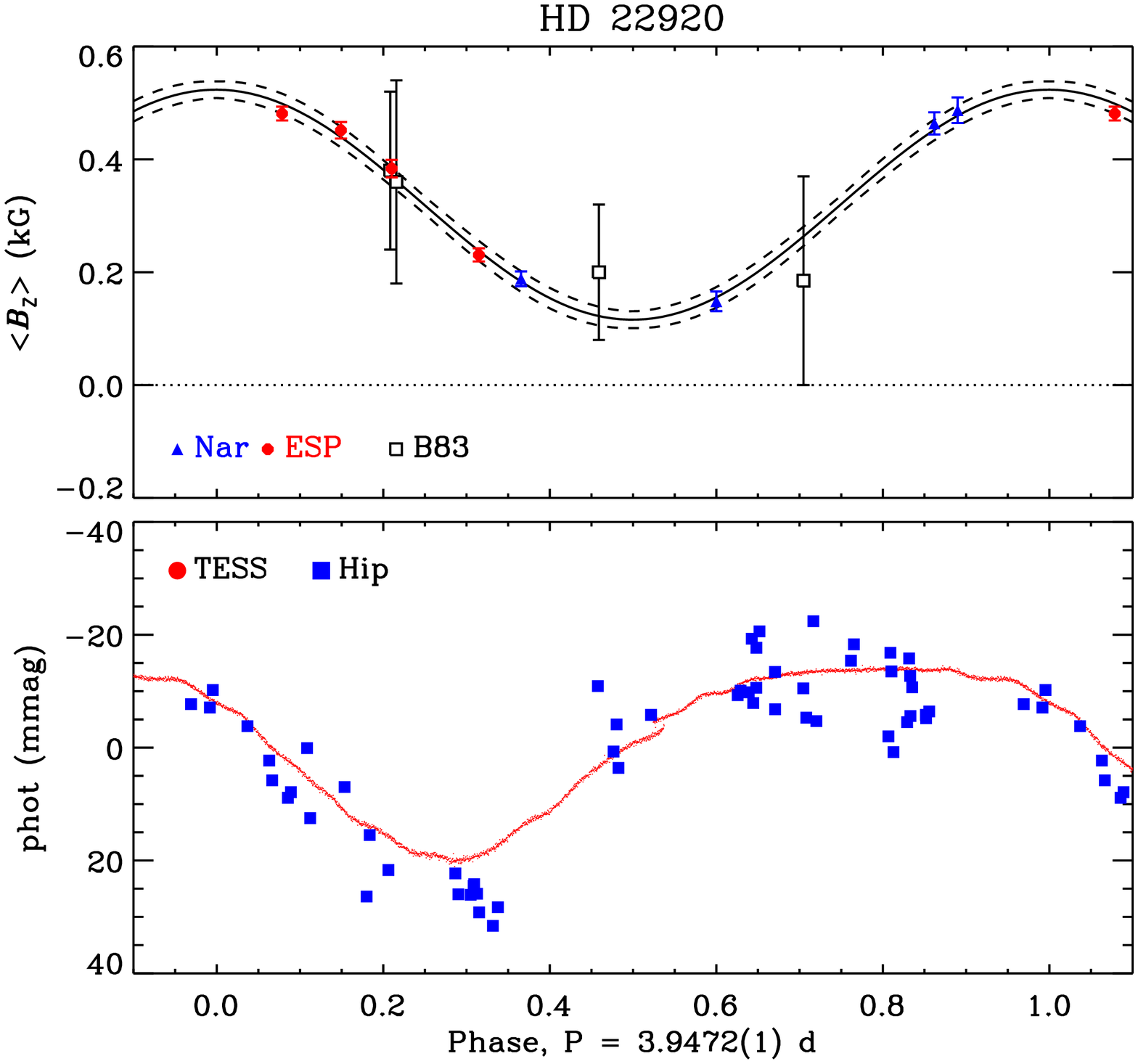}
      \caption[]{{\em Top}: \bz~measurements for HD\,22920, obtained from \protect\cite{1983ApJS...53..151B}, ESPaDOnS, and Narval, folded with the rotation period. Solid/dashed curves show first-order harmonic fit and uncertainty. {\em Bottom}: {\em Hipparcos} and {\em TESS} light curves folded with the rotation period.}
         \label{hd22920_bz}
   \end{figure}

\noindent {\bf HD\,22920}: This star is listed in the \cite{2009A&A...498..961R} catalogue as a B8 Si star. \cite{1998A&AS..127..421C} give a 3.95~d rotation period. \cite{2020MNRAS.493.3293B} provide a photometric period of 3.9489(3)~d. The {\em TESS} light curve gives a 3.956(3)~d period; combining it with {\em Hipparcos} yields 3.94724(1)~d. All 8 ESPaDOnS observations are definite detections (Fig.\ \ref{lsd_all}). This period gives a good phasing of the ESPaDOnS and Narval data; while it does not quite phase the modern measurements with the measurements published by \cite{1983ApJS...53..151B}, the latter have large uncertainties. The photometric and magnetic data are shown in Fig.\ \ref{hd22920_bz}.

\noindent {\bf HD\,23408}: This star is listed in the \cite{2009A&A...498..961R} catalogue as a B7 He-weak Mn star. No magnetic field was detected in the FORS observation by \cite{2015A&A...583A.115B}, with a 64 G error bar. No magnetic field was detected in 5 DAO measurements, with $B_{\rm rms} = 105$~G and $\langle \sigma_B \rangle = 93$~G. There are 3 ESPaDOnS measurements, all non-detections, with the smallest error bar of 5 G (Fig.\ \ref{lsd_all}). There are no obviously significant periodicities in the {\em Hipparcos} or {\em K2} light curves. This is not a magnetic star and was removed from the analysis.

   \begin{figure}
   \centering
   \includegraphics[width=0.45\textwidth]{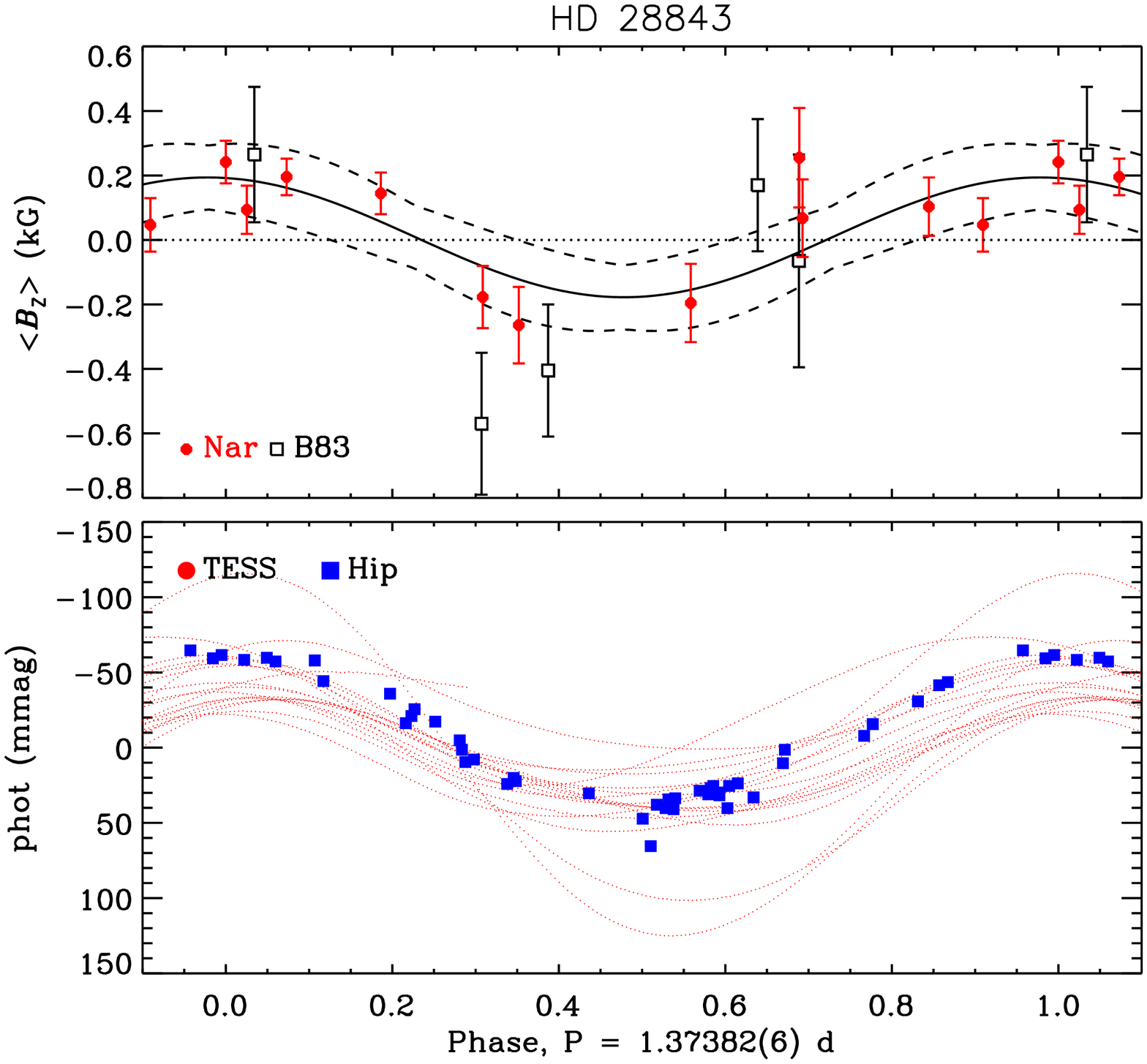}
      \caption[]{{\em Top}: \bz~measurements for HD\,28843, obtained from \protect\cite{1983ApJS...53..151B} and Narval, folded with the rotation period. Solid/dashed curves show first-order harmonic fit and uncertainty. {\em Bottom}: {\em Hipparcos} and {\em TESS} light curves folded with the rotation period.}
         \label{hd28843_bz}
   \end{figure}

\noindent {\bf HD\,28843}: This star is listed in the \cite{2009A&A...498..961R} catalogue as a B9 He-weak Si star. \cite{2005AA...430.1143B} inferred the \bz~curve from only 5 measurements from \cite{1983ApJS...53..151B}, using the 1.37~d period from \cite{1986A&AS...63..403M}. There are 11 Narval observations (Fig.\ \ref{lsd_all}), two yielding marginal detections and the remainder non-detections, with a mean error bar of 95 G. The 1.37~d periodicity is absolutely unambiguous in the {\em Hipparcos} light curve, and the \bz~measurements phase coherently with this period. The same is true of the TESS light curve, although there are large-scale variations in the amplitude of the TESS light curve (see the bottom panel of Fig.\ \ref{hd28843_bz}) which appear to be real and require further analysis to determine their origin. The evidence suggests either that this is a rapidly rotating star with a weak magnetic field, or that the period is related to pulsation. The highly variable LSD Stokes $I$ profiles are consistent with either hypothesis.

   \begin{figure}
   \centering
   \includegraphics[width=0.45\textwidth]{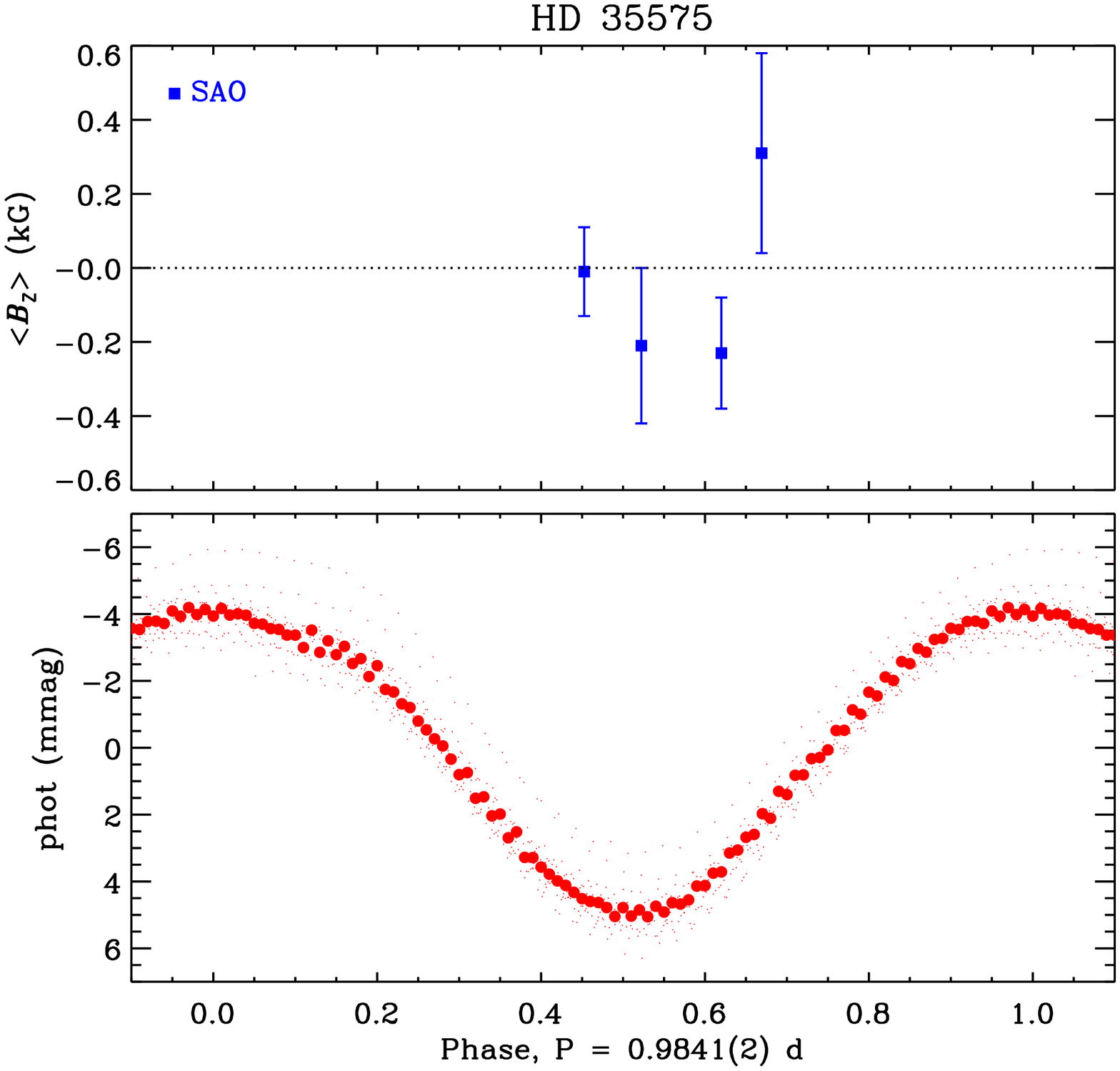}
      \caption[]{SAO \bz~measurements ({\em top}) and {\em TESS} photometry ({\em bottom}) of HD\,35575 phased with the rotation period. Larger points indicate phase-binned data.}
         \label{hd35575_bz_phot}
   \end{figure}

\noindent {\bf HD\,35575}: This star is listed in the \cite{2009A&A...498..961R} catalogue as a B3 He-weak star. The {\em TESS} light curve was obtained in sector 6. Despite some systematics, there is clear rotational modulation at 0.9841(2)~d, with several harmonics of the rotational frequency present. The light curve is shown folded with the rotation period in Fig.\ \ref{hd35575_bz_phot}. This is consistent with the star's large \vsini~$=150$~\kms~\citep{2019AstBu..74...55R}. Four magnetic measurements were reported by \cite{2019AstBu..74...55R}, all non-detections, however as can be seen in Fig.\ \ref{hd35575_bz_phot} they were all acquired at a similar rotation phase; it is possible that observation at other phases may lead to a successful detection of the field. 

   \begin{figure}
   \centering
   \includegraphics[width=0.45\textwidth]{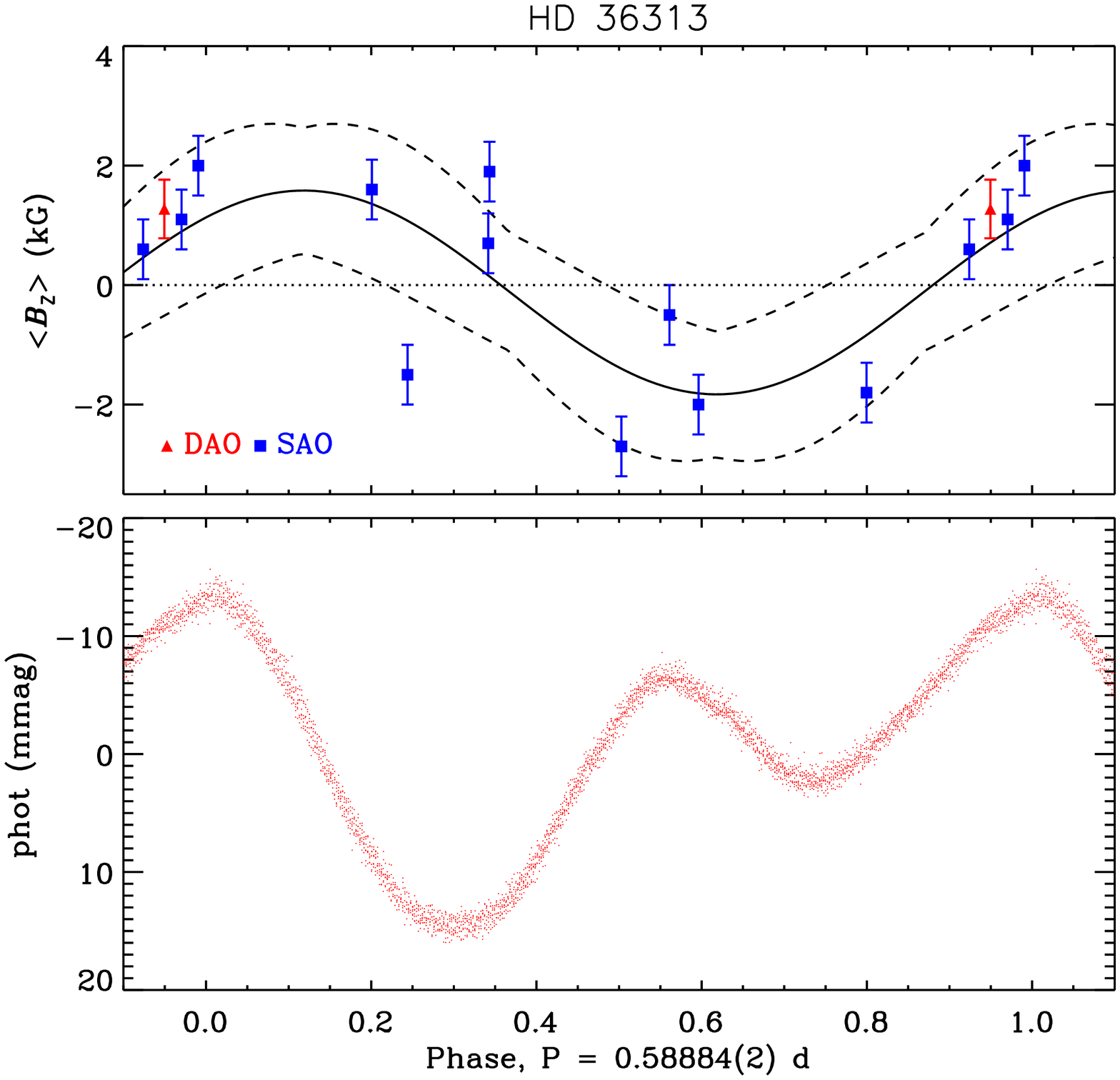}
      \caption[]{{\em Top}: SAO and DAO \bz~measurements from \cite{2021AstBu..76...39R} phased with the rotation period. {\em Bottom}: {\em TESS} photometry phased with the rotation period.}
         \label{HD36313_bz}
   \end{figure}

\noindent {\bf HD\,36313}: \cite{2021AstBu..76...39R} obtained several \bz~measurements from this SB2 star, and determined a rotation period of 0.58913~d from the TESS light curve. They noted that the magnetic field could only be detected in H lines, due to contamination of the low-resolution spectra by the non-magnetic secondary. The TESS period does not phase the \bz~measurements. \cite{2021AstBu..76...39R} noted the presence of other periods in the light curve, which they attributed to the rotation period of the cooler secondary; this periodicity appears to be only marginally significant. The light curve contains 11 harmonics of the rotational frequency; simultaneous fitting of the rotational frequency and all harmonics yields 0.58921(1)~d, which still does not provide a coherent phasing of \bz. The closest period that provides an acceptable phasing is 0.58884(2)~d, which is used to phase the \cite{2021AstBu..76...39R} and DAO \bz~measurements in Fig.\ \ref{HD36313_bz}. While this period is well outside the formal uncertainties from period analysis of the TESS data, it provides an almost equivalently good phasing of the photometry as determined by eye. This star should certainly be followed up with high-resolution spectropolarimetry in order to evaluate the accuracy of the existing magnetic data in light of the star's binarity. 


\noindent {\bf HD\,36429}:  This star is listed in the \cite{2009A&A...498..961R} catalogue as a B6 He-weak star. No magnetic field was detected by \cite{1981ApJ...249L..39B}. There is one ESPaDOnS measurement, a non-detection with a 67 G error bar. The LSD Stokes $I$ profile may be consistent with binarity (see Fig.\ \ref{lsd_all}). The {\em TESS} light curve is dominated by a 15.6(7)~d period, which could be rotational but is clearly inconsistent with the large \vsini~(77~\kms; Fig.\ \ref{lsd_all}) and is therefore related either to orbital motion or instrumental systematics. The light curve also demonstrates numerous high-frequency signals likely related to pulsation. This is probably not a magnetic chemically peculiar star. 

   \begin{figure}
   \centering
   \includegraphics[width=0.45\textwidth]{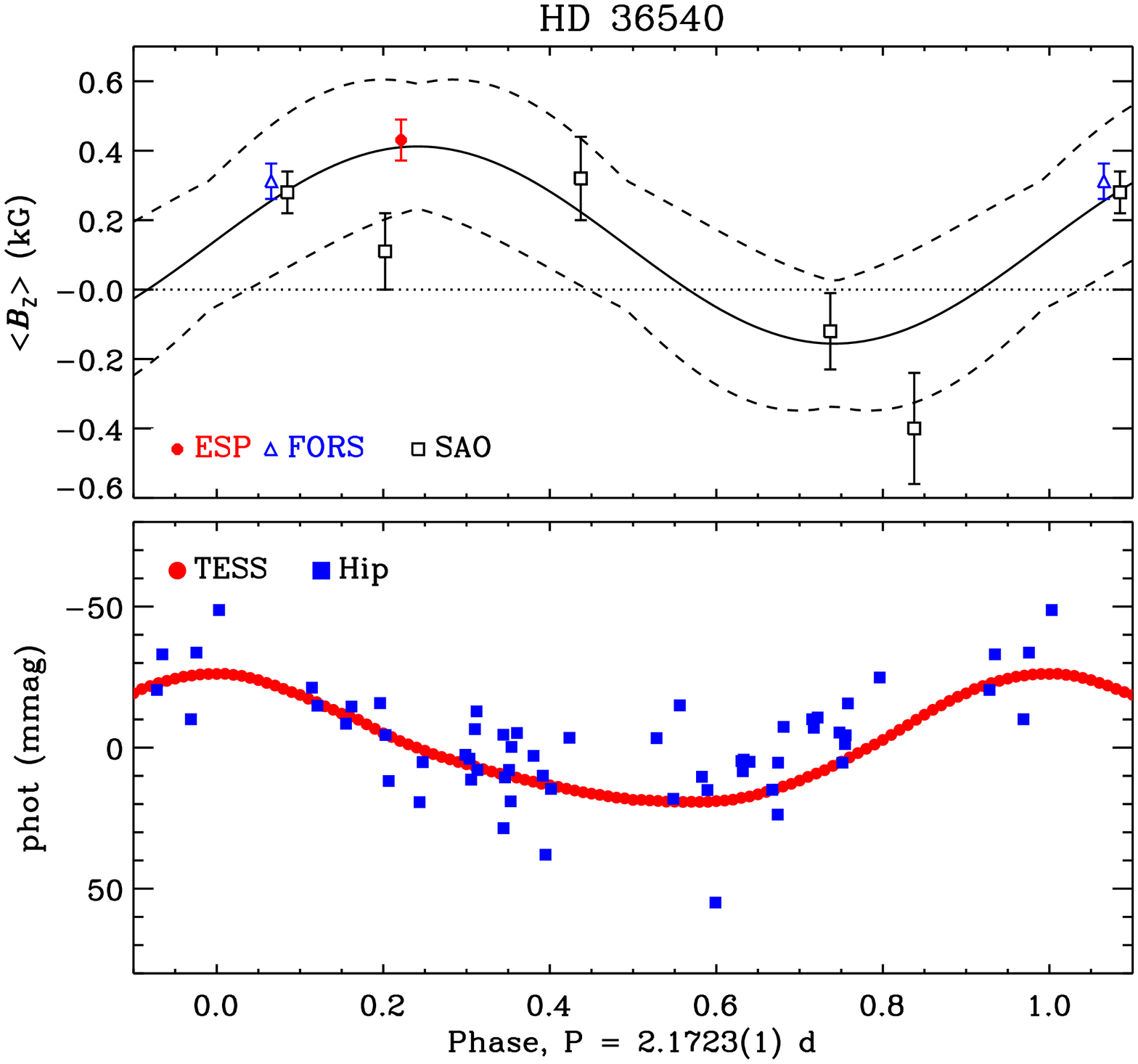}
      \caption[]{{\em Top}: \bz~measurements for HD\,36540, obtained from \protect\cite{2017AstBu..72..391R}, \protect\cite{2015A&A...583A.115B}, and ESPaDOnS, folded with the rotation period. Solid/dashed curves show first-order harmonic fit and uncertainty. {\em Bottom}: {\em Hipparcos} and {\em TESS} light curves folded with the rotation period. Larger points indicate phase-binned data.}
         \label{hd36540_bz}
   \end{figure}

\noindent {\bf HD\,36540}:  This star is listed in the \cite{2009A&A...498..961R} catalogue as a B7 He-weak star. The largest set of magnetic measurements were published by \cite{2017AstBu..72..391R}. The magnetic field was also detected by \cite{2015A&A...583A.115B}. There is 1 ESPaDOnS observation, which yields a definite detection in the LSD Stokes $V$ profile (see Fig.\ \ref{lsd_all}). By combining all available magnetic data together with the {\em Hipparcos} photometry, we get a period of 2.1725(4)~d, consistent with the {\em Hipparcos} period determined by \cite{2011MNRAS.414.2602D}. The {\em TESS} light curve gives a period of 2.1723(1)~d, formally consistent with the {\em Hipparcos} period, which is used to phase the data in Fig.\ \ref{hd36540_bz}.

\noindent {\bf HD\,36629}: This star was included in the \cite{2005AA...430.1143B} catalogue due to an old measurement by \cite{1970ApJ...159..723C}. However, its magnetic field was not detected by \cite{2015A&A...583A.115B} or by \cite{2017AstBu..72..391R}. There are 2 ESPaDOnS observations, both of which are non-detections with 4 G error bars (Fig.\ \ref{lsd_all}). The rotational period given by \cite{2005AA...430.1143B}, based on \bz~measurements, is therefore certainly spurious and this star was dropped from the analysis.

   \begin{figure}
   \centering
   \includegraphics[width=0.45\textwidth]{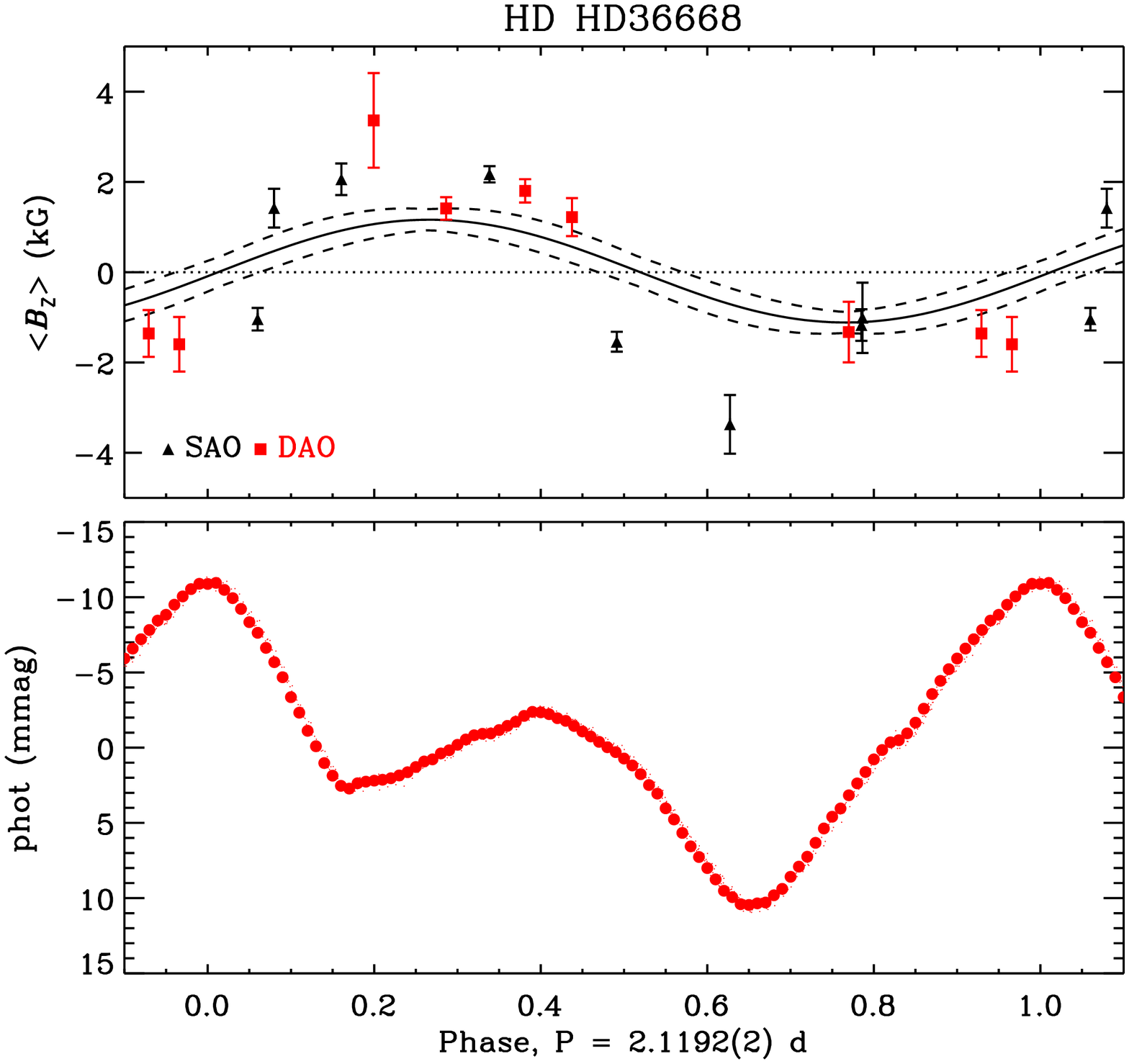}
      \caption[]{{\em Top} DAO and SAO \bz~measurements phased with the rotation period. {\em Bottom}: {\em TESS} light curve phased with the rotation period. Larger points indicate phase-binned data.}
         \label{HD36668_bz}
   \end{figure}

\noindent {\bf HD\,36668}: \cite{2021AstBu..76...39R} determined a period of 2.1204~d using the {\em TESS} light curve, however this utterly fails to phase either the SAO or DAO \bz~measurements, whether the datasets are treated individually or combined. Reanalysis of the {\em TESS} light curve yields 2.1192(2)~d. Fig.\ \ref{HD36668_bz} shows \bz~and the {\em TESS} data phased with the nearest period to the {\em TESS} period providing a reasonably coherent variation of both \bz~and {\em TESS} data, although there is considerable scatter. Additional magnetic observations, preferably at high resolution, are necessary in order to constrain both the magnetic and rotational properties. 

   \begin{figure}
   \centering
   \includegraphics[width=0.45\textwidth]{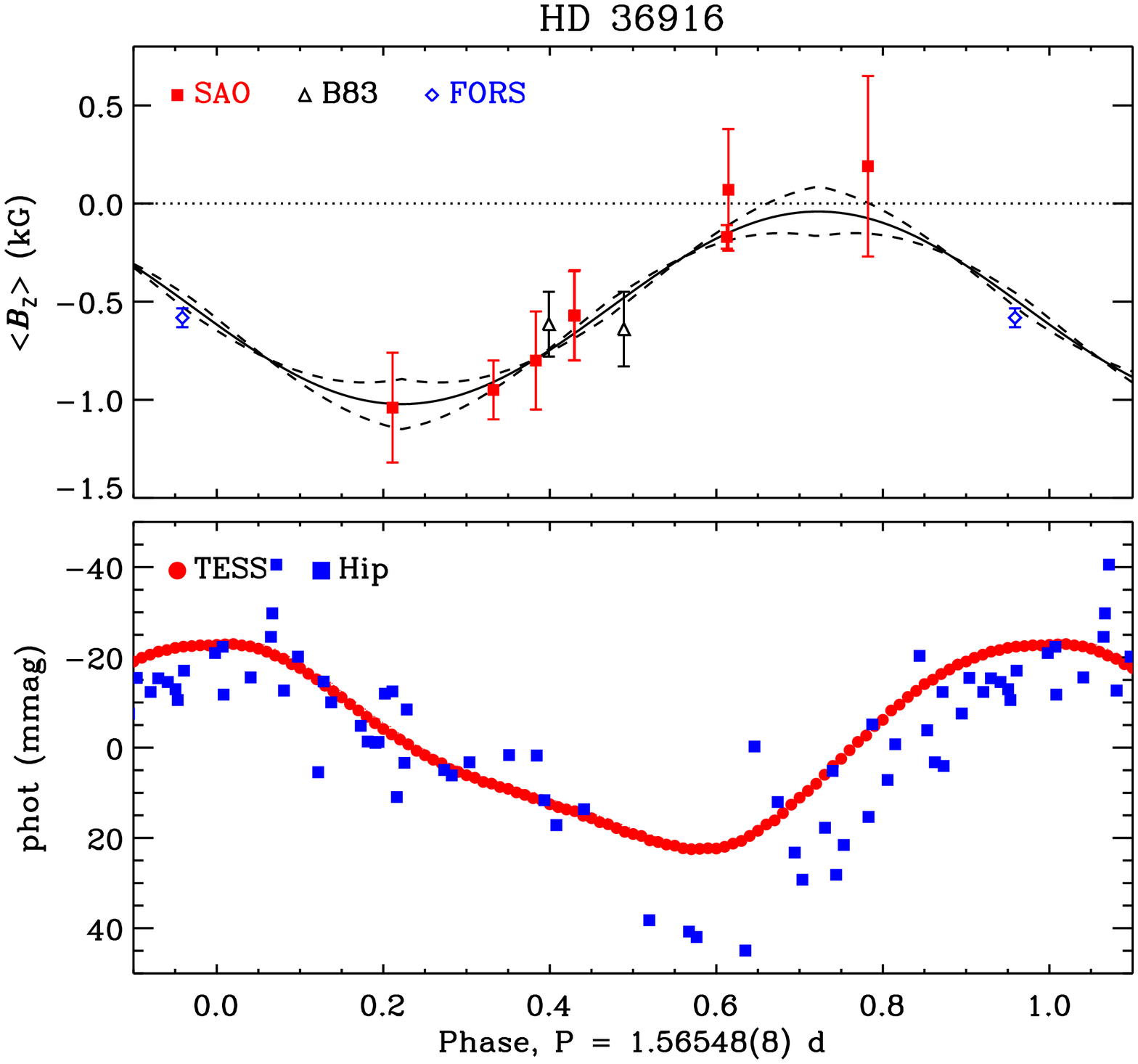}
      \caption[]{{\em Top}: \bz~measurements phased with the rotation period. {\em Bottom}: {\em Hipparcos} and {\em TESS} photometry phased with the rotation period. Larger points indicate phase-binned data.}
         \label{HD36916_bz}
   \end{figure}

\noindent {\bf HD\,36916}: The {\em Hipparcos} light curve yields a period of 1.5652(2)~d \citep{2017AstBu..72..165R}. There are \bz~measurements available from \cite{1983ApJS...53..151B}, \cite{2015AA...583A.115B}, and \cite{2017AstBu..72..165R,2018AstBu..73..178R,2020AstBu..75..294R}. The {\em TESS} period is 1.5652(1), identical to the {\em Hipparcos} period but slightly more precise. The photometric periods do not quite provide a coherent phasing of the magnetic data; the closest period which does is 1.56548(8)~d, which still provides a reasonable phasing of the photometry (Fig.\ \ref{HD36916_bz}). 

\noindent {\bf HD\,36960}: This star is listed in the \cite{2009A&A...498..961R} catalogue as a B0 Si star. No magnetic field has been detected either in FORS observations \citep{2015A&A...583A.115B} or with ESPaDOnS (10 G error bar, Fig.\ \ref{lsd_all}). Furthermore, a rotation period cannot be determined as there is no indication of regular variation in the TESS light curve. This star is probably not magnetic and was dropped from the sample.

\noindent {\bf HD\,37041}: This star was included in the \cite{2005AA...430.1143B} catalogue, however the rotation period was determined using old \bz~measurements \citep{1973ApJ...182L..43K,1975ApJ...196L.109B}, which have been superseded by modern low-resolution spectropolarimetry \citep{2015A&A...583A.115B} and by high-resolution MiMeS data \citep{2017MNRAS.465.2432G}, all non-detections. \cite{2019MNRAS.489.5669P} established an upper limit of 193 G on $B_{\rm d}$. This star is therefore not magnetic, the period is clearly spurious, and it was dropped from the analysis. 

\noindent {\bf HD\,37129}: This star is listed in the \cite{2009A&A...498..961R} catalogue as a B3 He-weak star. No magnetic field has been detected \citep{1983ApJS...53..151B}. There are numerous significant periods in the {\em TESS} light curve, which do not resemble rotation and are therefore likely pulsations. 

   \begin{figure}
   \centering
   \includegraphics[width=0.45\textwidth]{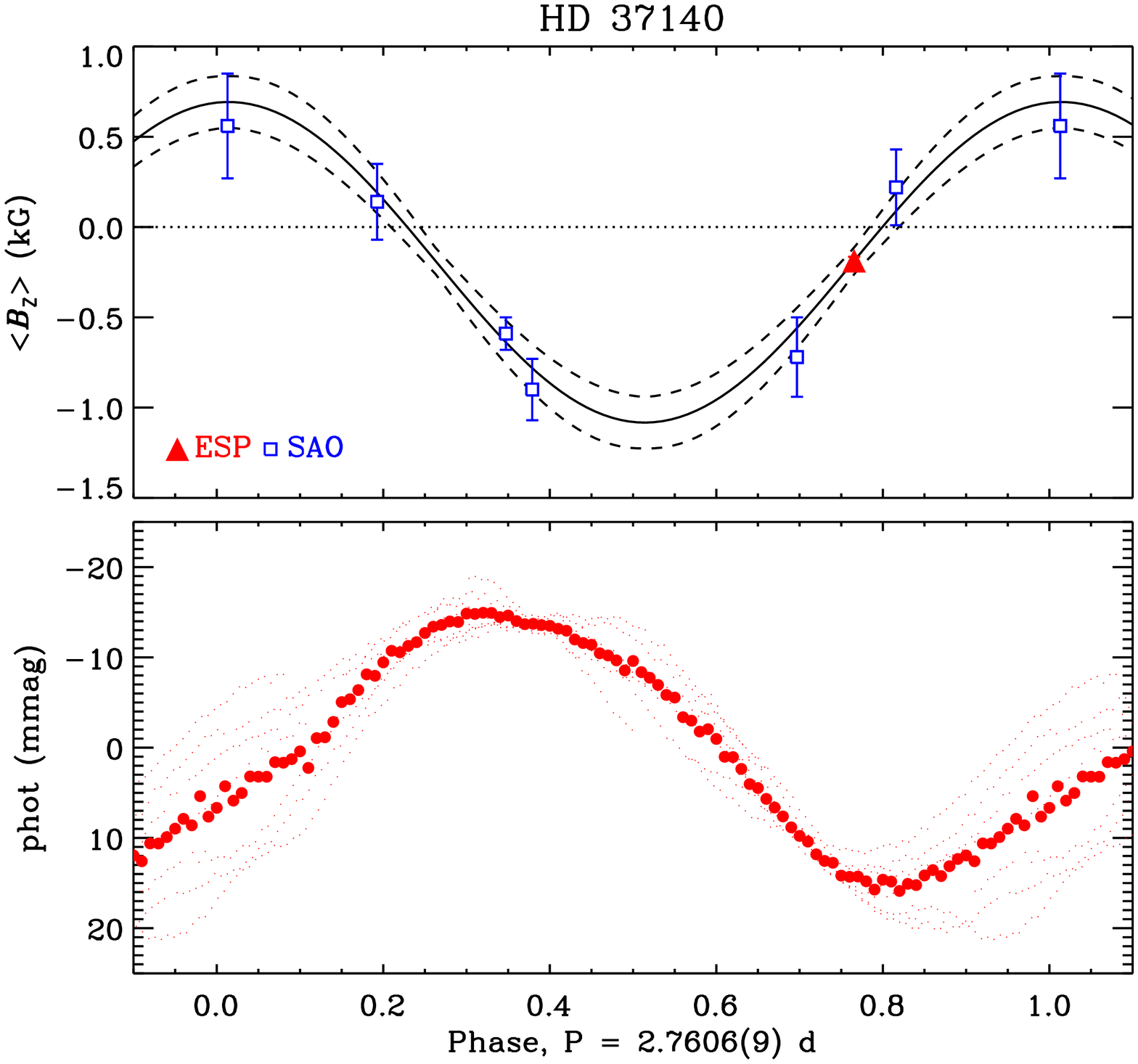}
      \caption[]{{\em Top}: SAO and ESPaDOnS \bz~measurements for HD\,37140. {\em Bottom}: TESS light curve folded with the rotation period. Larger points indicate phase-binned data.}
         \label{HD37140_bz}
   \end{figure}

\noindent {\bf HD\,37140}: This star is listed in the \cite{2009A&A...498..961R} catalogue as a B8 SiSr star. It is included in the \cite{2005AA...430.1143B} catalogue, where the 2.7~d period from \cite{1984AA...141..328N} was used to phase available \bz~measurements. More recent magnetic data have been provided by \cite{2017AstBu..72..391R,2020AstBu..75..294R}. There is one available ESPaDOnS observation, yielding a definite detection ($-186 \pm 22$~G; Fig.\ \ref{lsd_all}). \cite{2021AstBu..76...39R} used the TESS light curve to refine the period to 2.70418(1) d. However, this does not coherently phase the combined SAO and ESPaDOnS dataset. Our own analysis of the TESS data gives 2.702(2)~d, which also fails to phase the magnetic data. The closest period to the TESS period resulting in a smooth variation is 2.7606(9)~d (Fig.\ \ref{HD37140_bz}). The {\em TESS} data are severely affected by systematics (or other sources of variability beyond rotation) that may affect the accuracy of the photometric period; only one rotational cycle of {\em TESS} data are shown in Fig.\ \ref{HD37140_bz}.

\noindent {\bf HD\,37150}: This B3\,III star is not listed in the \cite{2009A&A...498..961R} catalogue, and no magnetic field was detected by \cite{1981ApJ...249L..39B}. There are 3 DAO \bz~measurements, all non-detections (see Table \ref{dao_table}). The {\em TESS} light curve has two closely spaced peaks at around 0.7 and 0.8 d, which are too short to be associated with rotation in the absence of other compelling evidence that the star is magnetic. This star was removed from the analysis.

   \begin{figure}
   \centering
   \includegraphics[width=0.45\textwidth]{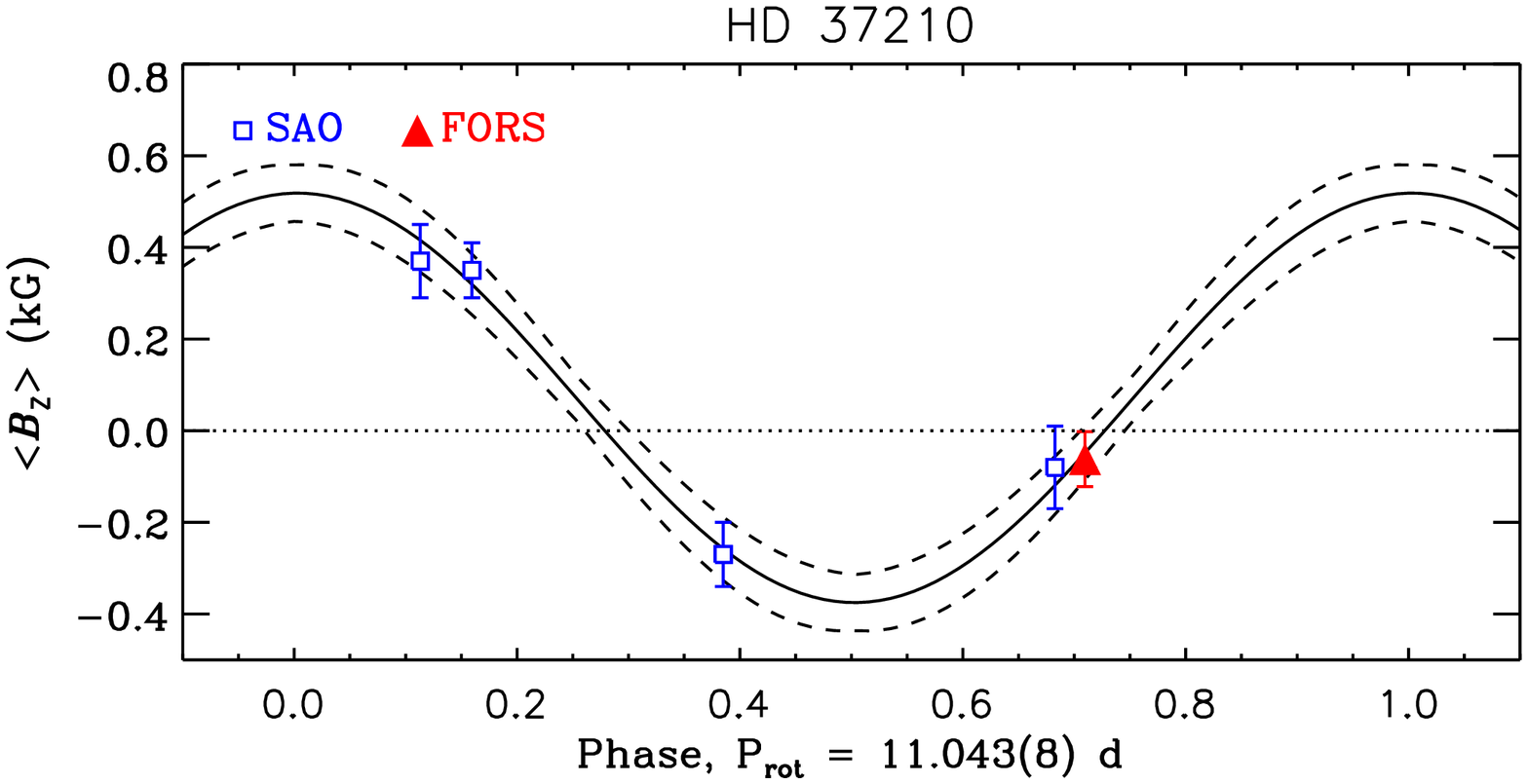}
      \caption[]{SAO and FORS \bz~measurements for HD\,37210.}
         \label{HD37210_bz}
   \end{figure}

\noindent {\bf HD\,37210}: The 11.0494(1)~d photometric period \citep{1998A&AS..127..421C} does not quite phase the available \bz~measurements from FORS \citep{2015A&A...583A.115B} and the SAO \citep{2021AstBu..76..163R}. The {\em TESS} light curve is heavily affected by systematics and cannot be used to determine the period. The closest period that coherently phases the data is 11.043(8)~d (Fig.\ \ref{HD37210_bz}).

\noindent {\bf HD\,37321}: This star is listed in the \cite{2009A&A...498..961R} catalogue as a B5 He-weak star. No magnetic measurements are available. The {\em TESS} light curve, acquired in sector 6, is dominated by high-frequencies (more than 2 c/d) that must be pulsation; there is no obvious rotational modulation in the time series. This star was removed from the sample.

   \begin{figure}
   \centering
   \includegraphics[width=0.45\textwidth]{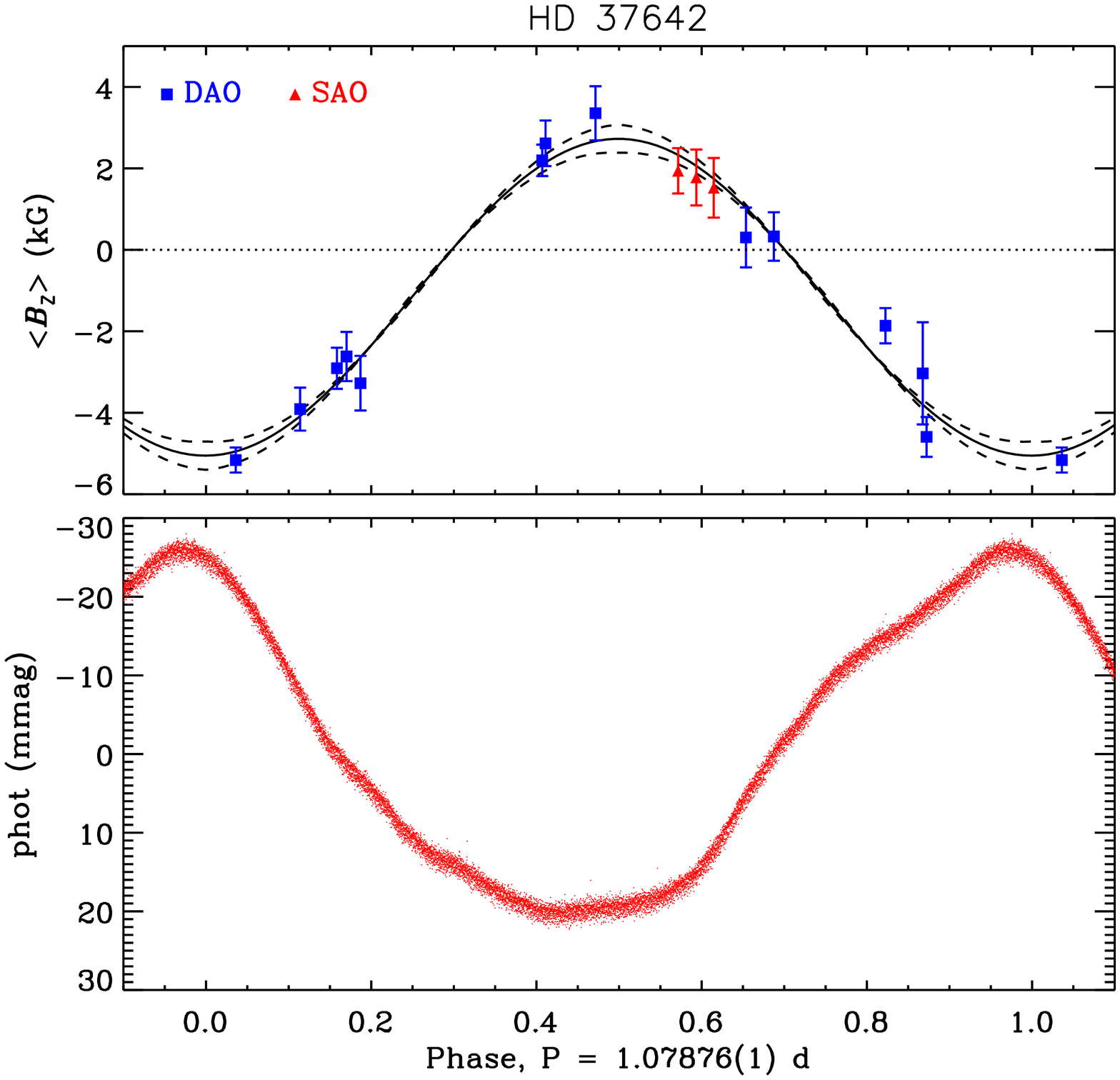}
      \caption[]{{\em Top}: DAO and SAO \bz~measurements for HD\,37642, phased with the TESS rotation period. {\em Bottom}: TESS photometry phased with the rotation period.}
         \label{HD37642_bz}
   \end{figure}

\noindent {\bf HD\,37642}: This star is listed in the \cite{2009A&A...498..961R} catalogue as a B9 He-weak Si star. The period determined from the {\em TESS} data is 1.07876(1)~d. \bz~measurements from DAO and SAO are phased with this period in Fig.\ \ref{HD37642_bz}.

   \begin{figure}
   \centering
   \includegraphics[width=0.45\textwidth]{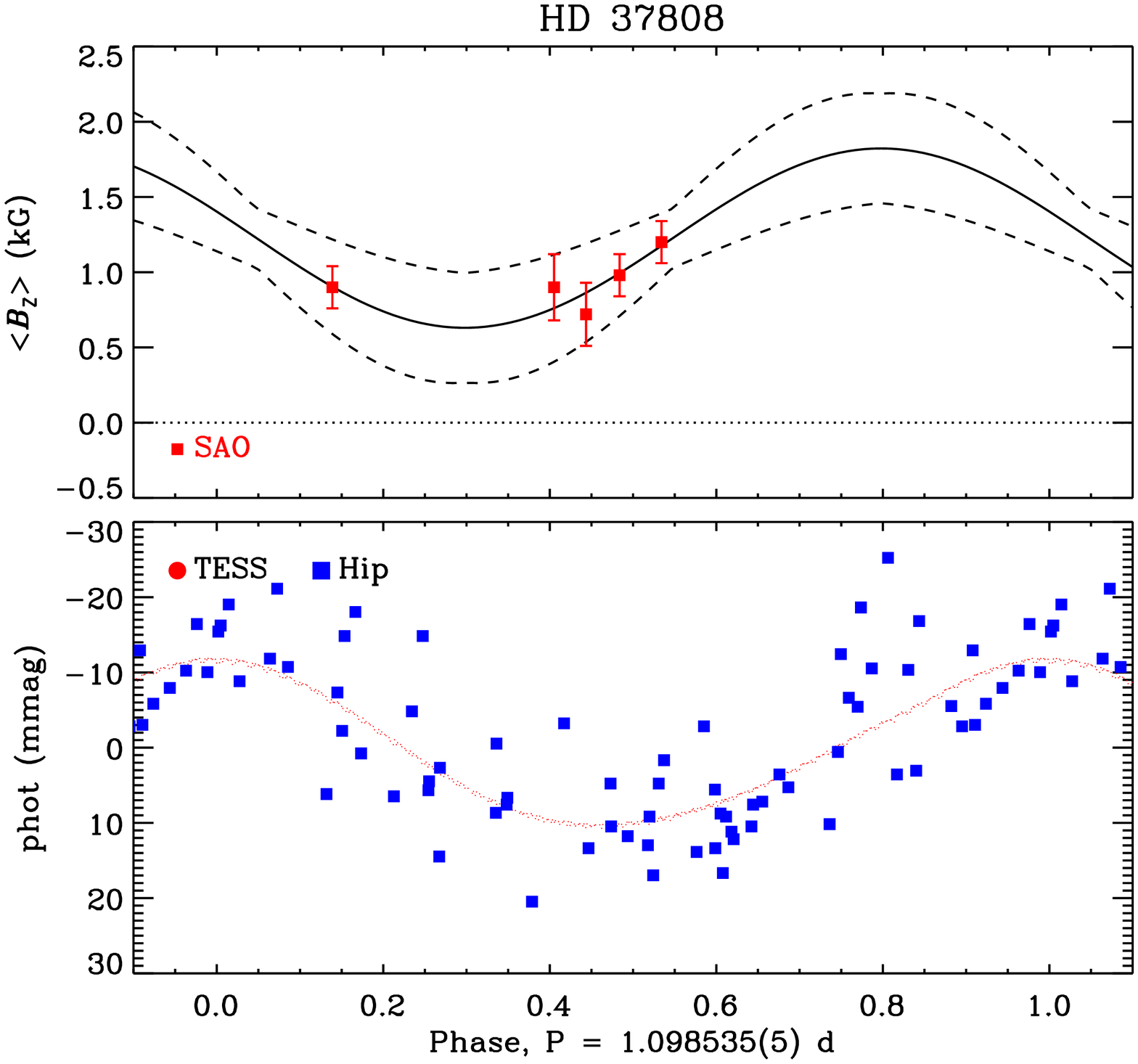}
      \caption[]{{\em Top}: DAO and SAO \bz~measurements for HD\,37808, phased with the TESS rotation period. {\em Bottom}: {\em Hipparcos} and {\em TESS} photometry phased with the rotation period.}
         \label{HD37808_bz}
   \end{figure}

\noindent {\bf HD\,37808}: This star is listed in the \cite{2009A&A...498..961R} catalogue as a B9 Si star. Clear rotational modulation is present in the {\em Hipparcos} light curve at 1.0989(1)~d \citep{2011MNRAS.414.2602D}; reanalysis of the {\em Hipparcos} data confirms this period. \cite{2020MNRAS.493.3293B} found a period of 1.09852(2) by combining KELT, ASAS-3, and MASCARA data. The {\em TESS} light curve gives a period of 1.09851(8)~d, consistent with previous results. Combining {\em TESS} and {\em Hipparcos} yields a coherent phasing with a period of 1.098535(1)~d, which is used to phase the data in Fig.\ \ref{HD37808_bz}. Five magnetic measurements were published by \cite{2021AstBu..76..163R}. This star is detected in radio, and is an obvious candidate for spectropolarimetric follow-up given the sparse magnetic coverage. 


   \begin{figure}
   \centering
   \includegraphics[width=0.45\textwidth]{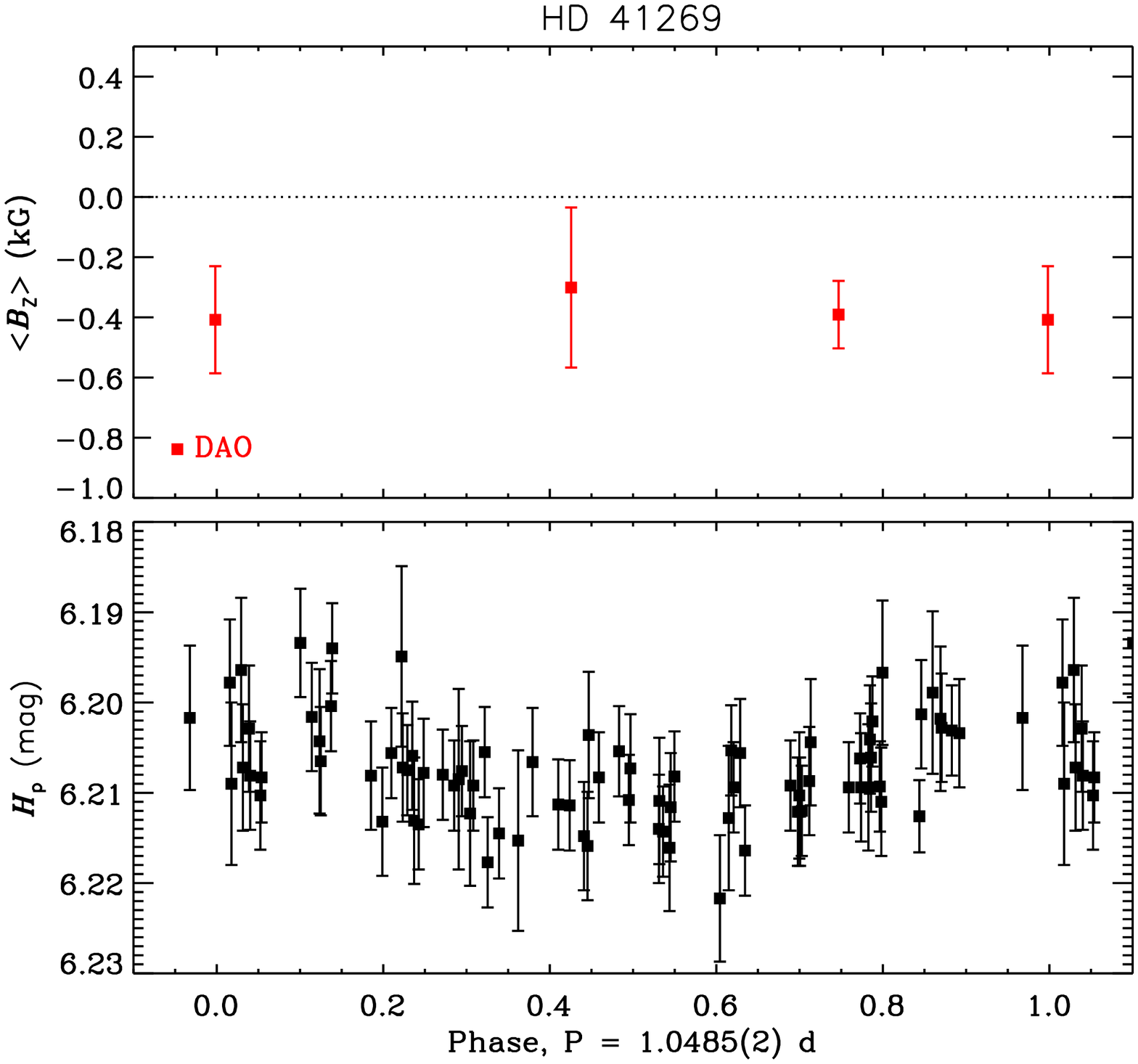}
      \caption[]{{\em Top}: DAO SAO \bz~measurements for HD\,41269, phased with the {\em Hipparcos} rotation period. {\em Bottom}: {\em Hipparcos} photometry phased with the rotation period.}
         \label{hd41269_bz_phot}
   \end{figure}

\noindent {\bf HD\,41269}: This star is listed in the \cite{2009A&A...498..961R} catalogue as a B9 Si star. The {\em Hipparcos} light curve gives possible periods of 1.0485(2)~d and 2.537(1)~d, both with FAPs of about 0.01. \cite{1998A&AS..127..421C} list possible periods of 1.68~d and 2.47~d, neither of which can be seen in the {\em Hipparcos} photometry. We adopt the shorter {\em Hipparcos} period, as the FAP is slightly lower than the longer. There are 3 DAO \bz~measurements, one of which is a 3$\sigma$ detection. As can be seen in Fig.\ \ref{hd41269_bz_phot}, all 3 measurements are negative, and while a fit cannot be performed they span the shorter {\em Hipparcos} period sufficiently to infer that only one magnetic pole is seen through a rotation cycle. 

\noindent {\bf HD\,45105}: This star is listed in the \cite{2009A&A...498..961R} catalogue as a B9 Si star. There are no magnetic data. There is no indication in the {\em TESS} light curve of rotational modulation. There is one peak in the periodogram at about 2.6~d, however it is exceedingly weak (0.015 mmag) and probably not associated with rotation.

\noindent {\bf HD\,49606}: \cite{2005AA...430.1143B} found a 1.1~d period using \bz~measurements collected from \cite{1993A&A...269..355B} and \cite{1997smf..proc..197B}. It is listed in the \cite{2009A&A...498..961R} catalogue as a B8 HgMn star, a class generally known to be non-magnetic \citep{2011A&A...525A..97M}. There are two ESPaDOnS observations, both non-detections with 7 G error bars (Fig.\ \ref{lsd_all}). The period determined from \bz~measurements is therefore certainly spurious. The TESS light curve give a period of 8.546(1)~d \citep{2021MNRAS.506.5328K}. 

\noindent {\bf HD\,50204}: This star is listed in the \cite{2009A&A...498..961R} catalogue as a B9 Si star. There are no magnetic data. There is no suggestion in the {\em TESS} light curve of rotational modulation. While there is a single peak in the periodogram at about 2.3~d, it is very weak (0.04 mmag) and therefore unlikely to be associated with rotation.

   \begin{figure}
   \centering
   \includegraphics[width=0.45\textwidth]{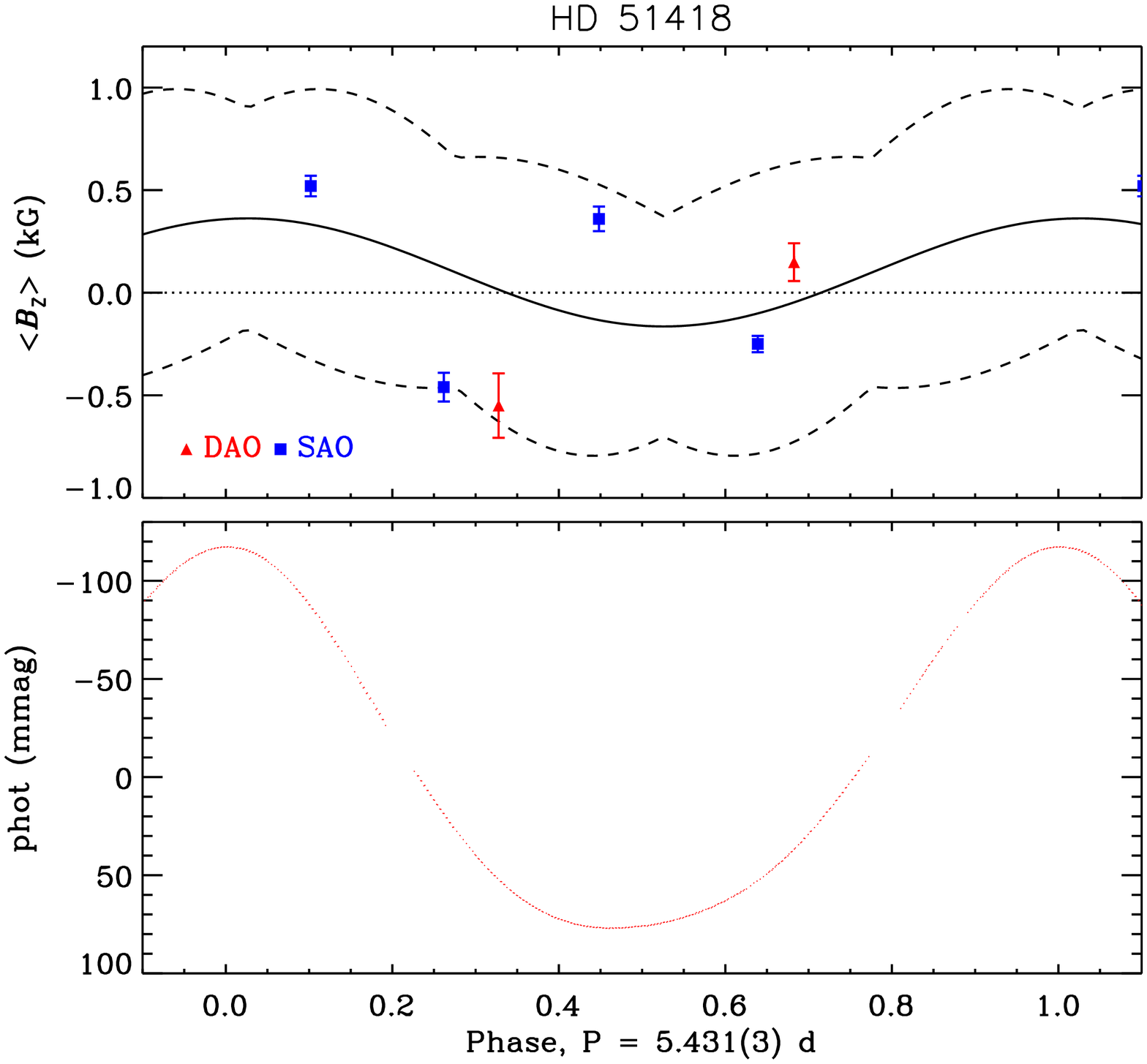}
      \caption[]{{\em Top}: DAO and SAO \bz~measurements for HD\,51418, phased with the TESS rotation period. {\em Bottom}: TESS photometry phased with the rotation period.}
         \label{HD51418_bz}
   \end{figure}

\noindent {\bf HD\,51418}: \cite{2021A&A...652A..31B} found a rotation period of 2.2908 d from the \cite{2017AstBu..72..391R} SAO magnetic measurements, in disagreement with the 5.4379 d photometric period from \cite{2001A&A...378..113R}. The {\em TESS} light curve shows clear rotational modulation with a period of 5.431(3)~d (Fig.\ \ref{HD51418_bz}). Combining the SAO \bz~measurements with DAO \bz~measurements yields a period of 2.3357(4)~d, with no power in the periodogram at higher periods. As can be seen in Fig.\ \ref{HD51418_bz}, the magnetic and photometric periods are completely incompatible assuming a single-wave variation in \bz. We adopt the photometric period, and rather than sinusoidal fitting parameters for \bz~use the mean and standard deviation of \bz~to approximate $B_0$ and $B_1$.

\noindent {\bf HD\,57219}: This star is listed in the \cite{2009A&A...498..961R} catalogue as a B3 He-variable star. There are no magnetic data. There is possible rotational modulation in the {\em TESS} light curve with a period of about 1.4~d, however the 0.4~mmag amplitude is much lower than is usually associated with chemical spots \citep{2019MNRAS.487.4695S}, there are several other closely spaced peaks, and the peaks are fairly broad, all of which suggest that this is pulsation rather than rotation. This star was removed from the analysis.

\noindent {\bf HD\,60344}: This star is listed in the \cite{2009A&A...498..961R} catalogue as a B3 He star. A magnetic field was detected by \cite{2018A&A...618L...2J}. {\em TESS} observations were acquired in sectors 7 and 34. As the star is in a crowded region, contamination by other sources is highly likely. The sector 34 light curve is heavily affected by systematics. The sector 7 light curve is affected by a long-term trend, likely systematic; while there is real variability on top of this, with an apparent period of about 9 days, it is irregular in amplitude and there is no reason to think it is related to the star's rotation.

\noindent {\bf HD\,78556A}: This star is listed in the \cite{2009A&A...498..961R} catalogue as a B9 Si star. There are no magnetic data. There is no indication in the {\em TESS} light curve of rotational modulation. 

\noindent {\bf HD\,89822}: \cite{2005AA...430.1143B} phased the available \bz~measurements \citep{1958ApJS....3..141B,1970ApJ...159..723C,1980ApJS...42..421B,1993A&A...269..355B,1993BCrAO..87...83P} with a 7.6 d period determined photometrically by \cite{1991A&A...244..327C}. The star is listed in the \cite{2009A&A...498..961R} catalogue as an A0 HgSiSr star. The 7.6 d period cannot be confirmed via {\em Hipparcos} photometry. Furthermore, all 5 of the available ESPaDOnS observations are non-detections, with 2 G error bars (Fig.\ \ref{lsd_all}). The rotation period is therefore spurious, the star is furthermore not magnetic, and it was consequently dropped from the sample. The system is an SB2, with a HgMn primary and an Am secondary in an 11.579~d orbit \citep{1994MNRAS.266...97A,2004A&A...424..727P}; \cite{2021MNRAS.506.5328K} found an 11.581~d periodicity in the {\em TESS} light curve, which they attributed to heartbeat variability.

   \begin{figure}
   \centering
   \includegraphics[width=0.45\textwidth]{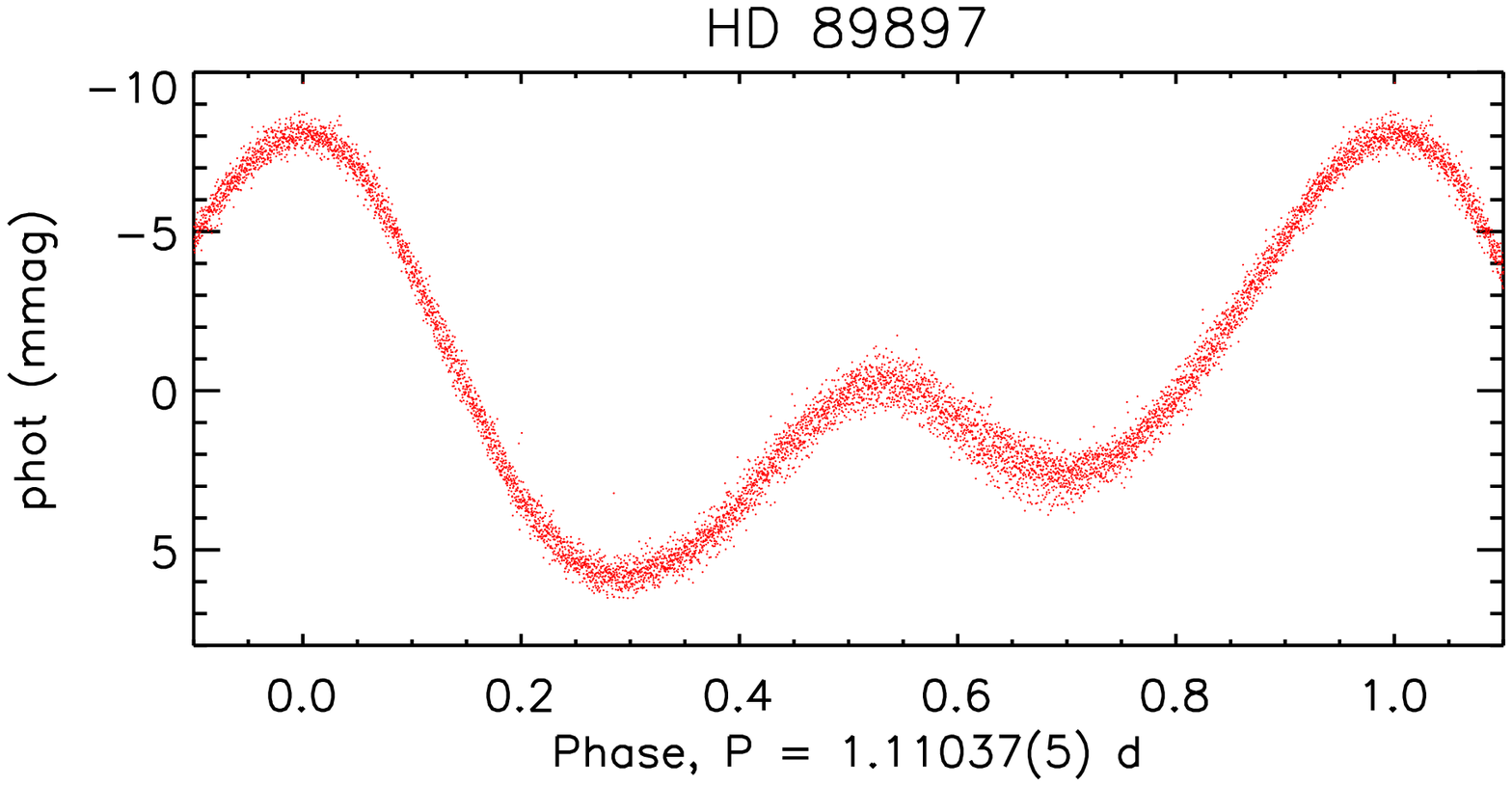}
      \caption[]{{\em TESS} light curve of HD\,89897 folded with the rotation period.}
         \label{HD89897_TESS}
   \end{figure}

\noindent {\bf HD\,89897}: This star is listed as a B9 SiCr star by \cite{2009A&A...498..961R}. No magnetic data are available. The {\em TESS} light curve yields a clear 1.11037(5)~d periodicity (see Fig.\ \ref{HD89897_TESS}). Despite the rapid rotation, the star is radio-dim, suggesting that its magnetic field must be weak if present.

\noindent {\bf HD\,120709}: This star is listed in the \cite{2009A&A...498..961R} catalogue as a B5\,He-weak star. The {\em TESS} light curve is dominated by a 0.27~d period and its 1st harmomic, almost certainly pulsation as this is much too fast to be rotation. Furthermore, all magnetic measurements are non-detections \citep{1983ApJS...53..151B,2015A&A...583A.115B}. The star was dropped from the analysis.

   \begin{figure}
   \centering
   \includegraphics[width=0.45\textwidth]{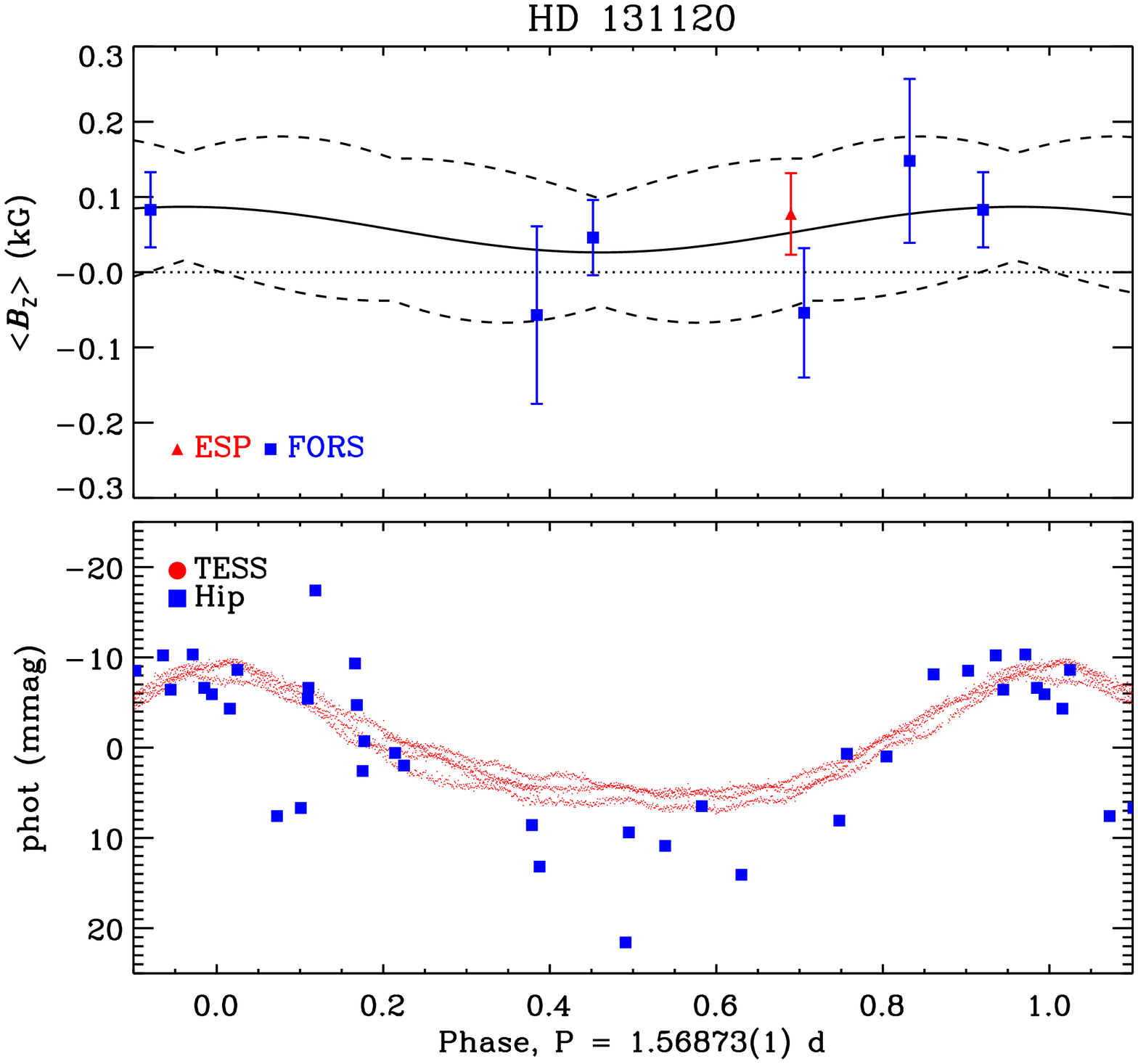}
      \caption[]{{\em Top}: \bz~measurements phased with the rotation period determined from photometric and magnetic data. {\em Bottom}: TESS photometry phased with the rotation period}
         \label{HD131120_bz}
   \end{figure}

\noindent {\bf HD\,131120}: This star is listed in the \cite{2009A&A...498..961R} catalogue as a B6 He-weak star. {\em Hipparcos} photometry yields a 1.5690(3)~d period \citep{2011MNRAS.414.2602D}. The {\em TESS} light curve gives a period of 1.5686(1)~d, formally compatible with the {\em Hipparcos} period. The {\em TESS} data shows some evidence of variability between cycles, which may indicate that the period is due to pulsation rather than rotation, or that it is also affected by pulsation (Fig.\ \ref{HD131120_bz}). Combining the {\em TESS} and {\em Hipparcos} data yields a coherent phasing with a period of 1.56873(1)~d. No magnetic field was detected in 5 FORS2 observations, with error bars of 50-100 G \citep{2015A&A...583A.115B}. There is 1 ESPaDOnS observation, also a non-detection with a 50 G error bar (Fig.\ \ref{lsd_all}). Asymmetry in the LSD profile is indicative of binarity. Under the assumption that the $\sim$1.5~d period is in fact rotational, the available \bz~measurements are phased with the {\em TESS} period in Fig.\ \ref{HD131120_bz}. Since the magnetic field was not detected, $B_0$ and $B_1$ were approximated using the mean and standard deviation of \bz. As can be seen, while no magnetic field is detected there is a tendency for \bz~to be systematically positive, which may be indicative that the star is indeed magnetic, even though no individual measurement is formally consistent with a detection.

   \begin{figure}
   \centering
   \includegraphics[width=0.45\textwidth]{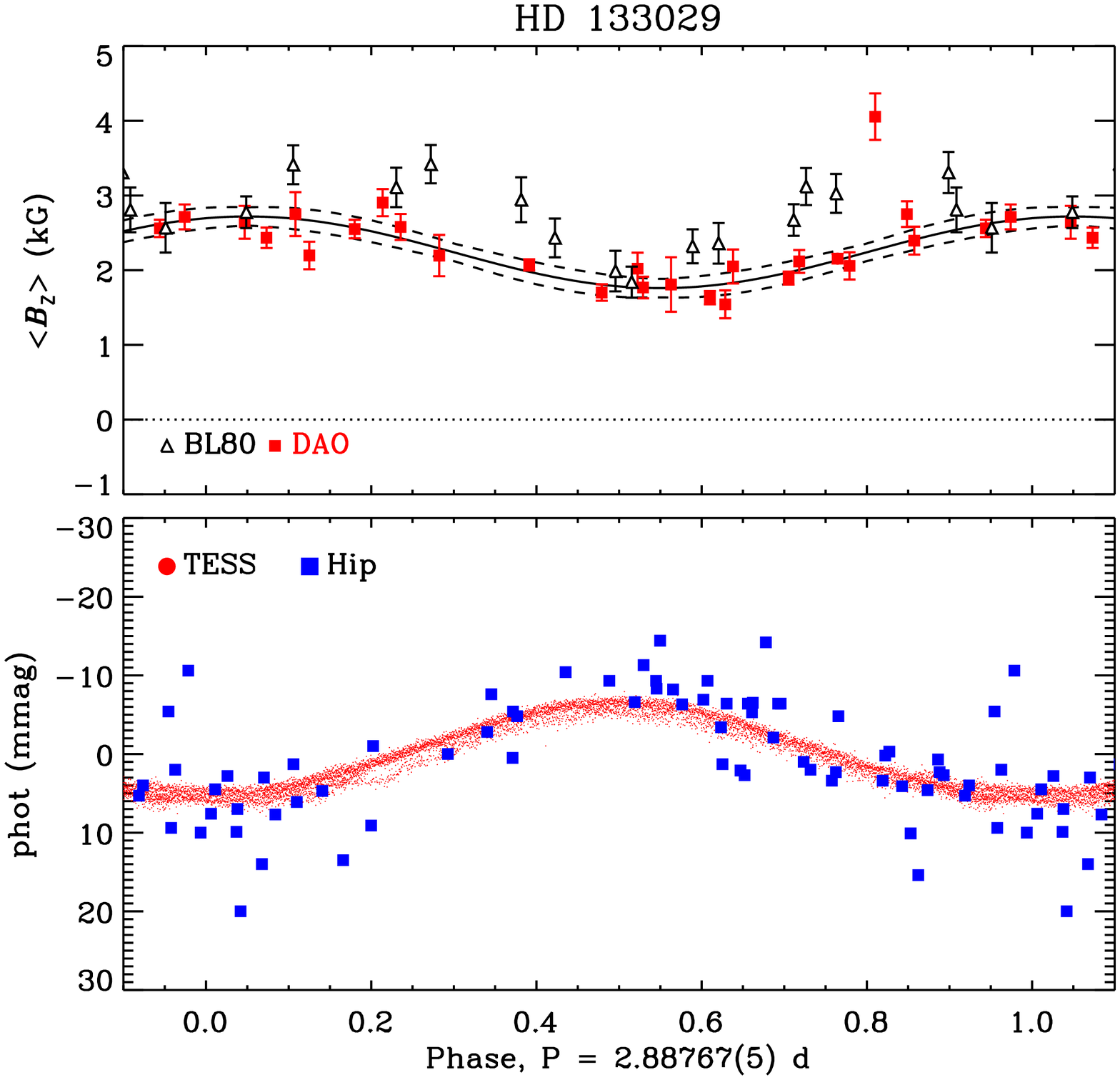}
      \caption[]{{\em Top}: \bz~measurements phased with the rotation period determined from photometric and magnetic data. {\em Bottom}: TESS photometry phased with the rotation period}
         \label{HD133029_bz}
   \end{figure}

\noindent {\bf HD\,133029}: The 2.8804(3)~d {\em TESS} period provides a poor phasing of the \bz~measurements. The period from \cite{2008PASP..120..595A} is similar (2.88756 d), but more precise, and while it does phase the individual \bz~datasets this period does not phase the combined DAO and \cite{1980ApJS...42..421B} magnetic datasets. The nearest period which approximately phases \bz~is 2.88767(6) d, which also gives a coherent phasing of the TESS and {\em Hipparcos} photometry, although this does not result in a perfect match in \bz (see Fig.\ \ref{HD133029_bz}). It is possible the period may not be constant. There are systematic differences between the two \bz~datasets; as the DAO dataset is larger and exhibits less scatter, it was used to fit \bz.

   \begin{figure}
   \centering
   \includegraphics[width=0.45\textwidth]{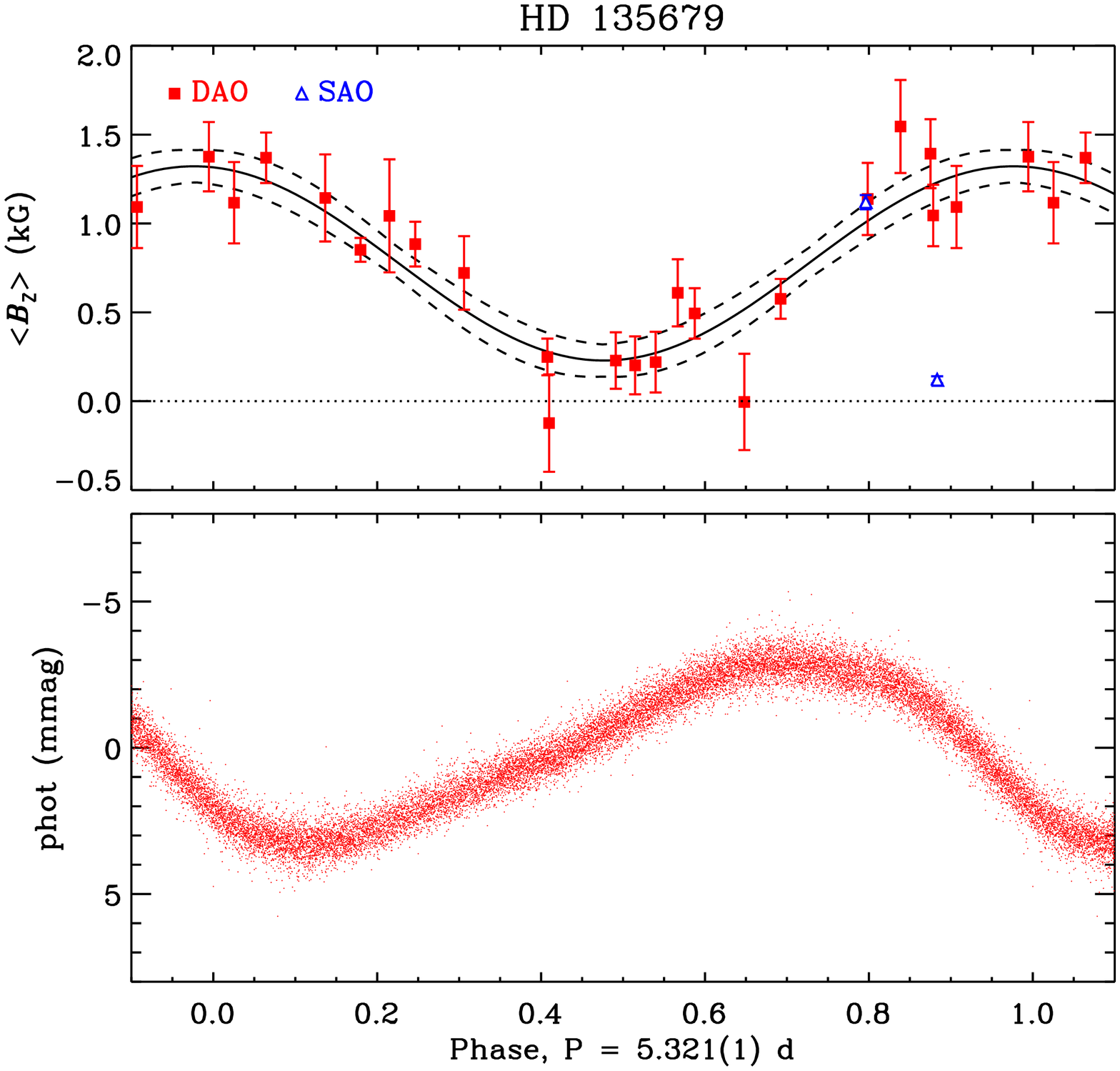}
      \caption[]{{\em Top}: \bz~measurements phased with the rotation period determined from photometric and magnetic data. {\em Bottom}: {\em TESS} photometry phased with the rotation period.}
         \label{HD135679_bz}
   \end{figure}

\noindent {\bf HD\,135679}: The period derived from the {\em TESS} light curve is 5.314(1) d, but this doesn't quite phase the \bz~measurements. Combining the DAO and SAO data, the nearest period that approximately phases the magnetic data is 5.321(1)~d, which also provides acceptable phasing of the {\em TESS} data (Fig.\ \ref{HD135679_bz}). While {\em Hipparcos} data are available, the amplitude of the photometric variation is smaller than the mean error bar of those measurements, which therefore cannot be used to improve the period.

   \begin{figure}
   \centering
   \includegraphics[width=0.45\textwidth]{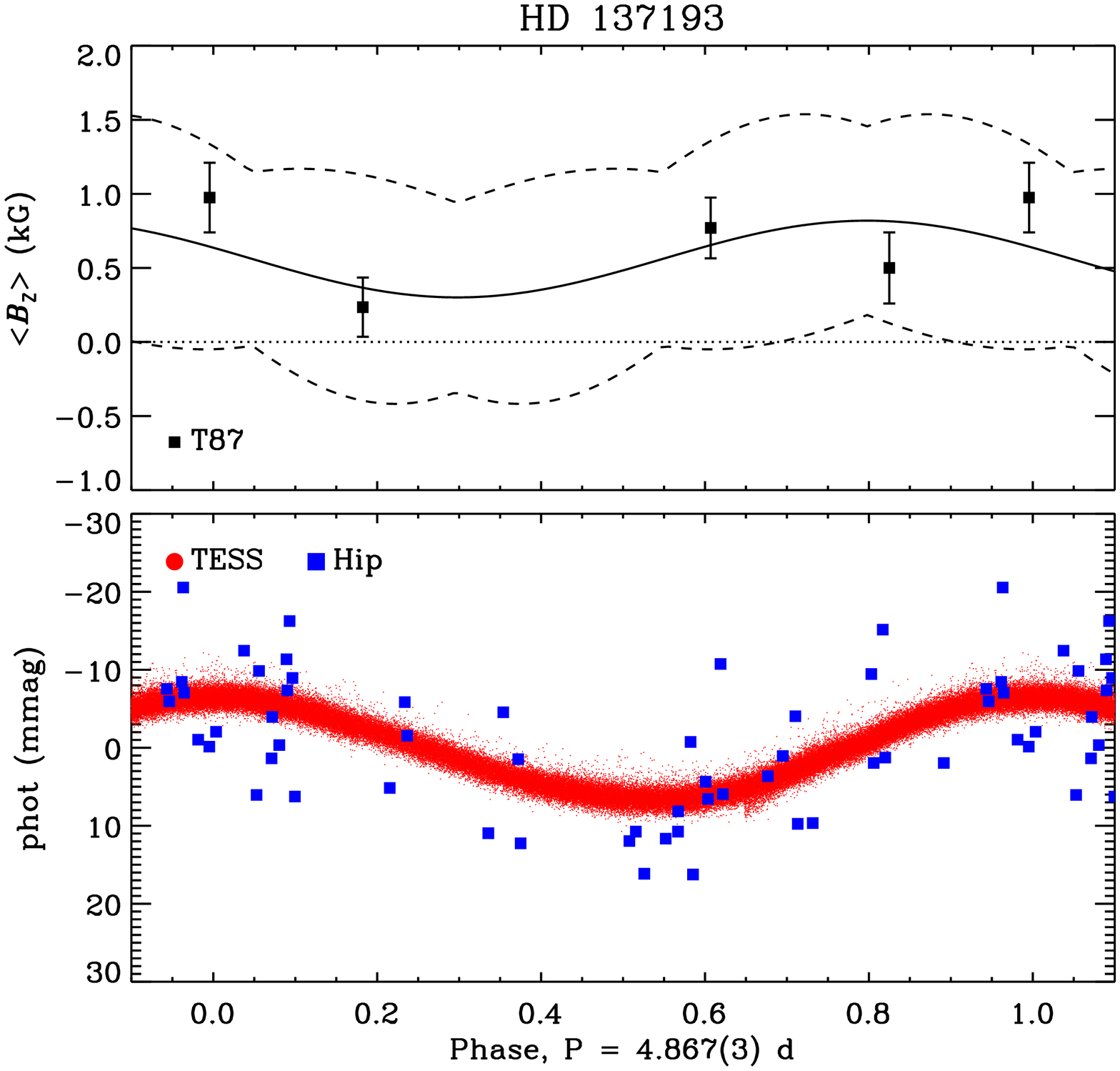}
      \caption[]{{\em Top}: \bz~measurements for HD\,137193, obtained from \protect\cite{1987ApJS...64..219T}, folded with the rotation period. Solid curve shows first-order harmonic fit; fit uncertainty is larger than the y-axis. {\em Bottom}: {\em Hipparcos} and {\em TESS} light curves folded with the rotation period.}
         \label{HD137193_bz}
   \end{figure}

\noindent {\bf HD\,137193}: This star is listed in the \cite{2009A&A...498..961R} catalogue as a B9 Si star. The {\em Hipparcos} light curve gives a period of 16.3 d. However, the {\em TESS} light curve gives a clear period of 4.867(3)~d, which also phases the {\em Hipparcos} data coherently. Since the amplitude of the modulation in the {\em TESS} light curve is comparable to the uncertainty in the {\em Hipparcos} data, it is probable that the {\em Hipparcos} result is spurious. Four magnetic measurements were published by \citep{1987ApJS...64..219T}, shown phased with the TESS period in Fig.\ \ref{HD137193_bz}. 

\noindent {\bf HD142250}: This star is listed in the \cite{2009A&A...498..961R} catalogue as a B6 He-weak star. No magnetic field was detected by \cite{1987ApJS...64..219T}. The {\em TESS} light curve is clearly dominated by pulsation, with several peaks around 0.4~d with similar amplitudes (0.3--0.5 mmag). There is one low-amplitude (0.14 mmag) peak at around 3.1~d which may be associated with rotation, however this could easily be $g$-mode pulsation.

\noindent {\bf HD\,142884}: This star is listed in the \cite{2009A&A...498..961R} catalogue as a B9 Si star. \cite{2017MNRAS.468.2745N} give a period of 0.8~d, as confirmed from {\em K2} data by \cite{2018AJ....155..196R}, and by \cite{2020MNRAS.493.3293B} from ASAS-3, KELT, and MASCARA photometry. No magnetic field was detected by \cite{1983ApJS...53..151B}. There is 1 Narval observation, a non-detection with a 150~G error bar (Fig.\ \ref{lsd_all}). The broad spectral lines (\vsini~$=130$~\kms) are consistent with an 0.8~d rotation period. 

   \begin{figure}
   \centering
   \includegraphics[width=0.45\textwidth]{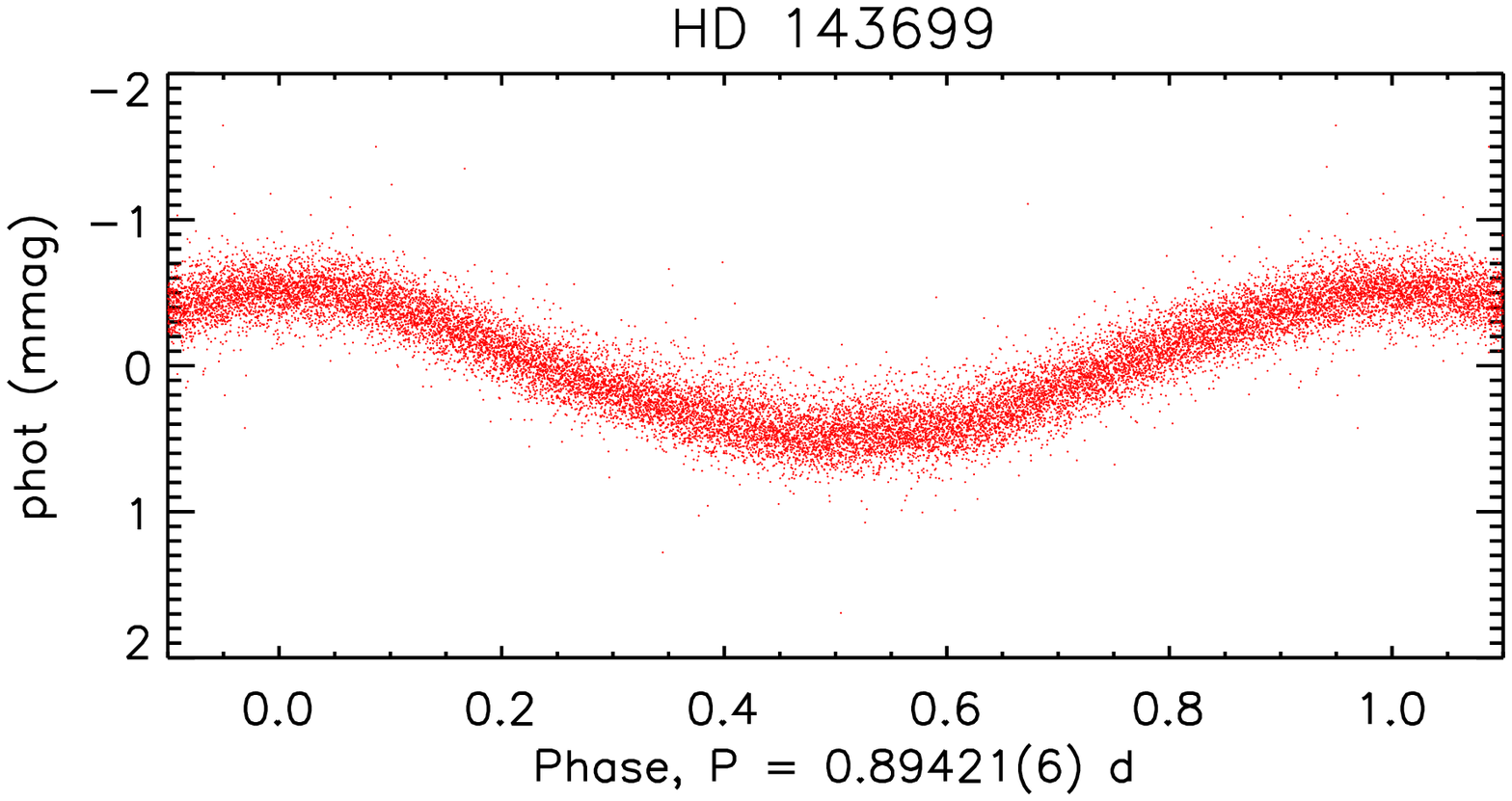}
      \caption[]{TESS light curve for HD\,143699, folded with the rotation period. The data have been pre-whitened with the probable pulsation frequency at 1.12 c/d.}
         \label{HD143699_TESS}
   \end{figure}

\noindent {\bf HD\,143699}: This star is listed in the \cite{2009A&A...498..961R} catalogue as a B6\,He-weak star. The strongest frequency in the TESS light curve corresponds to a period of 0.89421(6)~d, which is used to phase the light curve in Fig.\ \ref{HD143699_TESS}; however, there is also an unrelated peak in the periodogram corresponding to a possible rotational period of 1.3188(5)~d, with the first harmonic of this frequency also present. No magnetic field was detected by \cite{1983ApJS...53..151B}. There is one ESPaDOnS observation, a non-detection with a 76 G error bar. The LSD Stokes $I$ profile is highly asymmetric, possibly indicating binarity (Fig.\ \ref{lsd_all}). The broad lines (\vsini~$\sim 115$~\kms) are consistent with a short rotation period, however, the extremely low amplitude of the variation (below 1 mmag), and the presence of multiple unrelated frequencies, suggests star may also display $g$-mode pulsations, which may be related to one or all of the frequencies. However, it is a radio-bright star, indicating that it is probably magnetic.

\noindent {\bf HD\,144844}: This star is listed in the \cite{2009A&A...498..961R} catalogue as a B9\,MnPGa star, a higher-mass analogue to HgMn stars in which magnetic fields have not been detected \citep{2018MNRAS.475..839S}. \cite{2017MNRAS.468.2745N} give a photometric period of 2.69~d. No magnetic field was detected by \cite{1983ApJS...53..151B}. There is one ESPaDOnS observation, a ND with a 9 G error bar. The LSD profiles indicate the star is an SB2 (Fig.\ \ref{lsd_all}). Since the star does not appear to be magnetic it was dropped from the analysis.

   \begin{figure}
   \centering 
   \includegraphics[width=0.45\textwidth]{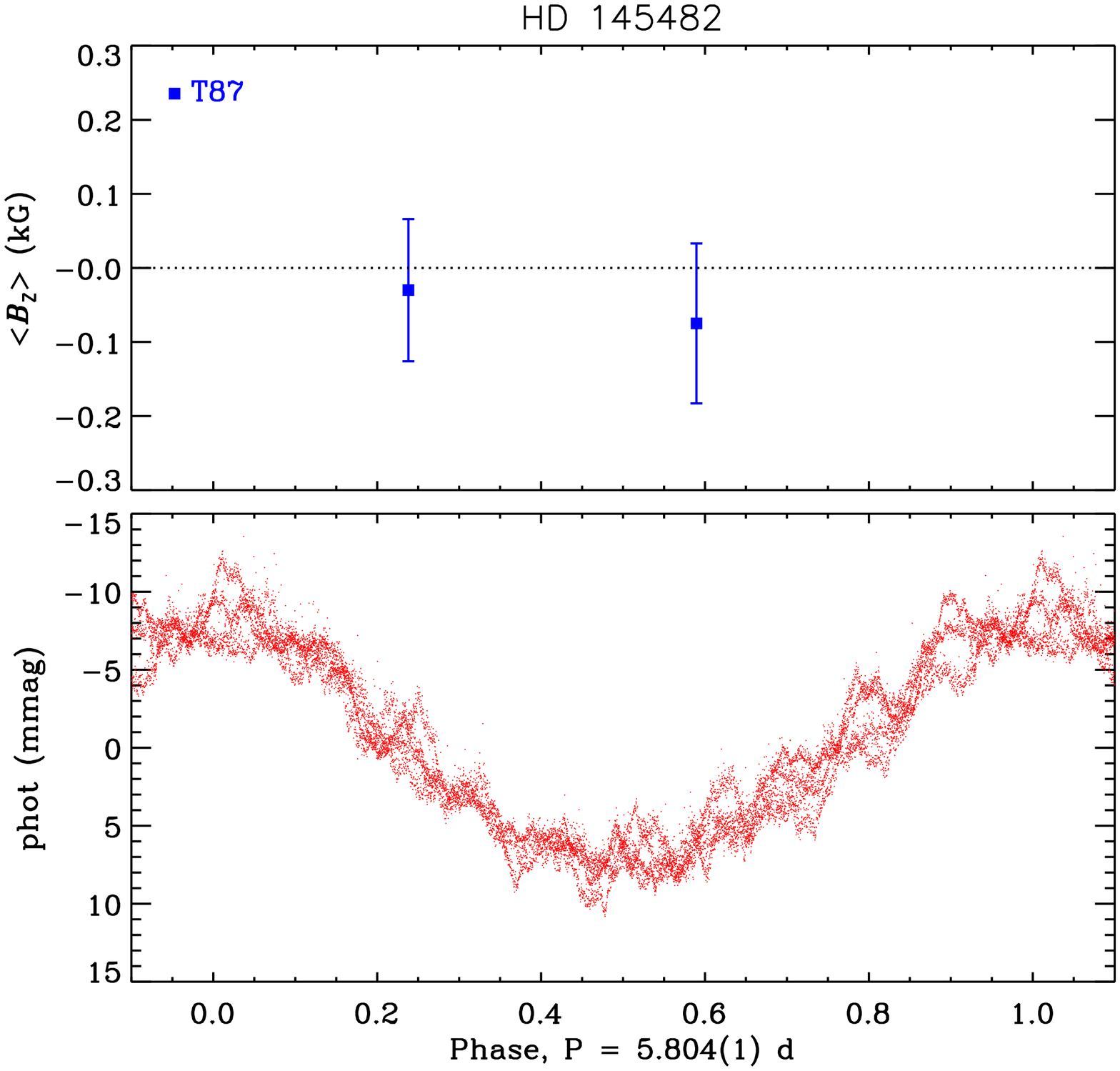}
      \caption[]{{\em Top} \bz~measurements phased with the rotation period. {\em Bottom}: TESS light curve for HD\,145482, folded with the rotation period. The light-curve also exhibits high-frequency pulsations.}
         \label{HD145482_TESS}
   \end{figure}

\noindent {\bf HD\,145482}: This star is not listed in the \cite{2009A&A...498..961R} catalogue. The 2 published magnetic measurements are non-detections \citep{1987ApJS...64..219T}. However, there is clear modulation in the TESS light curve at 5.804(1)~d, together with numerous low-amplitude, high-frequency (above about 3 c/d) signals, as can be seen superimposed on the probable rotational modulation in Fig.\ \ref{HD145482_TESS}.

\noindent {\bf HD\,146001}: This star is listed in the \cite{2009A&A...498..961R} catalogue as a B8\,He-weak star. \cite{2012MNRAS.420..757W} give two photometric periods, 3.6 d and 0.58612 d. There are 2 ESPaDOnS observations. The LSD profiles are both non-detections with 65 G error bars (Fig.\ \ref{lsd_all}). The broad lines (\vsini~$\sim 90$~\kms) are inconsistent with the longer \cite{2012MNRAS.420..757W} period, but are consistent with the short period. The LSD profiles exhibit line profile variability, which could be consistent with spots, binarity, or pulsations. There is no indication of a 0.58 d period in the {\em Hipparcos} light curve, and the lowest FAP in the periodogram is 0.57. However, despite the lack of clear rotational modulation in the {\em Hipparcos} data and the lack of a magnetic detection, this star is radio-bright. 

\noindent {\bf HD\,147084}: This star is not listed in the \cite{2009A&A...498..961R} catalogue. The published magnetic data are non-detections \citep{1987ApJS...64..219T,2015A&A...583A.115B}. There are 2 ESPaDOnS measurements, both NDs with 2 G error bars (Fig.\ \ref{lsd_all}). The {\em Hipparcos} period is 1.8611(2)~d, although as the FAP is 0.04 this signal is probably spurious. This is probably not a magnetic star and was therefore removed from the sample.

   \begin{figure}
   \centering
   \includegraphics[width=0.45\textwidth]{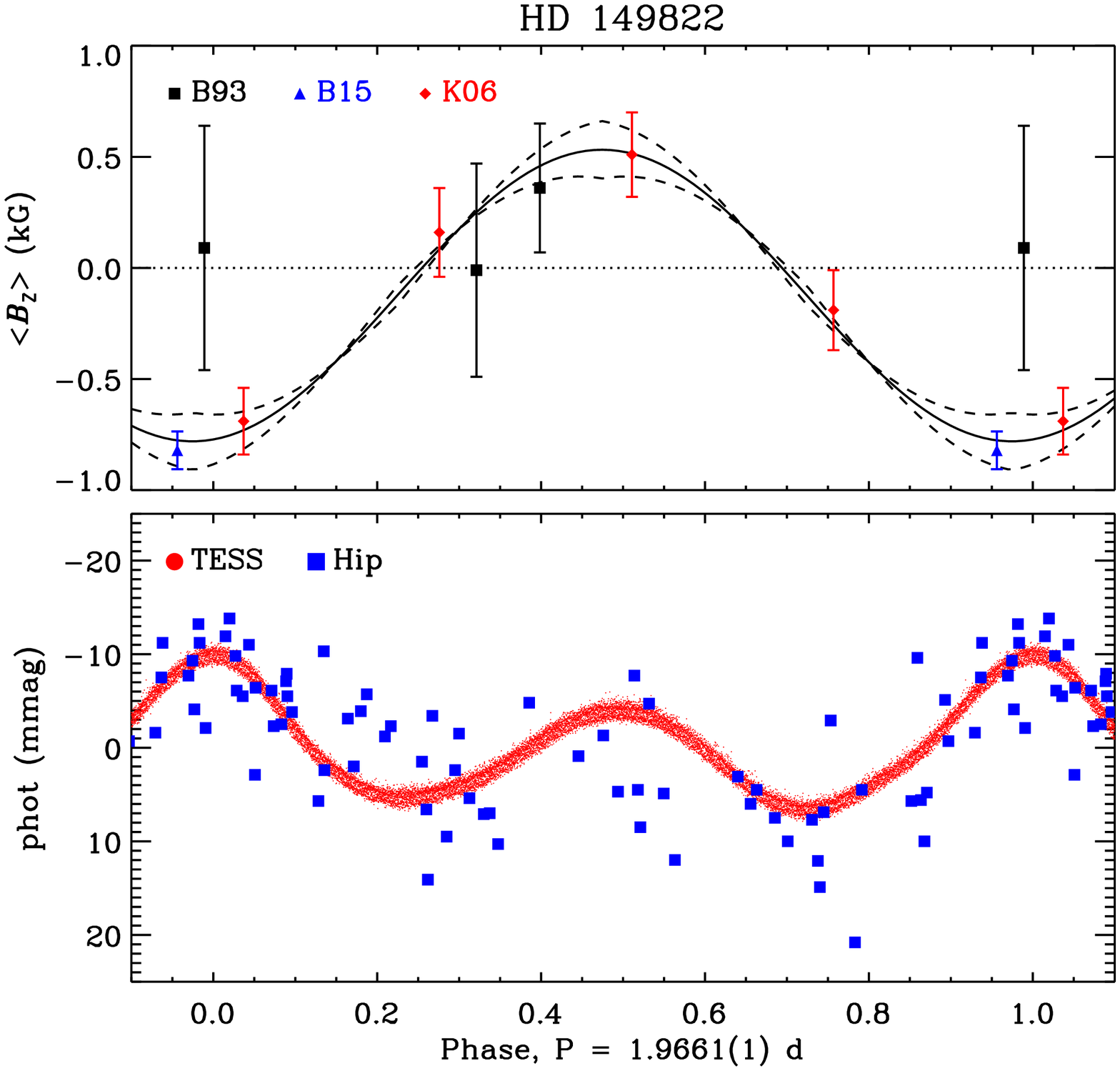}
      \caption[]{{\em Top}: \bz~measurements for HD\,149822, obtained from \protect\cite{1993A&A...269..355B}, \protect\cite{2006MNRAS.372.1804K}, and \protect\cite{2015A&A...583A.115B}, folded with the rotation period. Solid curve shows first-order harmonic fit. {\em Bottom}: {\em Hipparcos} and {\em TESS} light curves folded with the rotation period.}
         \label{HD149822_bz}
   \end{figure}

\noindent {\bf HD\,149822}: This star is listed in the \cite{2009A&A...498..961R} catalogue as a B9 SiCr star. There are several low-resolution magnetic measurements \citep{1993A&A...269..355B,2006MNRAS.372.1804K,2015A&A...583A.115B}. The {\em TESS} light curve gives a period of 1.9663(3)~d; a period of 1.9661(1)~d coherently phases the {\em TESS} and {\em Hipparcos} data, which provides a reasonable phasing of the sparse magnetic data (Fig. \ref{HD149822_bz}).

   \begin{figure}
   \centering
   \includegraphics[width=0.45\textwidth]{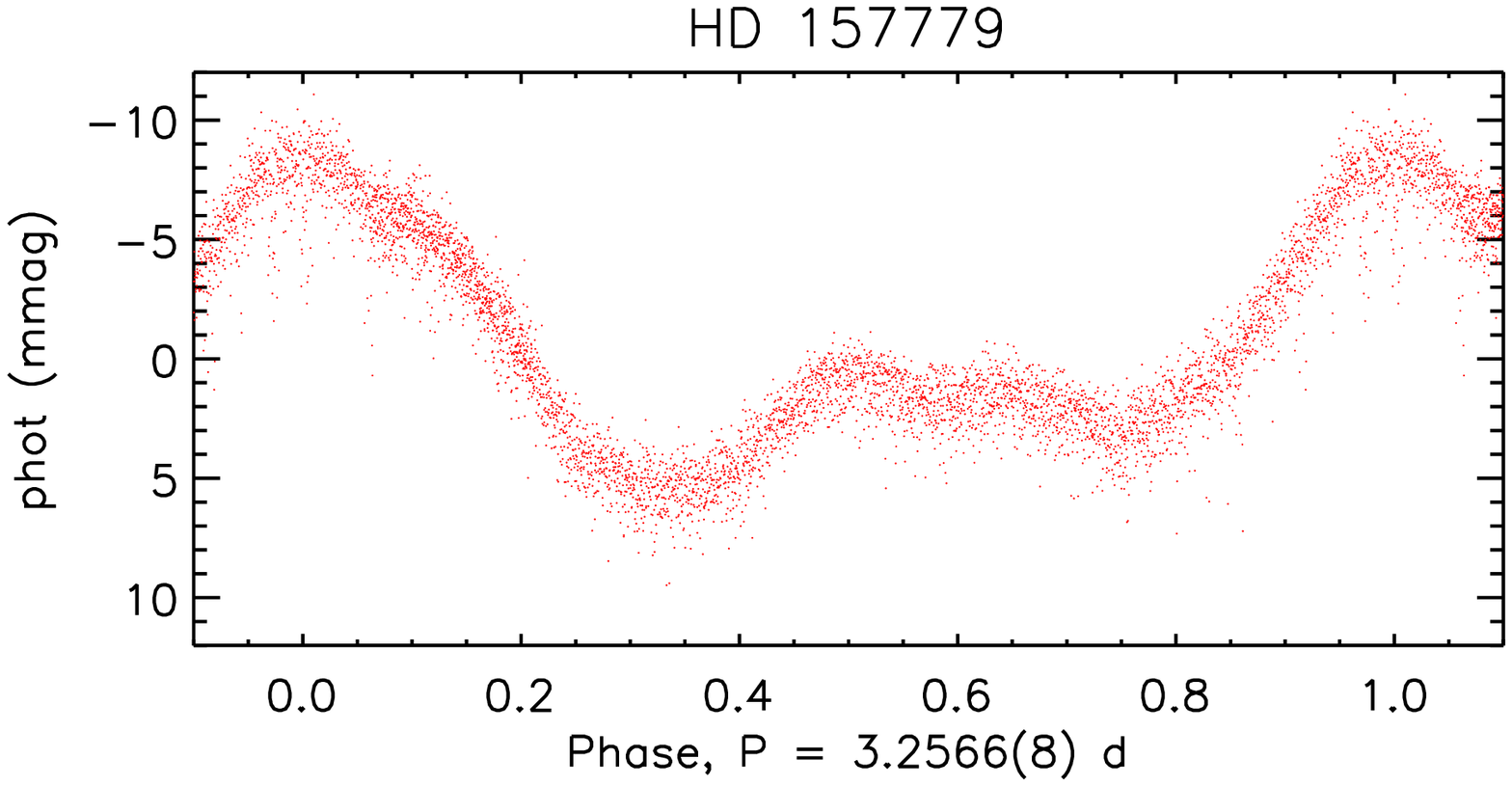}
      \caption[]{{\em TESS} light curve of HD\,157779 folded with the rotational period.}
         \label{HD157779_TESS}
   \end{figure}

\noindent {\bf HD\,157779}: This star is listed in the \cite{2009A&A...498..961R} catalogue as a B9 Si star. There are no magnetic data. The {\em TESS} light curve shows probable rotational modulation on a 3.2566(8)~d period (see Fig.\ \ref{HD157779_TESS}); the multiple harmonics of the rotational period are difficult to reconcile with other mechanisms such as pulsation.

\noindent {\bf HD\,162374}: This star is listed in the \cite{2009A&A...498..961R} catalogue as a B7 He-weak star. \cite{1998A&AS..127..421C} give a photometric period of 1.66~d. Its magnetic field was not detected by \cite{1983ApJS...53..151B}. There is one available ESPaDOnS measurement, a ND with a 27 G error bar. {\em Hipparcos} photometry does not confirm the 1.66~d period; the maximum amplitude is at 1.15~d, with a FAP of 0.09. Given the non-detection of the magnetic field and the low significance of the photometric period, the period is probably not a rotational period, this is probably not a magnetic star, and it was therefore removed from the sample. Asymmetry in the LSD Stokes $I$ profile is consistent with binarity (Fig.\ \ref{lsd_all}).

   \begin{figure}
   \centering
   \includegraphics[width=0.45\textwidth]{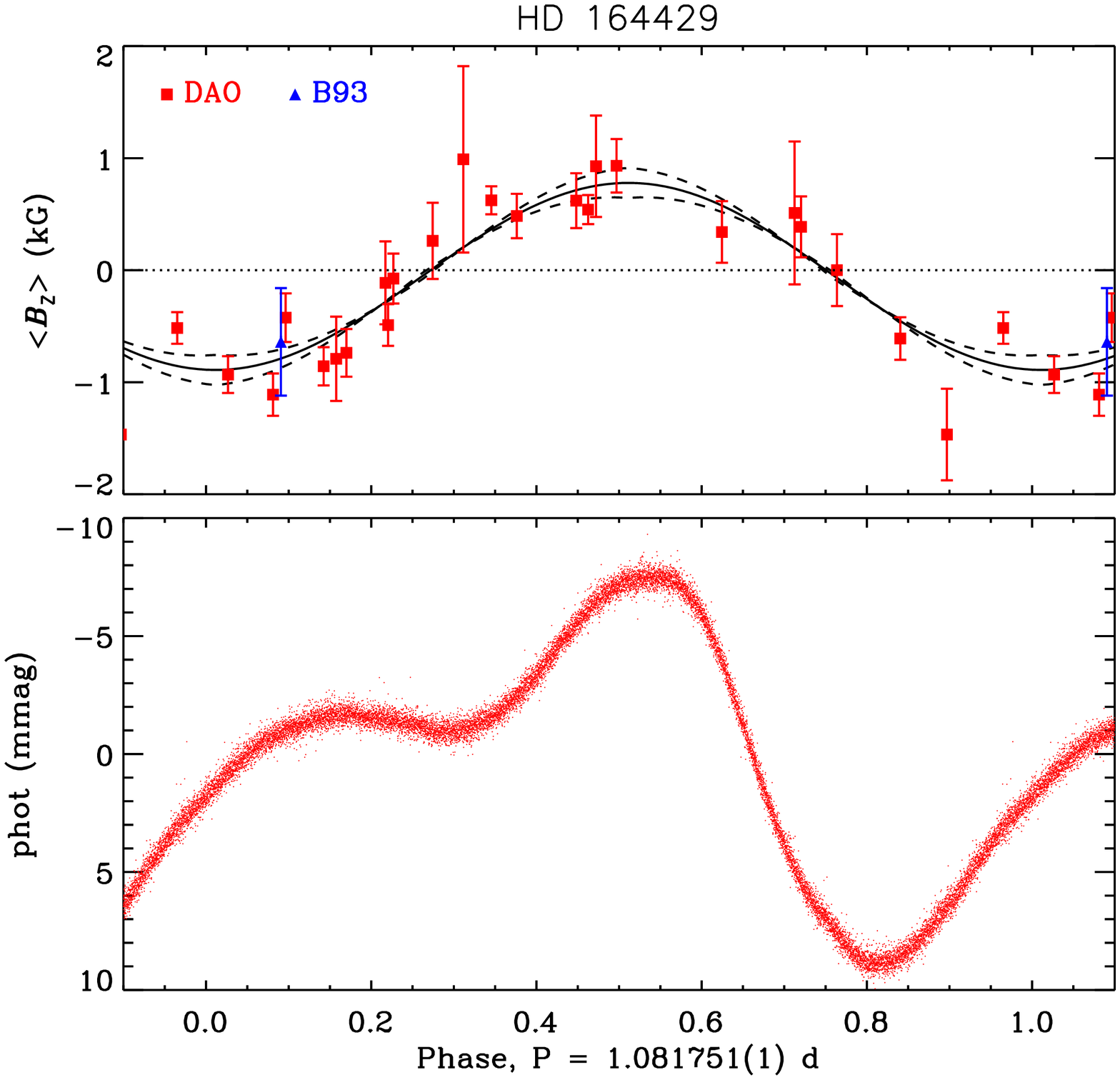}
      \caption[]{{\em Top}: \bz~measurements for HD\,164429, phased with the period determined from TESS data. {\em Bottom}: TESS light curve folded with the rotation period.}
         \label{HD164429_bz}
   \end{figure}

\noindent {\bf HD\,164429}: This star is listed in the \cite{2009A&A...498..961R} catalogue as a B9 SiCrSr star. There is only one published magnetic measurement, a non-detection \citep{1993A&A...269..355B}, however the magnetic field is clearly detected in the much larger DAO dataset. The {\em Hipparcos} light-curve gives a period of 1.0817(1) d, although as the FAP is 0.43 this period could not be confirmed. However, the photometric variation is clearly seen in the {\em TESS} light curve, which yields a period of 1.08175(1)~d, almost identical to the period recovered from the {\em Hipparcos} data. The DAO \bz~measurements are shown phased with this period in (Fig.\ \ref{HD164429_bz}). 

\noindent {\bf HD\,166182}: This is a chemically normal B2\,IV star. No magnetic field was detected by \cite{1982ApJ...258..639L}. The {\em TESS} light curve has two low-amplitude (0.06--0.1 mmag) periodicities at around 4~d, which are probably associated with pulsation. This star was dropped from the sample.

\noindent {\bf HD\,168785}: This star is listed in the \cite{2009A&A...498..961R} catalogue as a B3 He star. A magnetic field was detected by \cite{2018A&A...618L...2J}. It is in a crowded field, and the {\em TESS} light curve is probably contaminated by other stars; there is furthermore an artifact at the beginning of the time series, for which reason the first day was discarded. The light curve is dominated by a long-term trend, likely systematic. There are significant periods at 11.47~d, 4.98~d, and 6.43~d, which do not appear to be harmonically related and therefore likely do not reflect the star's rotation.

   \begin{figure}
   \centering
   \includegraphics[width=0.45\textwidth]{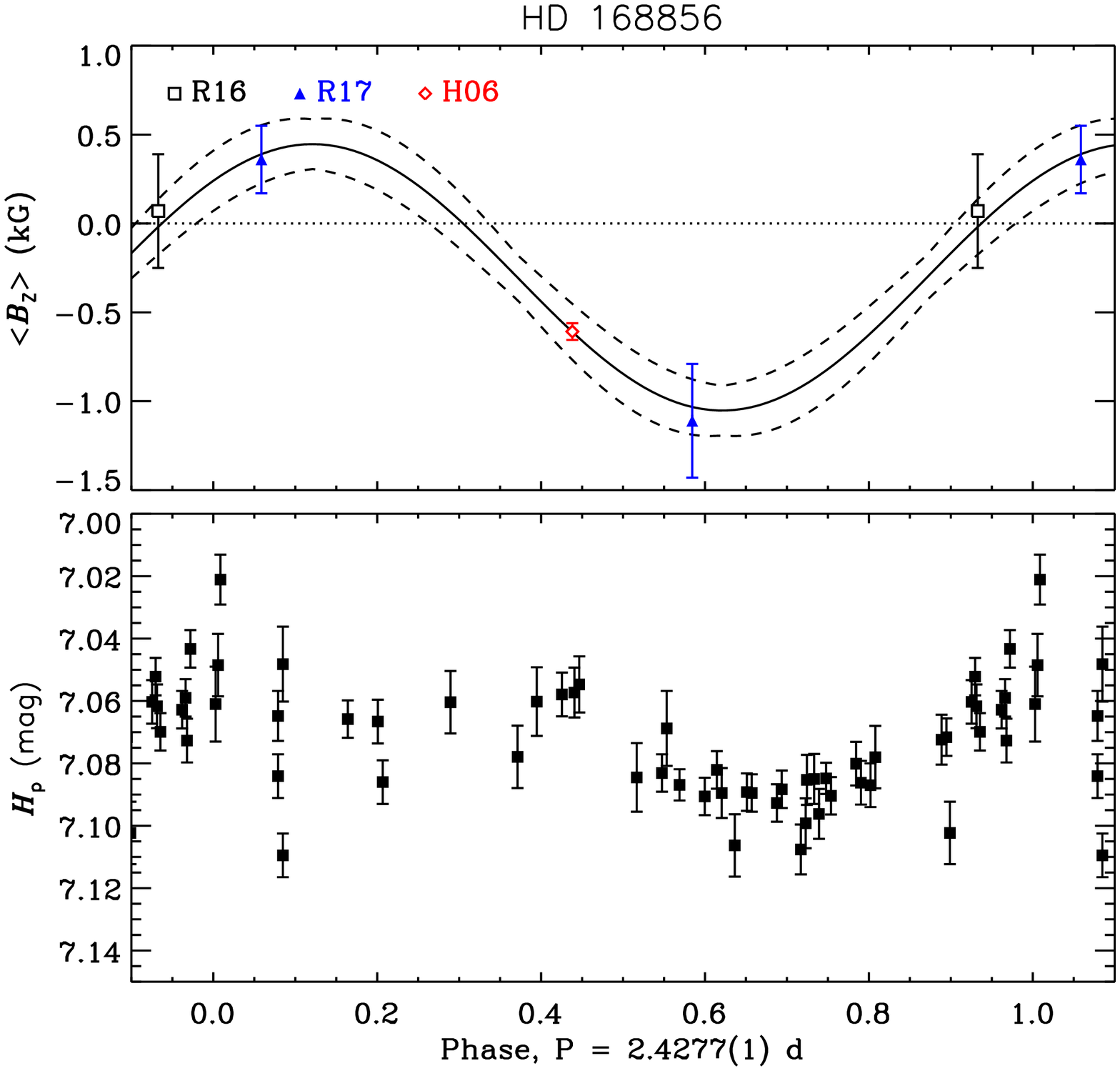}
      \caption[]{{\em Top}: \bz~measurements for HD\,168856, obtained from \protect\cite{2016AstBu..71..302R}, \protect\cite{2017AstBu..72..391R}, and \protect\cite{2006AN....327..289H}, folded with the rotation period. {\em Bottom}: {\em Hipparcos} light curve phased with the rotation period.}
         \label{HD168856_bz}
   \end{figure}

\noindent {\bf HD\,168856}: This star is listed in the \cite{2009A&A...498..961R} catalogue as a B9 Si star. A magnetic detection was reported by \cite{2017AstBu..72..391R}. The photometric period of 2.4277(1)~d was reported by \cite{2020MNRAS.493.3293B}, based on ASAS-3, KELT, and MASCARA data. Available \bz~and the {\em Hipparcos} light curve are shown folded with this period in Fig.\ \ref{HD168856_bz}.

   \begin{figure}
   \centering
   \includegraphics[width=0.45\textwidth]{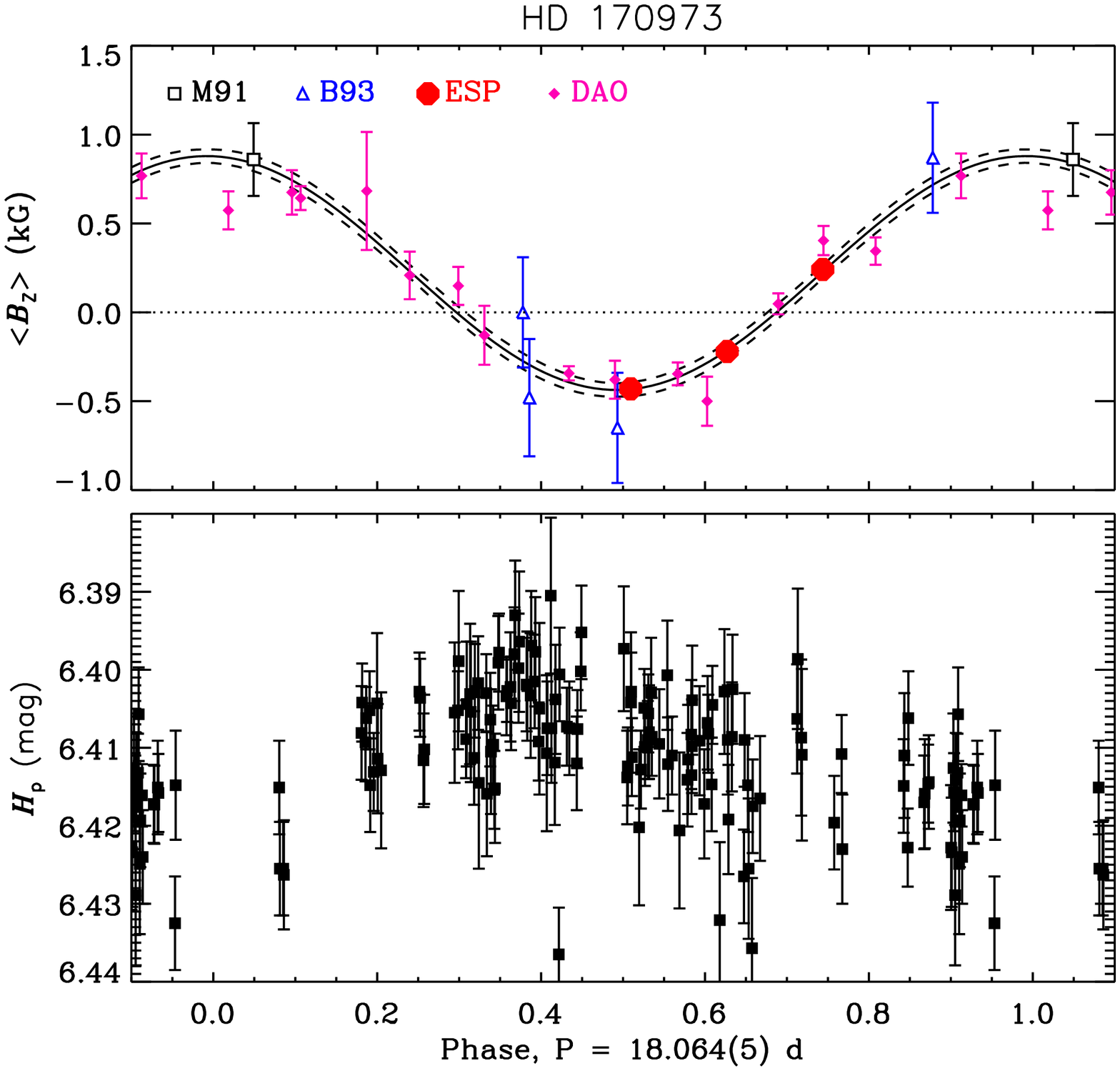}
      \caption[]{{\em Top}: \bz~measurements for HD\,170973, obtained from \protect\cite{1991A&AS...89..121M}, \protect\cite{1993A&A...269..355B}, and ESPaDOnS, folded with the rotation period. Solid curve shows first-order harmonic fit. {\em Bottom}: {\em Hipparcos} light curve folded with the rotation period.}
         \label{HD170973_bz}
   \end{figure}

\noindent {\bf HD\,170973}: This star is listed in the \cite{2009A&A...498..961R} catalogue as an A0 SiCrSr star. \cite{2005AA...430.1143B} gave a period of 18.52~d based on sparse \bz~measurements \citep{1991A&AS...89..121M,1993A&A...269..355B}. {\em Hipparcos} photometry gives a period of 18.09(3)~d, with a FAP of 0.02. There are 3 available ESPaDOnS measurements, all DDs (Fig.\ \ref{lsd_all}). The {\em Hipparcos} period does not phase the available \bz~measurements coherently; by combining them, a period of 18.064(5)~d is obtained (see Fig.\ \ref{HD170973_bz}). 

   \begin{figure}
   \centering
   \includegraphics[width=0.45\textwidth]{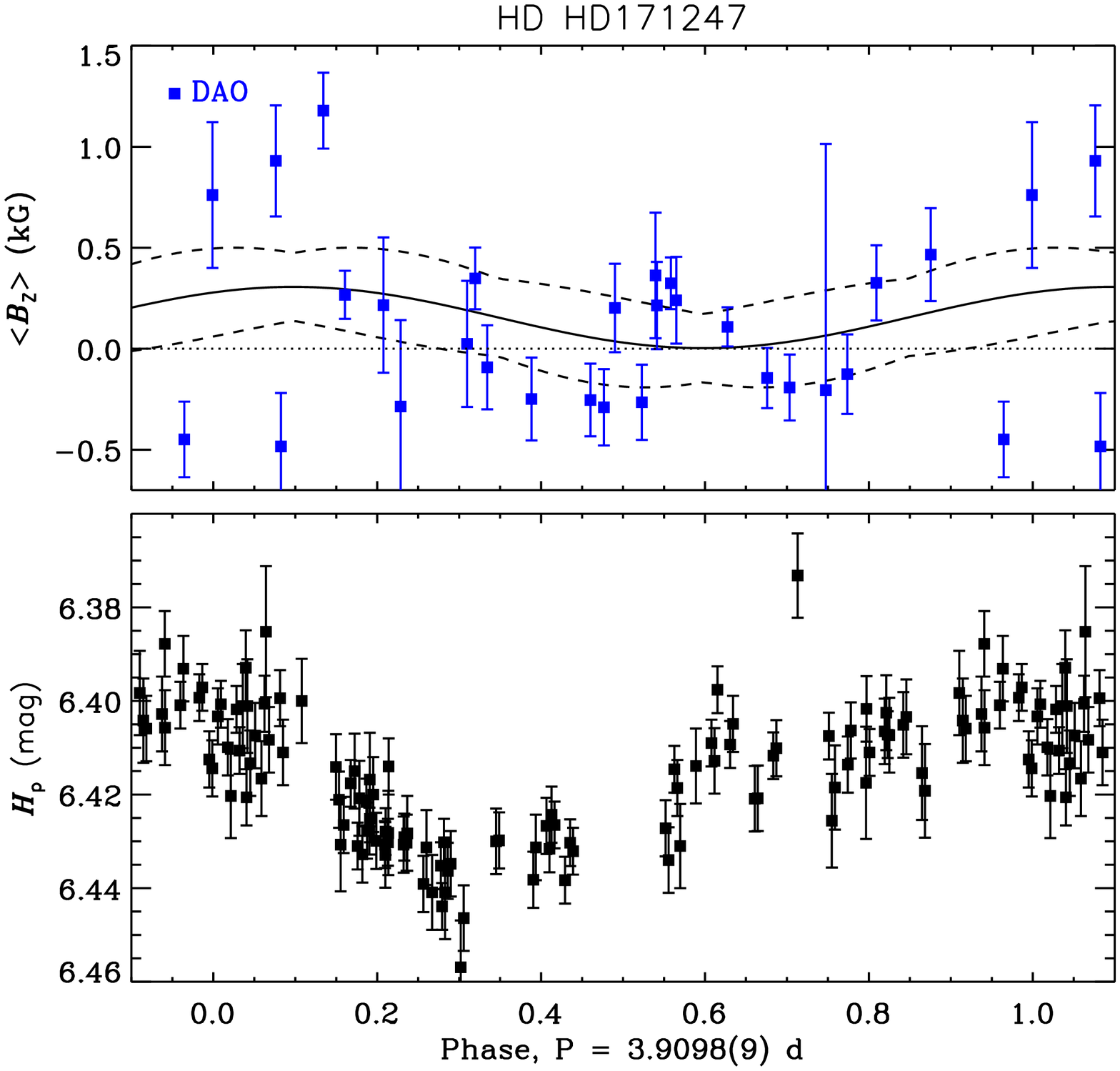}
      \caption[]{DAO \bz~measurements ({\em top}) and {\em Hipparcos} light curve ({\em bottom}) of HD\,171247 folded with the photometric period.}
         \label{hd171247_bz_phot}
   \end{figure}

\noindent {\bf HD\,171247}: This star is listed in the \cite{2009A&A...498..961R} catalogue as a B8 Si star. The star is notable for being one of the only radio-bright stars apparently in the second half of the main sequence, which is difficult to explain given that magnetospheric braking should have long ago spun the star down to the point where its radio emission should have disappeared, whereas to the contrary its radio luminosity is relatively high. There is clear rotational modulation in the {\em Hipparcos} light curve at 3.9124(8)~d, with a FAP of $10^{-15}$. This period does not provide a coherent phasing of \bz, period analysis of which yields 1.14482(6)~d, with a FAP of 0.009; however, this period fails to phase the photometry. The closest period that can approximately phase both datasets is 3.9098(9)~d (see Fig.\ \ref{hd171247_bz_phot}), a 3$\sigma$ difference from the best photometric period. 

   \begin{figure}
   \centering
   \includegraphics[width=0.45\textwidth]{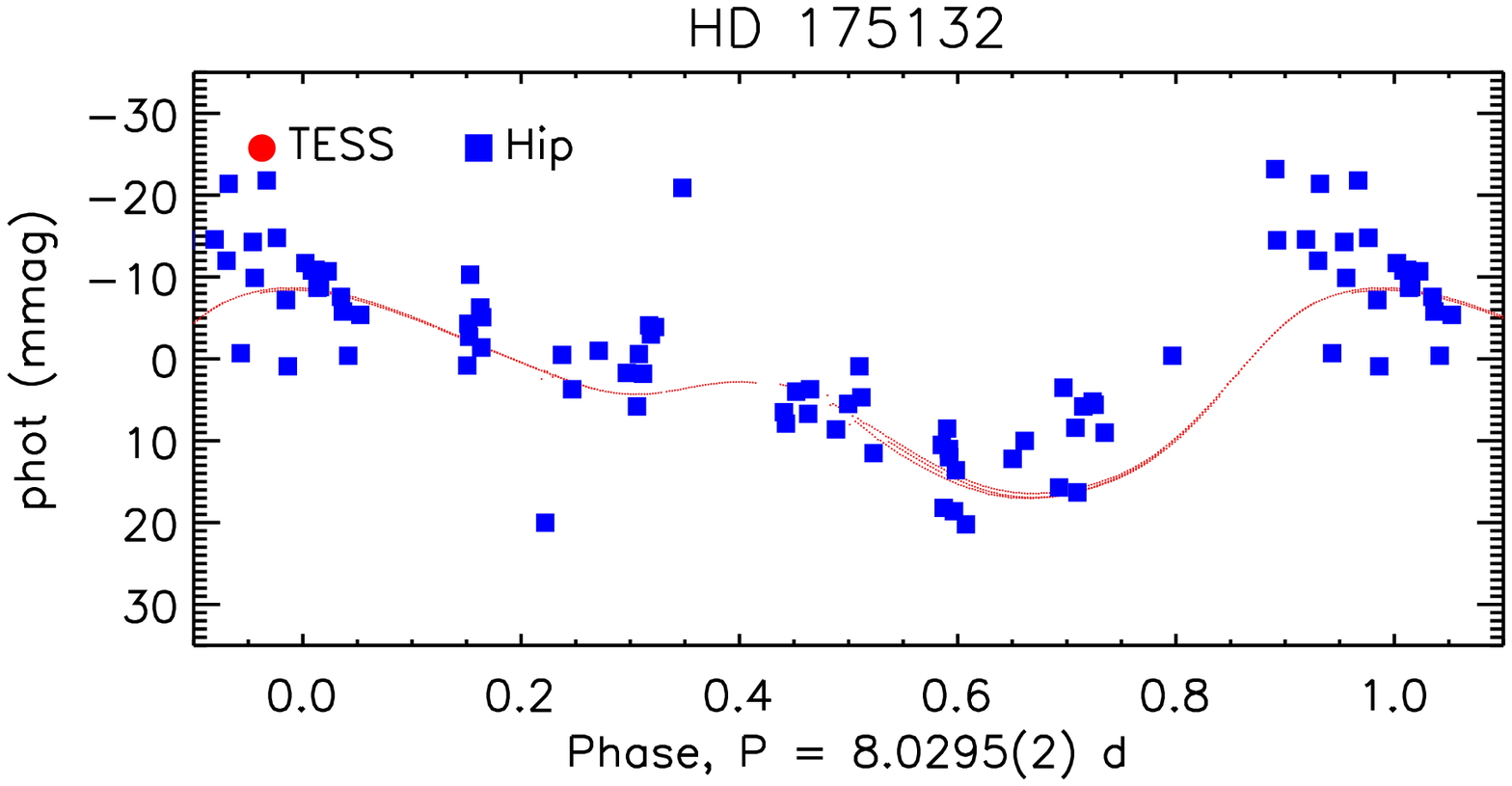}
      \caption[]{{\em Hipparcos} and {\em K2} light curves of HD\,175132 folded with the rotation period.}
         \label{HD175132_HIP}
   \end{figure}

\noindent {\bf HD\,175132}: This star is listed in the \cite{2009A&A...498..961R} catalogue as a B9 Si star. The magnetic field was reported by \cite{2003A&A...407..631B}, although only the root-mean-square \bz~was published, not the individual measurements. {\em Hipparcos} photometry gives a period of 8.031(6) d, with a FAP of $10^{-6}$. \cite{2018A&A...619A..98H} reported that a period could not be determined from the K2 light curve, however, examination of these data yields a clear periodicity of 8.033(5)~d, approximately consistent with the period inferred from {\em Hipparcos}. Combining the two gives a period of 8.0295(2)~d, which is used in Fig.\ \ref{HD175132_HIP}.

   \begin{figure}
   \centering
   \includegraphics[width=0.45\textwidth]{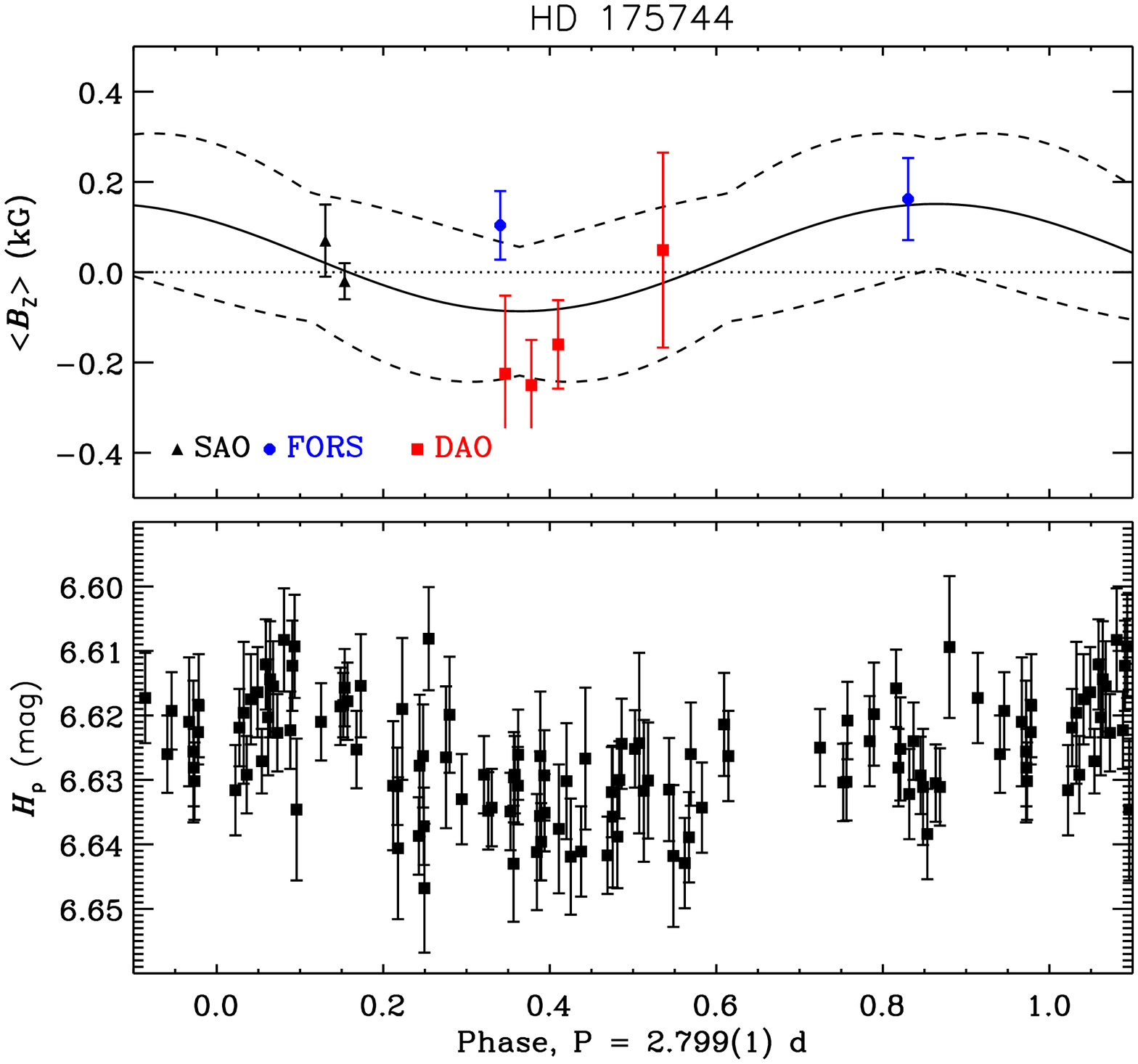}
      \caption[]{{\em Top}: \bz~measurements of HD\,175744 folded with the rotation period inferred from {\em Hipparcos} photometry ({\em Bottom}). The harmonic fit to \bz~is performed to the combined magnetic dataset.}
         \label{hd175744_bz_phot}
   \end{figure}

\noindent {\bf HD\,175744}: This star is listed in the \cite{2009A&A...498..961R} catalogue as a B9 Si star. \cite{1998A&AS..127..421C} gave a photometric period of 3.4~d. This period is not confirmed in Hipparcos photometry, which instead gives 2.799(1)~d, with a FAP of 0.005. No magnetic field has been detected in repeated observation by \cite{2003A&A...407..631B}, \cite{2006A&A...450..763K}, \cite{2020AstBu..75..294R}, or at the DAO (see Table \ref{dao_table}). The \bz~measurements and {\em Hipparcos} light curve are shown folded with the rotation period in Fig.\ \ref{hd175744_bz_phot}. The period of this star must be regarded as highly uncertain, however it is clear that if it is in fact magnetic, its magnetic field must be very weak.

   \begin{figure}
   \centering
   \includegraphics[width=0.45\textwidth]{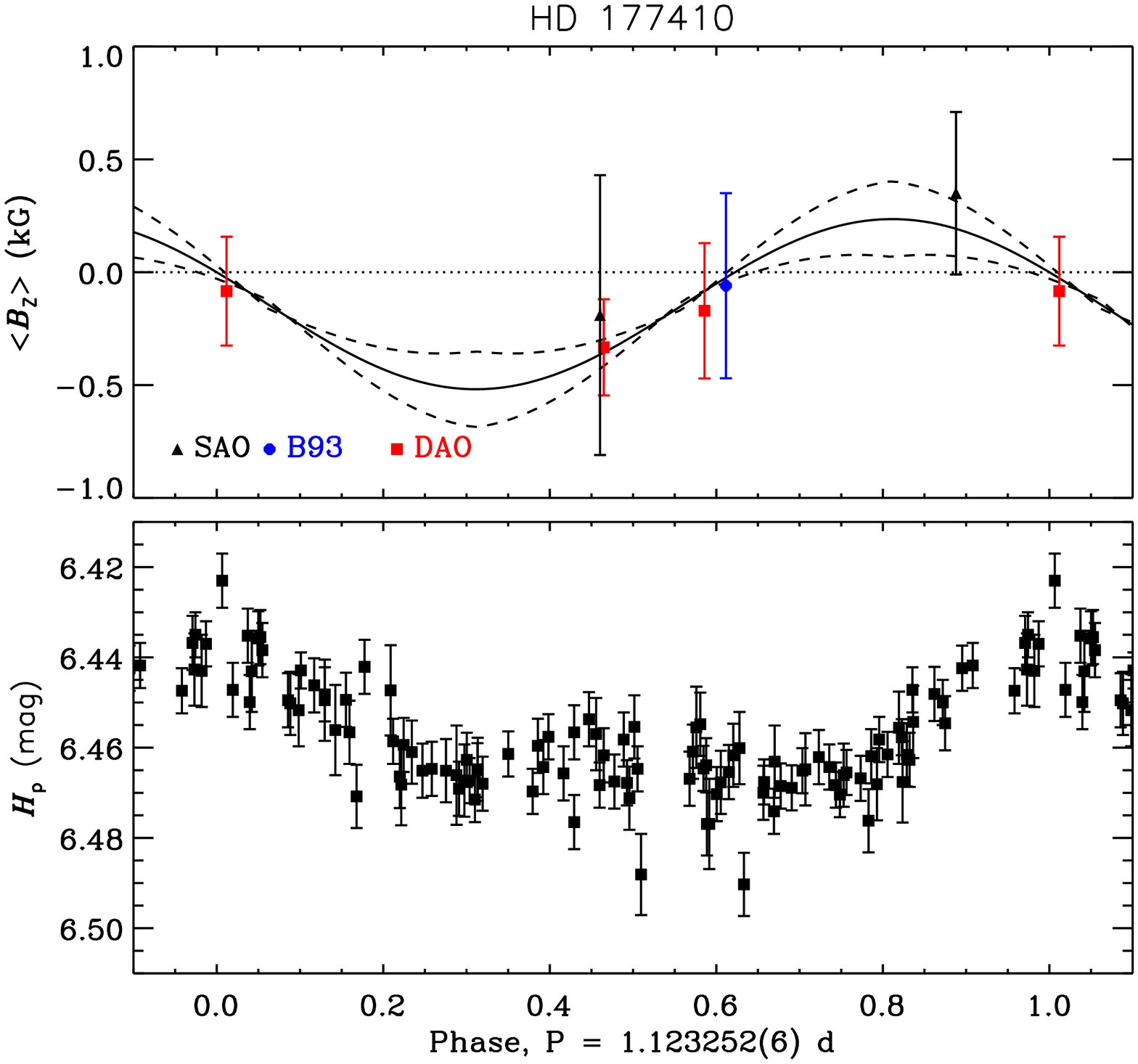}
      \caption[]{{\em Top}: \bz~measurements of HD\,177410 folded with the photometric rotation period {\em Bottom}: {\em Hipparcos} photometry folded with the rotation period. The harmonic fit to \bz~is performed to the combined magnetic dataset.}
         \label{hd175744_bz_phot}
   \end{figure}

\noindent {\bf HD\,177410}: This star is listed in the \cite{2009A&A...498..961R} catalogue as a B9 Si star. \cite{2009A&A...499..567K} determined a highly precise 1.1232524(6)~d rotation period from an extensive photometric dataset. Neither the published magnetic observations \citep{1993A&A...269..355B,2015AstBu..70..444R}, nor the data from the DAO, have detected the star's magnetic field, however as the photometric modulation is obvious this is most likely a magnetic star with a weak magnetic field. Fig.\ \ref{hd175744_bz_phot} shows the magnetic data and the {\em Hipparcos} light curve folded with the \cite{2009A&A...499..567K} period.

   \begin{figure}
   \centering
   \includegraphics[width=0.45\textwidth]{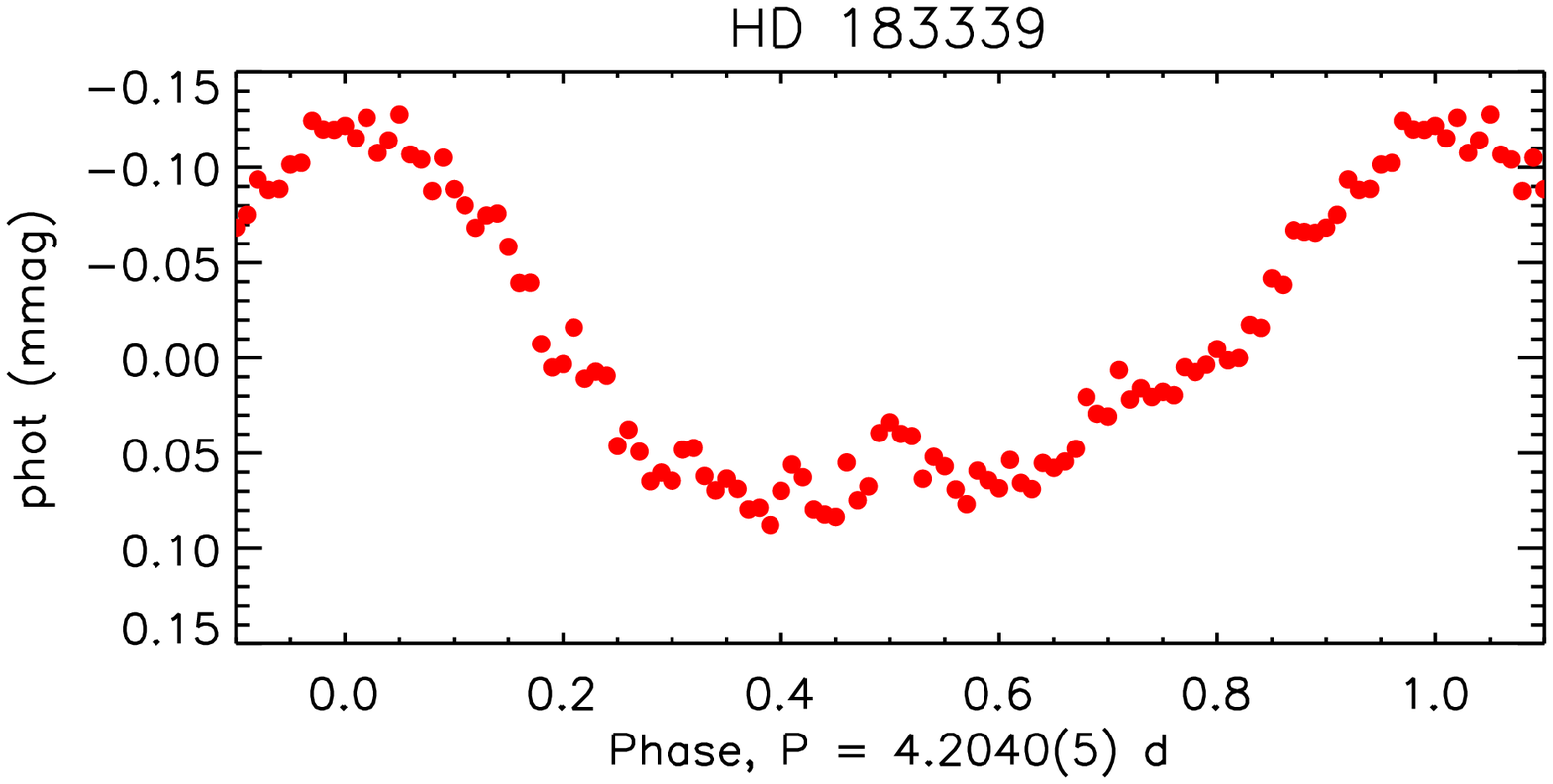}
      \caption[]{Phase-binned {\em TESS} light curve of HD\,183339, folded with the strongest period.}
         \label{HD183339_TESS}
   \end{figure}

\noindent {\bf HD\,183339}: This star is listed as a B8 He-weak star by \cite{2009A&A...498..961R}. \cite{2003A&A...407..631B} reported $B_{\rm rms} = 1300 \pm 465$~G based upon 8 measurements, although these were not individually published, and \cite{2021A&A...652A..31B} phased the available magnetic data with a 2.42~d period. The star was observed in 7 {\em TESS} sectors. There is no indication of the 2.42~d periodicity in the light curve. The strongest peak in the periodogram is at 4.2040(5)~d. The phase-binned light curve is shown folded with this period in Fig.\ \ref{HD183339_TESS}. The amplitude of the variation is extremely low (less than 0.1 mmag), and it is likely that this reflects $g-$mode pulsation rather than rotational modulation. Evaluating the \bz~curve published by \cite{2021A&A...652A..31B}, the uncertainties in the \bz~measurements used to obtain the magnetic model are very large (about 500 G). This star's magnetic field should be re-evaluated using modern instruments.

   \begin{figure}
   \centering
   \includegraphics[width=0.45\textwidth]{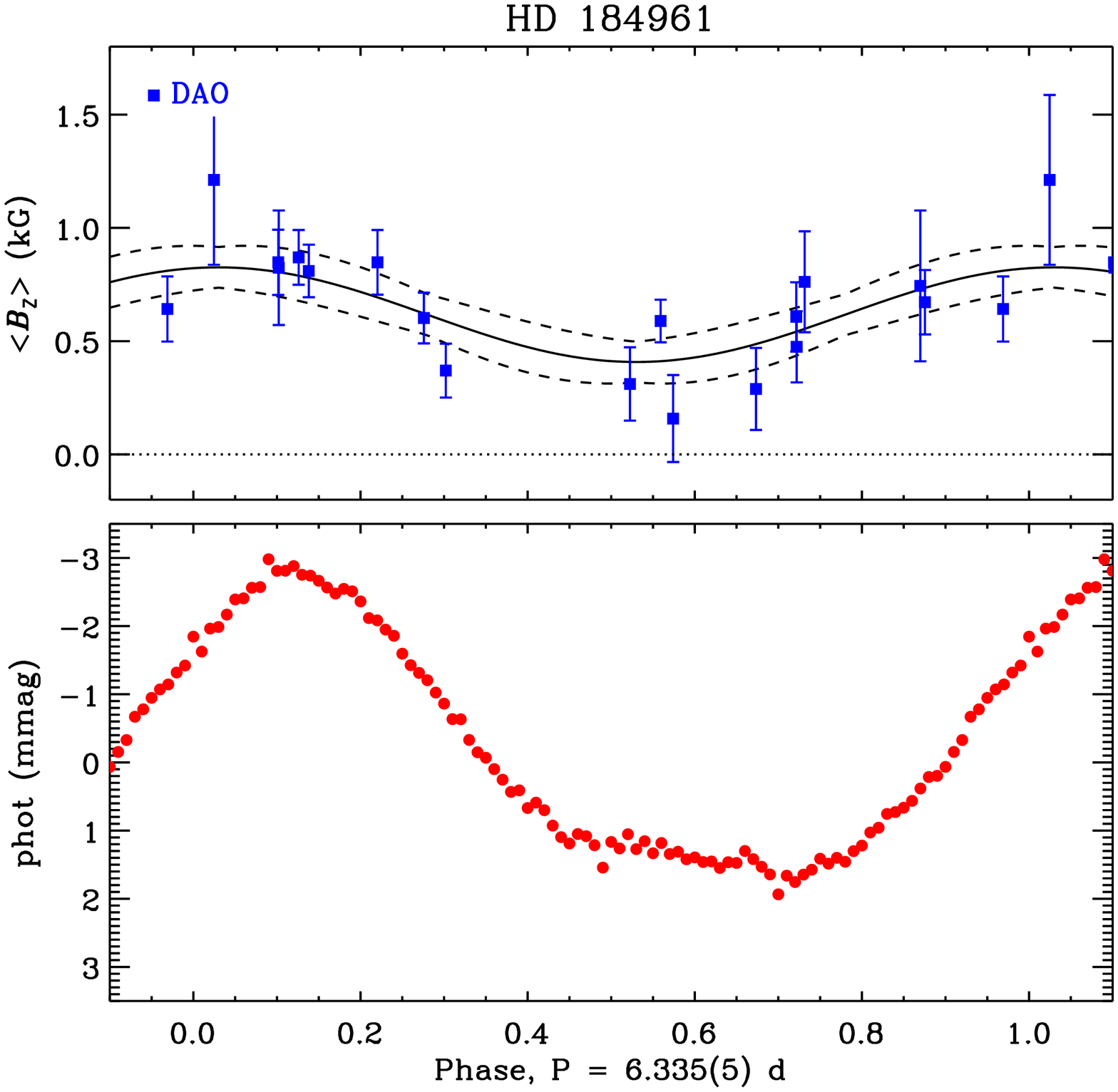}
      \caption[]{DAO \bz~measurements ({\em top}) and phase-binned {\em TESS} light curve ({\em bottom}) phased with the period determined from the magnetic data.}
         \label{HD184961_bz}
   \end{figure}

\noindent {\bf HD\,184961}: This star is listed in the \cite{2009A&A...498..961R} catalogue as a B9 CrSiEu star. The {\em Hipparcos} light curve yields a 6.016(3)~d period, although the FAP of 0.95 indicates this is probably not significant. The {\em TESS} light curve yields a period of 6.35(1) d, although this is uncertain due to the presence of multiple flares which may indicate contamination by a late-type star. DAO \bz~measurements yield a period of 6.692(6) d, which is similar to albeit formally different from the {\em TESS} period. The nearest period to the {\em TESS} period that provides a good phasing of \bz~is 6.335(5)~d, which is consistent within uncertainty with the {\em TESS} period. The DAO \bz~measurements and {\em TESS} photometry are shown phased with this period in Fig.\ \ref{HD184961_bz}.

   \begin{figure}
   \centering
   \includegraphics[width=0.45\textwidth]{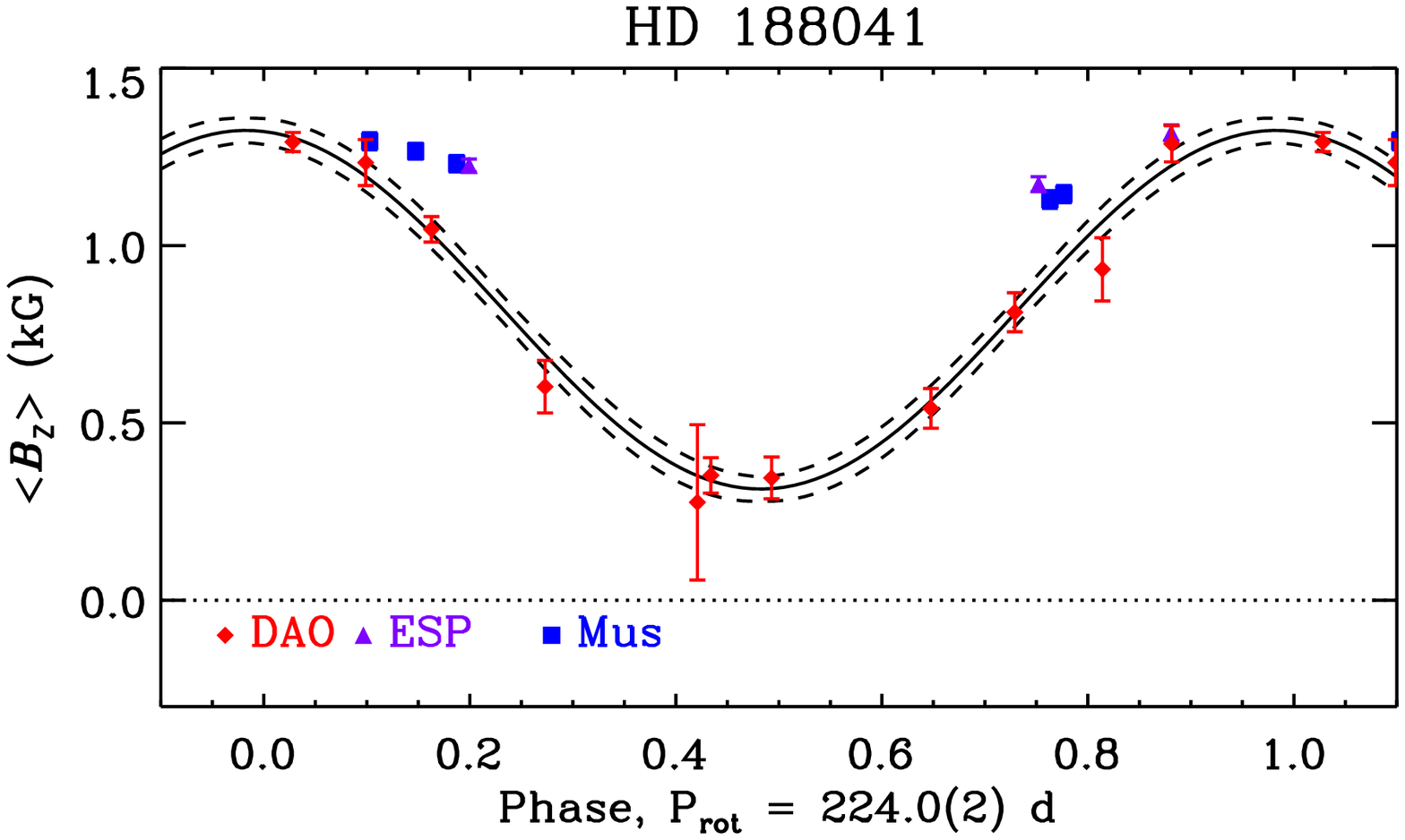}
      \caption[]{\bz~measurements of HD\,188041 phased with the period determined from the same measurements.}
         \label{HD188041_bz}
   \end{figure}

\noindent {\bf HD\,188041}: ESPaDOnS and MuSiCoS \bz~measurements were published by \cite{2019MNRAS.483.3127S}, but did not cover the entire rotation cycle. Fig.\ \ref{HD188041_bz} shows these measurements together with new DAO measurements, phased with the rotation period adopted by \cite{2019MNRAS.483.3127S}, with which we have re-determined the star's ORM parameters.

   \begin{figure}
   \centering
   \includegraphics[width=0.45\textwidth]{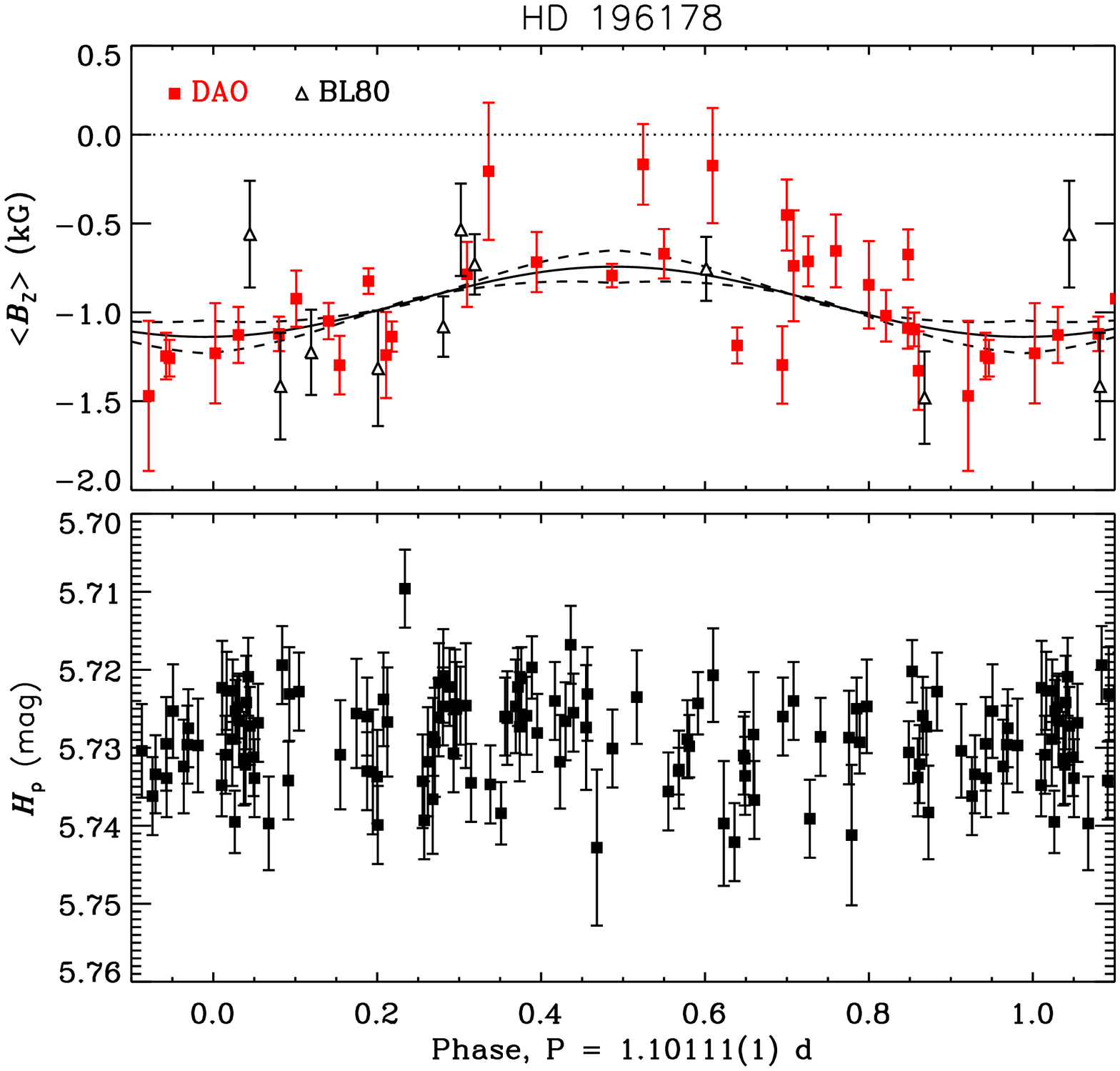}
      \caption[]{\bz~measurements of HD\,196178 ({\em top}) and {\em Hipparcos} light curve ({\em bottom}) phased with the period determined from the magnetic data.}
         \label{HD196178_bz}
   \end{figure}

\noindent {\bf HD\,196178}: This star is listed in the \cite{2009A&A...498..961R} catalogue as a B8 Si star. A photometric period of 1.92 d was given by \cite{2006SASS...25...47W}, however this cannot be confirmed with the {\em Hipparcos} photometry, and does not phase the magnetic dataset. No secure period can be obtained from the Hipparcos light curve. From the combined \bz~measurements, obtained at the DAO and published by \cite{1980ApJS...42..421B}, we find a period of 1.10111(1)~d with a FAP of 0.007 (Fig.\ \ref{HD196178_bz}).

\noindent {\bf HD\,207840}: This star is listed in the \cite{2009A&A...498..961R} catalogue as a B8 Si star. It was claimed to be magnetic by \cite{1971Obs....91...37G}, however the 2 available ESPaDOnS observations are non-detections with 4 G error bars (Fig.\ \ref{lsd_all}). No rotational period is known. This is probably not a magnetic star and was removed from the sample.

\noindent {\bf HD\,209339}: This star is listed in the \cite{2009A&A...498..961R} catalogue as a B0 He star. \cite{2003A&A...407..631B} reported $B_{\rm rms} = 493 \pm 545$~G based on 3 measurements, i.e.\ it was a magnetic non-detection. There are 6 DAO \bz~measurements, 5/6 of which are non-detections and 1 of which is a 4$\sigma$ detection, with a mean error bar across the dataset of 316~G, although given that this is the only detection out of 6 non-detections it must be regarded as highly uncertain. The star was observed by {\em TESS} in sectors 16, 17, and 24. The light curve exhibits numerous significant periods, but these are stochastic and likely related to pulsation rather than rotation. 

\noindent {\bf HD\,224128}: This star is listed as a B3 He star by \cite{2009A&A...498..961R}. No magnetic data are available. The {\em TESS} light curve yields several low-amplitude, closely spaced peaks near 0.5 c/d, which are likely due to $g$-mode pulsations rather than rotation. 

\noindent {\bf HD\,260858}: This star is listed in the \cite{2009A&A...498..961R} catalogue as a B6 He star. A magnetic field was detected by \cite{2018AstBu..73..178R}. The star was observed by {\em TESS} in sector 6. The light curve is strongly contaminated by nearby brighter sources. It is dominated by a monotonic long-term trend, with two significant periods at 13.59~d and 5.96~d, neither of which resemble rotation. 

\noindent {\bf HD\,264111}: This star is listed in the \cite{2009A&A...498..961R} catalogue as a B2 He star. No magnetic field was detected by \cite{2018AstBu..73..178R}. The star was observed by {\em TESS} in sector 6. There are several significant periods, none of which have any obvious harmonic relationship, suggesting that they are related to pulsations rather than rotation. 

\pagebreak

\clearpage

\section{Measured radio flux densities}\label{append:flux_measurements}

\newpage

\input fluxtab.tex

\end{document}

%% file: radio_partab.tex
\begin{table*}
\caption[]{Stellar parameters, rotational periods, dipolar magnetic field strengths, and radio luminosities for the sample. Stars with superscripts are listed in Appendix B if they have new radio ({\em f})lux density observations, and are discussed in Appendix C if new ({\em m})agnetic data or ({\em r})otational periods are available. Superscipt numbers in square brackets in other columns correspond to the reference key at the end of the table.}
\label{partab}
\begin{tabular}{l l l l l l l l l}
Star & $\log{\frac{L_{\rm bol}}{L_\odot}}$ & $\frac{T_{\rm eff}}{{\rm kK}}$ & $\frac{M_*}{{\rm M_\odot}}$ & $\frac{v\sin{i}}{{\rm km~s^{-1}}}$ & $\frac{P_{\rm rot}}{{\rm d}}$ & $\frac{\beta}{{\rm deg}}$ & $\frac{B_{\rm d}}{{\rm kG}}$ & $\log{\frac{L_{\rm rad}}{{\rm L_\odot}}}$ \\
\hline
ALS8988$^{r}$ &  4.05$\pm$0.27 & 27.3$\pm$ 1.4$^{[130]}$ & 10.6$\pm$ 0.6 &  23$^{[42]}$ &  -- &  -- & $>$1.5$^{[42]}$ & $<$-4.40$^{[88]}$ \\ [2.5 pt]
ALS9522 &  3.65$\pm$0.12 & 22.4$\pm$ 1.1$^{[121]}$ &  6.0$\pm$ 0.1 & 105$^{[121]}$ & 1.091$^{[121]}$ & 78$^{+ 2}_{- 3}$ & 11$^{+3}_{-1}$$^{[121]}$ & -5.29$\pm$0.04$^{[88]}$ \\ [2.5 pt]
CPD-271791 &  3.69$\pm$0.09 & 23.8$\pm$ 1.6$^{[130]}$ &  8.8$\pm$ 0.4 &  37$^{[92]}$ & 2.641$^{[92]}$ &  -- & $>$3.9$^{[92]}$ & $<$-4.71$^{[6]}$ \\ [2.5 pt]
HD3360$^{f}$ &  3.82$\pm$0.06 & 20.8$\pm$ 0.2$^{[107]}$ &  8.6$\pm$ 0.1 &  19$^{[96]}$ & 5.370$^{[27]}$ & 82$^{+ 1}_{- 1}$ & 0.15$^{+0.03}_{-0.03}$$^{[110]}$ & $<$-7.79$^{[88,130]}$ \\ [2.5 pt]
HD5737 &  3.33$\pm$0.16 & 13.9$\pm$ 0.4$^{[43]}$ &  5.0$\pm$ 0.4 &  17$^{[109]}$ &  21.7$^{[25]}$ & 73$^{+ 6}_{- 7}$ & 1.8$^{+0.2}_{-0.3}$$^{[4,10,21,130]}$ & $<$-6.44$^{[6]}$ \\ [2.5 pt]
HD11503$^{f}$ &  1.52$\pm$0.03 & 10.1$\pm$ 0.2$^{[105]}$ &  2.35$\pm$ 0.03 &  54$^{[105]}$ & 1.610$^{[106]}$ & 84$^{+ 4}_{-37}$ & 2.3$^{+3.7}_{-0.7}$$^{[106]}$ & -7.85$\pm$0.04$^{[14,130]}$ \\ [2.5 pt]
HD12447 &  1.73$\pm$0.06 & 10.0$\pm$ 0.2$^{[105]}$ &  2.49$\pm$ 0.05 &  70$^{[105]}$ & 1.491$^{[106]}$ & 87$^{+ 1}_{- 9}$ & 1.7$^{+0.5}_{-0.4}$$^{[106,130]}$ & -7.40$\pm$0.04$^{[36]}$ \\ [2.5 pt]
HD12767 &  2.43$\pm$0.15 & 13.0$\pm$ 0.3$^{[43]}$ &  4.0$\pm$ 0.1 &  40$^{[32]}$ & 1.892$^{[25]}$ & 89$^{+ 0}_{- 3}$ & 2.0$^{+0.5}_{-0.6}$$^{[9,116,130]}$ & $<$-6.88$^{[128]}$ \\ [2.5 pt]
HD19832 &  2.08$\pm$0.16 & 12.4$\pm$ 0.4$^{[43]}$ &  3.3$\pm$ 0.1 & 160$^{[99]}$ & 0.728$^{[114]}$ & 89$^{+ 0}_{- 3}$ & 2.7$^{+0.6}_{-0.3}$$^{[114]}$ & -6.70$\pm$0.04$^{[36]}$ \\ [2.5 pt]
HD21699$^{mr}$ &  2.78$\pm$0.04 & 16.0$\pm$ 0.1$^{[43]}$ &  4.7$\pm$ 0.2 &  35$^{[109]}$ & 2.492$^{[25,130]}$ & 78$^{+ 2}_{- 2}$ & 2.8$^{+0.5}_{-0.1}$$^{[5,130]}$ & $<$-6.49$^{[6]}$ \\ [2.5 pt]
HD22470 &  2.43$\pm$0.13 & 13.8$\pm$ 0.3$^{[43]}$ &  3.8$\pm$ 0.2 &  62$^{[32]}$ & 1.929$^{[114]}$ & 87$^{+ 1}_{- 2}$ & 7.5$^{+1.2}_{-0.5}$$^{[114]}$ & $<$-6.87$^{[6]}$ \\ [2.5 pt]
HD22920$^{mr}$ &  2.64$\pm$0.12 & 13.6$\pm$ 0.2$^{[70]}$ &  4.2$\pm$ 0.1 &  34$^{[32]}$ & 3.947$^{[22,130]}$ & 28$^{+ 6}_{- 7}$ & 1.6$^{+1.1}_{-0.0}$$^{[4,130]}$ & $<$-6.77$^{[128]}$ \\ [2.5 pt]
HD25267 &  2.32$\pm$0.03 & 12.6$\pm$ 0.2$^{[119]}$ &  3.5$\pm$ 0.2 &  20$^{[105]}$ & 3.823$^{[119]}$ & 17$^{+ 7}_{- 8}$ & 1.0$^{+0.1}_{-0.1}$$^{[119]}$ & $<$-7.25$^{[128]}$ \\ [2.5 pt]
HD27309 &  1.88$\pm$0.02 & 11.2$\pm$ 0.3$^{[105]}$ &  2.81$\pm$ 0.04 &  56$^{[105]}$ & 1.569$^{[106]}$ &  3$^{+ 5}_{- 3}$ & 1.9$^{+2.6}_{-0.6}$$^{[106]}$ & -7.20$\pm$0.04$^{[36]}$ \\ [2.5 pt]
HD28843$^{mr}$ &  2.51$\pm$0.07 & 14.8$\pm$ 0.2$^{[43]}$ &  4.20$\pm$ 0.06 &  91$^{[130]}$ & 1.374$^{[25,130]}$ & 87$^{+ 1}_{- 5}$ & 0.93$^{+0.30}_{-0.24}$$^{[4,130]}$ & $<$-6.85$^{[6]}$ \\ [2.5 pt]
HD32633 &  1.94$\pm$0.12 & 12.5$\pm$ 0.5$^{[90]}$ &  3.50$\pm$ 0.09 &  25$^{[99]}$ & 6.430$^{[20]}$ & 76$^{+ 3}_{- 3}$ & 17$^{+1}_{-2}$$^{[60,76]}$ & $<$-6.74$^{[14]}$ \\ [2.5 pt]
HD34452 &  2.45$\pm$0.27 & 13.8$\pm$ 0.8$^{[43]}$ &  4.2$\pm$ 0.1 &  53$^{[99]}$ & 2.469$^{[25]}$ & 35$^{+13}_{-16}$ & 3.6$^{+1.1}_{-1.4}$$^{[3,15,130]}$ & -6.69$\pm$0.04$^{[6,14]}$ \\ [2.5 pt]
HD35298 &  2.40$\pm$0.14 & 15.8$\pm$ 0.8$^{[107]}$ &  4.3$\pm$ 0.2 &  60$^{[96]}$ & 1.855$^{[96]}$ & 77$^{+ 2}_{- 2}$ & 11$^{+1}_{-1}$$^{[110]}$ & -5.35$\pm$0.04$^{[14,31,36]}$ \\ [2.5 pt]
HD35456 &  2.88$\pm$0.29 & 13.5$\pm$ 1.4$^{[130]}$ &  4.1$\pm$ 0.5 &  22$^{[80]}$ & 4.951$^{[80]}$ & 15$^{+ 9}_{-11}$ & 2.2$^{+0.2}_{-0.3}$$^{[80,130]}$ & $<$-6.19$^{[128]}$ \\ [2.5 pt]
HD35502$^{f}$ &  2.95$\pm$0.12 & 18.4$\pm$ 0.6$^{[107]}$ &  5.8$\pm$ 0.2 &  78$^{[96]}$ & 0.854$^{[81]}$ & 70$^{+ 1}_{- 1}$ & 7.3$^{+0.5}_{-0.5}$$^{[110]}$ & -5.05$\pm$0.04$^{[36,130]}$ \\ [2.5 pt]
HD35575$^{r}$ &  3.11$\pm$0.09 & 16.7$\pm$ 1.3$^{[130]}$ &  5.8$\pm$ 0.3 & 150$^{[102]}$ & 0.984$^{[130]}$ &  -- & $<$1.3$^{[102,130]}$ & $<$-7.09$^{[17]}$ \\ [2.5 pt]
HD36313$^{r}$ &  2.04$\pm$0.19 & 13.0$\pm$ 0.5$^{[117]}$ &  3.4$\pm$ 0.2 & 160$^{[117]}$ & 0.589$^{[117,130]}$ & 88$^{+ 1}_{- 5}$ & 9.0$^{+1.8}_{-1.3}$$^{[117,130]}$ & -6.10$\pm$0.04$^{[36]}$ \\ [2.5 pt]
HD36429$^{mr}$ &  2.42$\pm$0.02 & 13.8$\pm$ 0.1$^{[130]}$ &  3.87$\pm$ 0.03 &  77$^{[130]}$ &  15.6$^{[130]}$ &  -- & $<$0.20$^{[28,130]}$ & $<$-6.13$^{[128]}$ \\ [2.5 pt]
HD36485 &  3.10$\pm$0.20 & 20.0$\pm$ 2.0$^{[107]}$ &  6.3$\pm$ 0.2 &  33$^{[96]}$ & 1.478$^{[50]}$ &  3$^{+ 1}_{- 1}$ & 8.9$^{+0.2}_{-0.2}$$^{[110]}$ & -5.43$\pm$0.04$^{[6,14,31,75]}$ \\ [2.5 pt]
HD36526$^{f}$ &  2.30$\pm$0.24 & 15.0$\pm$ 2.0$^{[107]}$ &  4.3$\pm$ 0.2 &  59$^{[96]}$ & 1.542$^{[96]}$ & 56$^{+ 1}_{- 2}$ & 11$^{+0}_{-0}$$^{[110]}$ & $<$-6.29$^{[14,130]}$ \\ [2.5 pt]
HD36540$^{mr}$ &  2.73$\pm$0.15 & 14.9$\pm$ 0.7$^{[90]}$ &  4.54$\pm$ 0.09 &  80$^{[86]}$ & 2.173$^{[130]}$ &  9$^{+26}_{- 8}$ & 1.4$^{+3.2}_{-0.1}$$^{[86,130]}$ & $<$-5.99$^{[128]}$ \\ [2.5 pt]
HD36668$^{mr}$ &  2.40$\pm$0.19 & 13.5$\pm$ 0.2$^{[117]}$ &  3.81$\pm$ 0.05 &  60$^{[117]}$ & 2.119$^{[117,130]}$ & 80$^{+ 5}_{- 6}$ & 4.5$^{+2.3}_{-2.7}$$^{[117,130]}$ & $<$-6.05$^{[128]}$ \\ [2.5 pt]
HD36916$^{mr}$ &  2.10$\pm$0.20 & 14.7$\pm$ 0.2$^{[86]}$ &  4.2$\pm$ 0.1 &  78$^{[86]}$ & 1.565$^{[86,130]}$ & 30$^{+ 8}_{-10}$ & 3.4$^{+3.3}_{-0.2}$$^{[4,69,86,94,111,130]}$ & $<$-6.92$^{[14,17]}$ \\ [2.5 pt]
HD37017 &  3.42$\pm$0.25 & 21.0$\pm$ 2.0$^{[107]}$ &  8.4$\pm$ 0.5 & 134$^{[96]}$ & 0.901$^{[96]}$ & 56$^{+ 2}_{- 3}$ & 6.2$^{+0.9}_{-0.9}$$^{[110]}$ & -5.11$\pm$0.04$^{[6,14,18,31,75,85,127]}$ \\ [2.5 pt]
HD37058 &  2.90$\pm$0.11 & 18.6$\pm$ 0.6$^{[107]}$ &  5.8$\pm$ 0.2 &  11$^{[96]}$ &  14.6$^{[96]}$ & 55$^{+11}_{-13}$ & 2.5$^{+0.5}_{-0.5}$$^{[110]}$ & $<$-5.95$^{[14]}$ \\ [2.5 pt]
HD37061$^{f}$ &  3.30$\pm$0.30 & 22.0$\pm$ 1.0$^{[107]}$ &  7.7$\pm$ 0.4 & 100$^{[96]}$ & 1.095$^{[104]}$ & 59$^{+ 4}_{- 4}$ & 9.2$^{+1.0}_{-1.0}$$^{[110]}$ & -5.68$\pm$0.04$^{[85,130]}$ \\ [2.5 pt]
HD37140$^{mr}$ &  2.12$\pm$0.09 & 13.5$\pm$ 0.2$^{[117]}$ &  3.46$\pm$ 0.05 &  25$^{[130]}$ & 2.761$^{[117,130]}$ & 80$^{+ 3}_{- 4}$ & 3.9$^{+0.7}_{-0.4}$$^{[117,130]}$ & $<$-6.08$^{[128]}$ \\ [2.5 pt]
HD37151 &  2.08$\pm$0.08 & 13.5$\pm$ 0.9$^{[130]}$ &  3.4$\pm$ 0.2 &  -- &  -- & 87$^{+ 1}_{-12}$ & 1.4$^{+1.2}_{-1.4}$$^{[118,130]}$ & $<$-6.39$^{[128]}$ \\ [2.5 pt]
HD37210$^{mr}$ &  2.51$\pm$0.05 & 13.5$\pm$ 0.6$^{[130]}$ &  3.92$\pm$ 0.09 &  20$^{[118]}$ &  11.0$^{[22,130]}$ & 78$^{+ 7}_{- 8}$ & 1.8$^{+0.9}_{-0.1}$$^{[68,118,130]}$ & $<$-5.71$^{[128]}$ \\ [2.5 pt]
HD37479 &  3.51$\pm$0.21 & 23.0$\pm$ 2.0$^{[107]}$ &  7.9$\pm$ 0.2 & 145$^{[96]}$ & 1.191$^{[49]}$ & 37$^{+ 7}_{-10}$ & 10$^{+11}_{-0}$$^{[74]}$ & -4.73$\pm$0.04$^{[6,14,18,31,75]}$ \\ [2.5 pt]
HD37642$^{mr}$ &  2.42$\pm$0.10 & 16.0$\pm$ 0.5$^{[118]}$ &  4.3$\pm$ 0.1 &  85$^{[118]}$ & 1.079$^{[118,130]}$ & 74$^{+ 3}_{- 3}$ & 18$^{+1}_{-0}$$^{[118,130]}$ & -5.69$\pm$0.04$^{[36]}$ \\ [2.5 pt]
HD37752 &  2.63$\pm$0.14 & 15.0$\pm$ 0.7$^{[43]}$ &  4.5$\pm$ 0.2 &  35$^{[32]}$ & 1.305$^{[90]}$ &  -- & $<$2.4$^{[28]}$ & $<$-7.08$^{[17]}$ \\ [2.5 pt]
HD37776$^{f}$ &  3.30$\pm$0.15 & 22.0$\pm$ 1.0$^{[107]}$ &  8.3$\pm$ 0.3 & 101$^{[96]}$ & 1.539$^{[84]}$ & 47$^{+ 8}_{-10}$ & 6.1$^{+0.7}_{-0.7}$$^{[56,110]}$ & $<$-6.35$^{[6,130]}$ \\ [2.5 pt]
HD37808$^{mr}$ &  2.28$\pm$0.10 & 14.5$\pm$ 0.2$^{[118]}$ &  3.90$\pm$ 0.08 &  30$^{[99]}$ & 1.099$^{[112,118,130]}$ & 45$^{+20}_{-24}$ & 3.2$^{+1.0}_{-0.3}$$^{[118,130]}$ & -6.16$\pm$0.04$^{[17,31]}$ \\ [2.5 pt]
HD40312 &  2.33$\pm$0.01 & 10.2$\pm$ 0.1$^{[105]}$ &  3.11$\pm$ 0.06 &  55$^{[99]}$ & 3.619$^{[106]}$ & 68$^{+13}_{-15}$ & 1.3$^{+0.4}_{-0.2}$$^{[100]}$ & -7.78$\pm$0.04$^{[14]}$ \\ [2.5 pt]
HD41269$^{mr}$ &  2.30$\pm$0.08 & 12.9$\pm$ 0.9$^{[130]}$ &  3.5$\pm$ 0.2 &  85$^{[32]}$ & 1.048$^{[130]}$ &  0$^{+16}_{- 0}$ & 1.3$^{+0.5}_{-0.3}$$^{[130]}$ & $<$-7.20$^{[17]}$ \\ [2.5 pt]
HD43819 &  2.15$\pm$0.20 & 10.9$\pm$ 0.4$^{[40]}$ &  3.1$\pm$ 0.2 &  10$^{[40]}$ &  15.0$^{[40]}$ & 47$^{+16}_{-19}$ & 2.6$^{+75.7}_{-0.1}$$^{[40]}$ & $<$-6.79$^{[17]}$ \\ [2.5 pt]
HD45583 &  2.07$\pm$0.12 & 13.3$\pm$ 0.3$^{[44]}$ &  3.35$\pm$ 0.09 &  70$^{[99]}$ & 1.177$^{[114]}$ & 69$^{+ 2}_{- 2}$ & 9.1$^{+0.3}_{-0.3}$$^{[114]}$ & -6.24$\pm$0.04$^{[36,127]}$ \\ [2.5 pt]
HD46328 &  4.49$\pm$0.11 & 27.0$\pm$ 1.0$^{[107]}$ & 14.4$\pm$ 0.8 &   8$^{[96]}$ & 30.0 (yr)$^{[91,98,123]}$ & 89$^{+ 0}_{- 9}$ & 1.2$^{+0.6}_{-0.1}$$^{[110]}$ & $<$-7.03$^{[88]}$ \\ [2.5 pt]
HD47777 &  3.42$\pm$0.15 & 22.0$\pm$ 1.0$^{[107]}$ &  7.9$\pm$ 0.4 &  60$^{[96]}$ & 2.640$^{[63]}$ & 82$^{+ 5}_{- 5}$ & 3.3$^{+0.7}_{-0.7}$$^{[110]}$ & $<$-6.40$^{[88]}$ \\ [2.5 pt]
HD49333 &  2.73$\pm$0.04 & 15.8$\pm$ 0.1$^{[43]}$ &  4.8$\pm$ 0.3 &  65$^{[32]}$ & 2.180$^{[25]}$ & 85$^{+ 3}_{-12}$ & 3.6$^{+1.0}_{-1.2}$$^{[15,130]}$ & $<$-6.49$^{[14]}$ \\ [2.5 pt]
\hline
\hline
\end{tabular}
\end{table*}
 
\begin{table*}
\contcaption{}
\label{partab:continued}
\begin{tabular}{l l l l l l l l l}
Star & $\log{\frac{L_{\rm bol}}{L_\odot}}$ & $\frac{T_{\rm eff}}{{\rm kK}}$ & $\frac{M_*}{{\rm M_\odot}}$ & $\frac{v\sin{i}}{{\rm km~s^{-1}}}$ & $\frac{P_{\rm rot}}{{\rm d}}$ & $\frac{\beta}{{\rm deg}}$ & $\frac{B_{\rm d}}{{\rm kG}}$ & $\log{\frac{L_{\rm rad}}{{\rm L_\odot}}}$ \\
\hline
HD49606$^{mr}$ &  2.59$\pm$0.06 & 13.5$\pm$ 0.1$^{[43]}$ &  4.08$\pm$ 0.06 &  19$^{[99]}$ & 8.546$^{[126]}$ &  -- & $<$0.040$^{[130]}$ & $<$-6.25$^{[128]}$ \\ [2.5 pt]
HD51418$^{mr}$ &  1.80$\pm$0.04 &  9.5$\pm$ 0.8$^{[103]}$ &  2.48$\pm$ 0.09 &  28$^{[87]}$ & 5.431$^{[116,130]}$ & 89$^{+ 0}_{- 0}$ & 3.5$^{+1.0}_{-0.4}$$^{[87,130]}$ & $<$-6.65$^{[128]}$ \\ [2.5 pt]
HD55522$^{f}$ &  3.00$\pm$0.18 & 17.4$\pm$ 0.4$^{[107]}$ &  5.9$\pm$ 0.2 &  70$^{[96]}$ & 2.729$^{[96]}$ & 89$^{+ 0}_{- 1}$ & 3.1$^{+0.4}_{-0.4}$$^{[110]}$ & $<$-6.16$^{[130]}$ \\ [2.5 pt]
HD58260 &  3.22$\pm$0.26 & 19.3$\pm$ 1.3$^{[107]}$ &  6.2$\pm$ 0.5 &   3$^{[96]}$ &  -- &  0$^{+ 5}_{- 0}$ & 6.5$^{+0.2}_{-0.2}$$^{[110]}$ & $<$-6.69$^{[6,88]}$ \\ [2.5 pt]
HD60344$^{r}$ &  3.58$\pm$0.09 & 21.0$\pm$ 0.3$^{[43]}$ &  7.9$\pm$ 0.2 &  55$^{[109]}$ &  -- &  -- & $>$1.2$^{[92,130]}$ & $<$-4.76$^{[6]}$ \\ [2.5 pt]
HD61556$^{f}$ &  3.12$\pm$0.24 & 18.5$\pm$ 0.8$^{[73]}$ &  6.1$\pm$ 0.3 &  58$^{[73]}$ & 1.909$^{[73]}$ & 58$^{+ 6}_{- 7}$ & 2.8$^{+0.3}_{-0.3}$$^{[110]}$ & -5.93$\pm$0.04$^{[120,130]}$ \\ [2.5 pt]
HD64740$^{f}$ &  3.81$\pm$0.15 & 24.5$\pm$ 1.0$^{[107]}$ & 10.1$\pm$ 0.5 & 135$^{[96]}$ & 1.330$^{[96]}$ & 71$^{+ 5}_{- 5}$ & 3.0$^{+0.5}_{-0.5}$$^{[110]}$ & -7.19$\pm$0.04$^{[130]}$ \\ [2.5 pt]
HD65339 &  1.45$\pm$0.02 &  8.5$\pm$ 0.1$^{[105]}$ &  2.09$\pm$ 0.02 &  13$^{[99]}$ & 8.027$^{[106]}$ & 89$^{+ 0}_{- 1}$ & 15$^{+2}_{-1}$$^{[30]}$ & $<$-7.50$^{[6,14]}$ \\ [2.5 pt]
HD66665 &  4.69$\pm$0.23 & 28.5$\pm$ 1.0$^{[107]}$ & 15.9$\pm$ 1.1 &   8$^{[96]}$ &  24.5$^{[96]}$ & 75$^{+ 3}_{- 3}$ & 0.56$^{+0.10}_{-0.10}$$^{[110]}$ & $<$-4.96$^{[88]}$ \\ [2.5 pt]
HD66765$^{f}$ &  3.44$\pm$0.24 & 20.0$\pm$ 2.0$^{[65]}$ &  7.2$\pm$ 0.6 &  58$^{[96]}$ & 1.608$^{[96]}$ & 73$^{+ 5}_{- 5}$ & 2.8$^{+0.5}_{-0.5}$$^{[110]}$ & $<$-6.23$^{[130]}$ \\ [2.5 pt]
HD79158 &  2.61$\pm$0.06 & 13.3$\pm$ 0.1$^{[43]}$ &  4.3$\pm$ 0.1 &  49$^{[99]}$ & 3.835$^{[95]}$ & 87$^{+ 1}_{- 4}$ & 3.1$^{+0.4}_{-0.4}$$^{[95]}$ & -6.55$\pm$0.04$^{[6,36]}$ \\ [2.5 pt]
HD90044 &  1.66$\pm$0.10 & 10.0$\pm$ 0.2$^{[90]}$ &  2.8$\pm$ 0.3 &  23$^{[105]}$ & 4.379$^{[25]}$ & 89$^{+ 0}_{- 0}$ & 4.4$^{+1.2}_{-1.5}$$^{[15,24,130]}$ & $<$-7.71$^{[17]}$ \\ [2.5 pt]
HD105382 &  3.04$\pm$0.16 & 18.0$\pm$ 0.5$^{[26]}$ &  5.8$\pm$ 0.2 &  74$^{[96]}$ & 1.295$^{[26]}$ & 51$^{+ 7}_{- 8}$ & 2.6$^{+0.1}_{-0.1}$$^{[110]}$ & -6.07$\pm$0.04$^{[120]}$ \\ [2.5 pt]
HD112413 &  1.97$\pm$0.02 & 11.3$\pm$ 0.2$^{[105]}$ &  2.93$\pm$ 0.03 &  15$^{[99]}$ & 5.469$^{[106]}$ & 88$^{+ 1}_{- 8}$ & 3.5$^{+0.8}_{-0.2}$$^{[66,67]}$ & -8.13$\pm$0.04$^{[6,36]}$ \\ [2.5 pt]
HD118022 &  1.53$\pm$0.04 &  9.4$\pm$ 0.1$^{[105]}$ &  2.26$\pm$ 0.02 &  12$^{[105]}$ & 3.722$^{[106]}$ & 65$^{+15}_{-17}$ & 3.2$^{+8.0}_{-0.2}$$^{[106]}$ & -7.41$\pm$0.04$^{[124]}$ \\ [2.5 pt]
HD122532 &  2.37$\pm$0.12 & 11.9$\pm$ 0.5$^{[90]}$ &  2.99$\pm$ 0.04 &  -- & 3.681$^{[12]}$ & 89$^{+ 0}_{- 3}$ & 3.0$^{+0.7}_{-0.9}$$^{[7,13,15,116,130]}$ & $<$-7.00$^{[14]}$ \\ [2.5 pt]
HD124224 &  1.93$\pm$0.01 & 12.3$\pm$ 0.2$^{[105]}$ &  3.02$\pm$ 0.01 & 169$^{[105]}$ & 0.521$^{[84]}$ & 87$^{+ 2}_{-11}$ & 4.0$^{+0.3}_{-0.2}$$^{[64]}$ & -6.24$\pm$0.04$^{[17,18,31,36,120,125]}$ \\ [2.5 pt]
HD125248 &  1.39$\pm$0.23 &  9.7$\pm$ 0.3$^{[43]}$ &  2.42$\pm$ 0.07 &  11$^{[99]}$ & 9.300$^{[41]}$ & 89$^{+ 0}_{- 3}$ & 9.0$^{+1.1}_{-1.3}$$^{[78]}$ & $<$-7.17$^{[128]}$ \\ [2.5 pt]
HD125823 &  3.16$\pm$0.20 & 19.0$\pm$ 2.0$^{[107]}$ &  5.9$\pm$ 0.2 &  16$^{[96]}$ & 8.817$^{[96]}$ & 75$^{+ 4}_{- 6}$ & 1.8$^{+0.2}_{-0.2}$$^{[110]}$ & $<$-6.54$^{[6]}$ \\ [2.5 pt]
HD126515 &  1.36$\pm$0.16 &  8.9$\pm$ 0.2$^{[130]}$ &  2.65$\pm$ 0.07 &  16$^{[130]}$ & 129.9$^{[23]}$ & 84$^{+ 4}_{-13}$ & 13$^{+1}_{-0}$$^{[23]}$ & $<$-7.13$^{[14]}$ \\ [2.5 pt]
HD131120$^{mr}$ &  3.15$\pm$0.09 & 19.4$\pm$ 1.5$^{[130]}$ &  6.3$\pm$ 0.3 &  57$^{[32]}$ & 1.569$^{[57,130]}$ &  -- & $<$0.17$^{[68]}$ & $<$-6.97$^{[14]}$ \\ [2.5 pt]
HD133029$^{mr}$ &  1.98$\pm$0.08 & 11.8$\pm$ 0.9$^{[43]}$ &  2.8$\pm$ 0.1 &  30$^{[32]}$ & 2.888$^{[46,130]}$ & 12$^{+ 4}_{- 4}$ & 9.0$^{+5.0}_{-0.3}$$^{[3,130]}$ & $<$-6.89$^{[14]}$ \\ [2.5 pt]
HD133652 &  2.02$\pm$0.10 & 12.8$\pm$ 0.5$^{[90]}$ &  3.27$\pm$ 0.09 &  48$^{[99]}$ & 2.304$^{[25]}$ & 65$^{+ 8}_{-11}$ & 7.6$^{+0.8}_{-1.0}$$^{[15,130]}$ & -7.16$\pm$0.04$^{[36]}$ \\ [2.5 pt]
HD133880 &  1.73$\pm$0.06 & 10.7$\pm$ 0.1$^{[43]}$ &  3.08$\pm$ 0.08 & 103$^{[99]}$ & 0.877$^{[83]}$ & 83$^{+ 1}_{- 1}$ & 12$^{+0}_{-1}$$^{[83]}$ & -5.44$\pm$0.04$^{[75,36]}$ \\ [2.5 pt]
HD135679$^{mr}$ &  2.46$\pm$0.19 & 15.1$\pm$ 2.8$^{[130]}$ &  2.86$\pm$ 0.06 &   1$^{[87]}$ & 5.321$^{[130]}$ & 14$^{+ 6}_{- 8}$ & 4.4$^{+9.3}_{-0.2}$$^{[87,94,130]}$ & $<$-7.04$^{[17]}$ \\ [2.5 pt]
HD137193$^{mr}$ &  1.91$\pm$0.13 & 10.6$\pm$ 0.7$^{[90]}$ &  3.5$\pm$ 0.2 &  -- & 4.867$^{[130]}$ &  0$^{+25}_{- 0}$ & 3.5$^{+16.1}_{-0.5}$$^{[8,130]}$ & $<$-6.51$^{[14]}$ \\ [2.5 pt]
HD137909 &  1.46$\pm$0.01 &  7.5$\pm$ 0.1$^{[105]}$ &  1.98$\pm$ 0.03 &   3$^{[99]}$ &  18.5$^{[106]}$ & 89$^{+ 0}_{- 2}$ & 5.2$^{+7.1}_{-1.0}$$^{[106,130]}$ & $<$-7.90$^{[6]}$ \\ [2.5 pt]
HD138764 &  2.62$\pm$0.17 & 15.7$\pm$ 2.7$^{[130]}$ &  4.3$\pm$ 0.4 &  19$^{[32]}$ & 1.259$^{[90]}$ &  -- & $<$0.049$^{[48]}$ & $<$-7.63$^{[17]}$ \\ [2.5 pt]
HD142184 &  2.85$\pm$0.13 & 18.5$\pm$ 0.5$^{[107]}$ &  5.7$\pm$ 0.1 & 288$^{[96]}$ & 0.508$^{[58]}$ &  8$^{+ 3}_{- 3}$ & 9.0$^{+2.0}_{-2.0}$$^{[110]}$ & -4.25$\pm$0.04$^{[97,120]}$ \\ [2.5 pt]
HD142301 &  2.56$\pm$0.07 & 15.9$\pm$ 0.2$^{[43]}$ &  4.46$\pm$ 0.05 &  78$^{[32]}$ & 1.459$^{[114]}$ & 47$^{+ 9}_{-11}$ & 12$^{+9}_{-0}$$^{[114]}$ & -5.62$\pm$0.04$^{[14,17,18,31]}$ \\ [2.5 pt]
HD142884$^{mr}$ &  2.17$\pm$0.12 & 14.3$\pm$ 0.5$^{[90]}$ &  3.7$\pm$ 0.1 & 130$^{[32]}$ & 0.803$^{[112]}$ &  -- & $<$0.75$^{[4,130]}$ & $<$-6.84$^{[14]}$ \\ [2.5 pt]
HD142990 &  2.93$\pm$0.13 & 18.0$\pm$ 0.5$^{[107]}$ &  5.6$\pm$ 0.2 & 122$^{[96]}$ & 0.979$^{[108]}$ & 83$^{+ 2}_{- 3}$ & 4.7$^{+0.4}_{-0.4}$$^{[110]}$ & -5.77$\pm$0.04$^{[14,17,18,31,101]}$ \\ [2.5 pt]
HD143473 &  1.87$\pm$0.10 & 12.4$\pm$ 1.0$^{[90]}$ &  2.25$\pm$ 0.08 &  25$^{[32]}$ & 2.843$^{[25]}$ & 18$^{+ 5}_{- 6}$ & 18$^{+2}_{-2}$$^{[13,15,130]}$ & -6.56$\pm$0.04$^{[36]}$ \\ [2.5 pt]
HD143699$^{mr}$ &  2.65$\pm$0.12 & 15.5$\pm$ 0.4$^{[43]}$ &  4.6$\pm$ 0.2 & 115$^{[130]}$ & 0.894$^{[130]}$ &  -- & $<$0.60$^{[4,39,130]}$ & -7.21$\pm$0.04$^{[14,128]}$ \\ [2.5 pt]
HD144334 &  2.34$\pm$0.12 & 14.8$\pm$ 0.4$^{[43]}$ &  4.0$\pm$ 0.1 &  82$^{[32]}$ & 1.495$^{[114]}$ & 55$^{+ 7}_{- 8}$ & 3.6$^{+0.3}_{-0.3}$$^{[114]}$ & -6.79$\pm$0.04$^{[14,17,31]}$ \\ [2.5 pt]
HD145102 &  2.02$\pm$0.12 & 10.8$\pm$ 0.5$^{[90]}$ &  3.0$\pm$ 0.1 &  84$^{[32]}$ & 1.418$^{[59]}$ &  -- & $<$0.98$^{[8]}$ & $<$-6.90$^{[128]}$ \\ [2.5 pt]
HD145482$^{r}$ &  3.59$\pm$0.17 & 24.2$\pm$ 4.3$^{[130]}$ &  8.1$\pm$ 0.8 & 166$^{[32]}$ & 5.804$^{[130]}$ &  -- & $<$0.38$^{[8]}$ & $<$-6.94$^{[14]}$ \\ [2.5 pt]
HD145501C &  2.46$\pm$0.15 & 14.5$\pm$ 0.5$^{[90]}$ &  4.0$\pm$ 0.2 &  70$^{[32]}$ & 1.026$^{[114]}$ & 89$^{+ 0}_{- 2}$ & 5.8$^{+0.3}_{-0.3}$$^{[114]}$ & -6.08$\pm$0.04$^{[14,17,127]}$ \\ [2.5 pt]
HD146001$^{mr}$ &  2.38$\pm$0.14 & 13.8$\pm$ 0.3$^{[43]}$ &  3.61$\pm$ 0.06 &  90$^{[32]}$ & 0.586$^{[59]}$ &  -- & $<$0.50$^{[4,130]}$ & -6.87$\pm$0.04$^{[14]}$ \\ [2.5 pt]
HD147010 &  1.65$\pm$0.08 & 12.5$\pm$ 0.6$^{[90]}$ &  2.45$\pm$ 0.09 &  15$^{[99]}$ & 3.921$^{[62]}$ & 10$^{+ 4}_{- 5}$ & 19$^{+0}_{-0}$$^{[19,116,130]}$ & $<$-7.10$^{[14]}$ \\ [2.5 pt]
HD147890 &  2.43$\pm$0.12 & 11.3$\pm$ 0.5$^{[90]}$ &  3.7$\pm$ 0.2 &  65$^{[32]}$ & 4.336$^{[59]}$ &  -- & $<$0.90$^{[8]}$ & $<$-6.24$^{[128]}$ \\ [2.5 pt]
HD147932 &  2.50$\pm$0.20 & 17.0$\pm$ 1.0$^{[114]}$ &  4.8$\pm$ 0.3 & 140$^{[54]}$ & 0.864$^{[93]}$ &  0$^{+ 6}_{- 0}$ & 7.6$^{+9.6}_{-0.5}$$^{[114,129]}$ & -5.14$\pm$0.04$^{[115]}$ \\ [2.5 pt]
HD147933 &  3.30$\pm$0.17 & 20.8$\pm$ 0.5$^{[113]}$ &  7.3$\pm$ 0.2 & 200$^{[113]}$ & 0.747$^{[113]}$ & 75$^{+ 9}_{-11}$ & 4.7$^{+0.5}_{-0.4}$$^{[113,129]}$ & -5.55$\pm$0.04$^{[113]}$ \\ [2.5 pt]
HD148112 &  1.85$\pm$0.02 &  9.2$\pm$ 0.1$^{[105]}$ &  2.52$\pm$ 0.01 &  44$^{[40]}$ & 3.044$^{[106]}$ &  0$^{+ 9}_{- 0}$ & 0.76$^{+0.41}_{-0.11}$$^{[106,130]}$ & $<$-7.19$^{[6]}$ \\ [2.5 pt]
HD148199 &  1.88$\pm$0.12 & 11.7$\pm$ 0.7$^{[90]}$ &  2.5$\pm$ 0.1 &  15$^{[32]}$ & 7.726$^{[25]}$ & 63$^{+ 9}_{-12}$ & 5.0$^{+0.6}_{-0.7}$$^{[8,15,130]}$ & $<$-6.54$^{[14]}$ \\ [2.5 pt]
\hline
\hline
\end{tabular}
\end{table*}
 
\begin{table*}
\contcaption{}
\label{partab:continued}
\begin{tabular}{l l l l l l l l l}
Star & $\log{\frac{L_{\rm bol}}{L_\odot}}$ & $\frac{T_{\rm eff}}{{\rm kK}}$ & $\frac{M_*}{{\rm M_\odot}}$ & $\frac{v\sin{i}}{{\rm km~s^{-1}}}$ & $\frac{P_{\rm rot}}{{\rm d}}$ & $\frac{\beta}{{\rm deg}}$ & $\frac{B_{\rm d}}{{\rm kG}}$ & $\log{\frac{L_{\rm rad}}{{\rm L_\odot}}}$ \\
\hline
HD149438 &  4.47$\pm$0.13 & 32.0$\pm$ 1.0$^{[107]}$ & 17.5$\pm$ 0.9 &   7$^{[96]}$ &  41.0$^{[37]}$ & 75$^{+ 7}_{- 8}$ & 0.31$^{+0.10}_{-0.01}$$^{[77]}$ & $<$-7.83$^{[88]}$ \\ [2.5 pt]
HD149822$^{mr}$ &  1.85$\pm$0.18 & 10.8$\pm$ 0.3$^{[43]}$ &  2.7$\pm$ 0.2 &  60$^{[32]}$ & 1.966$^{[130]}$ & 87$^{+ 2}_{- 3}$ & 4.0$^{+2.1}_{-1.1}$$^{[16,38,68,130]}$ & $<$-7.36$^{[17]}$ \\ [2.5 pt]
HD151346 &  2.44$\pm$0.17 & 13.7$\pm$ 1.1$^{[90]}$ &  3.8$\pm$ 0.2 &  46$^{[32]}$ & 2.180$^{[90]}$ &  -- & $<$1.7$^{[4]}$ & $<$-6.66$^{[14]}$ \\ [2.5 pt]
HD152107 &  1.47$\pm$0.01 &  8.8$\pm$ 0.1$^{[105]}$ &  2.15$\pm$ 0.01 &  21$^{[105]}$ & 3.857$^{[106]}$ & 19$^{+ 4}_{- 5}$ & 4.2$^{+0.5}_{-0.2}$$^{[106,130]}$ & $<$-8.01$^{[6,14]}$ \\ [2.5 pt]
HD156424 &  3.00$\pm$0.40 & 18.0$\pm$ 3.0$^{[122]}$ &  4.6$\pm$ 0.3 &   7$^{[96]}$ & 0.524$^{[122]}$ & 39$^{+18}_{-21}$ & 8.0$^{+12.0}_{-2.0}$$^{[122]}$ & -4.91$\pm$0.04$^{[88]}$ \\ [2.5 pt]
HD163472 &  3.81$\pm$0.10 & 25.2$\pm$ 1.1$^{[107]}$ & 10.3$\pm$ 0.5 &  62$^{[96]}$ & 3.639$^{[29]}$ & 46$^{+13}_{-14}$ & 1.1$^{+0.5}_{-0.5}$$^{[110]}$ & $<$-6.83$^{[88]}$ \\ [2.5 pt]
HD164429$^{mr}$ &  1.85$\pm$0.04 & 12.0$\pm$ 0.5$^{[130]}$ &  2.99$\pm$ 0.06 &  90$^{[32]}$ & 1.082$^{[130]}$ & 88$^{+ 1}_{-12}$ & 3.0$^{+0.7}_{-0.3}$$^{[15,130]}$ & -6.91$\pm$0.04$^{[17,36]}$ \\ [2.5 pt]
HD165474 &  2.06$\pm$0.26 & 13.2$\pm$ 3.4$^{[130]}$ &  1.82$\pm$ 0.08 &  18$^{[32]}$ &  9.0 (yr)$^{[82]}$ & 89$^{+ 0}_{- 0}$ & 3.0$^{+4.9}_{-1.8}$$^{[82,130]}$ & $<$-6.96$^{[128]}$ \\ [2.5 pt]
HD168785$^{r}$ &  3.48$\pm$0.08 & 23.0$\pm$ 1.2$^{[130]}$ &  8.1$\pm$ 0.4 &  14$^{[109]}$ &  -- &  -- & $>$4.0$^{[92]}$ & $<$-5.25$^{[128]}$ \\ [2.5 pt]
HD168856$^{mr}$ &  2.10$\pm$0.09 & 11.9$\pm$ 1.2$^{[130]}$ &  3.0$\pm$ 0.1 &  73$^{[79]}$ & 2.428$^{[112]}$ & 55$^{+15}_{-16}$ & 3.5$^{+3.7}_{-0.6}$$^{[79,87,35,130]}$ & $<$-6.77$^{[128]}$ \\ [2.5 pt]
HD170000 &  2.36$\pm$0.01 & 11.6$\pm$ 0.0$^{[105]}$ &  3.47$\pm$ 0.02 &  83$^{[105]}$ & 1.716$^{[106]}$ & 70$^{+10}_{-12}$ & 1.8$^{+0.0}_{-0.1}$$^{[106]}$ & -7.11$\pm$0.04$^{[36]}$ \\ [2.5 pt]
HD170973$^{mr}$ &  2.32$\pm$0.13 & 10.8$\pm$ 0.2$^{[43]}$ &  3.39$\pm$ 0.06 &   4$^{[99]}$ &  18.1$^{[130]}$ & 83$^{+ 2}_{- 3}$ & 4.8$^{+11.2}_{-0.5}$$^{[15,13,130]}$ & $<$-6.56$^{[128]}$ \\ [2.5 pt]
HD171247$^{mr}$ &  2.79$\pm$0.16 & 12.2$\pm$ 0.3$^{[43]}$ &  4.0$\pm$ 0.2 &  68$^{[99]}$ & 3.910$^{[130]}$ & 85$^{+ 3}_{-34}$ & 4.1$^{+3.1}_{-0.4}$$^{[130]}$ & -5.16$\pm$0.04$^{[17,31]}$ \\ [2.5 pt]
HD175132$^{r}$ &  2.78$\pm$0.04 & 13.2$\pm$ 0.5$^{[130]}$ &  4.0$\pm$ 0.2 &  40$^{[32]}$ & 8.030$^{[130]}$ &  -- & $>$3.5$^{[28]}$ & $<$-6.69$^{[17,128]}$ \\ [2.5 pt]
HD175362$^{f}$ &  2.64$\pm$0.12 & 17.6$\pm$ 0.4$^{[107]}$ &  5.3$\pm$ 0.2 &  34$^{[96]}$ & 3.674$^{[96]}$ & 68$^{+ 5}_{- 6}$ & 17$^{+0}_{-0}$$^{[110]}$ & -6.72$\pm$0.04$^{[14,17,18,130]}$ \\ [2.5 pt]
HD175744$^{mr}$ &  2.65$\pm$0.09 & 12.6$\pm$ 0.2$^{[43]}$ &  4.00$\pm$ 0.08 &  50$^{[32]}$ & 2.799$^{[130]}$ &  -- & $<$0.50$^{[33,111,130]}$ & $<$-7.25$^{[17,128]}$ \\ [2.5 pt]
HD176582 &  2.90$\pm$0.15 & 17.0$\pm$ 1.0$^{[107]}$ &  5.6$\pm$ 0.2 & 103$^{[96]}$ & 1.582$^{[55]}$ & 89$^{+ 0}_{- 1}$ & 5.4$^{+0.2}_{-0.2}$$^{[110]}$ & -6.07$\pm$0.04$^{[36]}$ \\ [2.5 pt]
HD177003 &  3.05$\pm$0.10 & 17.7$\pm$ 0.7$^{[90]}$ &  5.7$\pm$ 0.2 &  12$^{[32]}$ & 1.800$^{[90]}$ &  -- & $<$0.81$^{[28]}$ & $<$-7.19$^{[17]}$ \\ [2.5 pt]
HD177410$^{m}$ &  2.30$\pm$0.15 & 14.5$\pm$ 0.5$^{[34]}$ &  3.6$\pm$ 0.1 & 100$^{[34]}$ & 1.123$^{[47]}$ &  -- & $<$1.7$^{[15,71,130]}$ & $<$-7.23$^{[17]}$ \\ [2.5 pt]
HD179527 &  2.63$\pm$0.16 & 10.4$\pm$ 0.3$^{[40]}$ &  3.39$\pm$ 0.05 &  33$^{[40]}$ & 7.098$^{[40]}$ & 88$^{+ 1}_{-12}$ & 0.52$^{+0.71}_{-0.11}$$^{[40]}$ & $<$-6.54$^{[128]}$ \\ [2.5 pt]
HD182180$^{f}$ &  3.09$\pm$0.18 & 19.8$\pm$ 1.4$^{[107]}$ &  6.5$\pm$ 0.2 & 306$^{[96]}$ & 0.521$^{[52,51]}$ & 81$^{+ 3}_{- 4}$ & 9.5$^{+0.6}_{-0.6}$$^{[110]}$ & -4.65$\pm$0.04$^{[89,130]}$ \\ [2.5 pt]
HD183056 &  2.69$\pm$0.11 & 11.7$\pm$ 0.4$^{[40]}$ &  4.0$\pm$ 0.2 &  35$^{[99]}$ & 2.992$^{[40]}$ & 82$^{+ 3}_{- 4}$ & 1.6$^{+2.4}_{-0.4}$$^{[40]}$ & $<$-7.14$^{[17]}$ \\ [2.5 pt]
HD183339$^{r}$ &  2.70$\pm$0.05 & 14.0$\pm$ 0.6$^{[130]}$ &  4.3$\pm$ 0.1 &  41$^{[32]}$ & 4.204$^{[130]}$ &  -- & $>$4.5$^{[28]}$ & $<$-6.39$^{[128]}$ \\ [2.5 pt]
HD184927$^{f}$ &  3.59$\pm$0.16 & 22.0$\pm$ 1.0$^{[107]}$ &  8.4$\pm$ 0.5 &   8$^{[96]}$ & 9.531$^{[72]}$ & 67$^{+ 3}_{- 4}$ & 8.8$^{+1.4}_{-0.5}$$^{[72]}$ & $<$-6.06$^{[17,128,130]}$ \\ [2.5 pt]
HD184961$^{mr}$ &  2.35$\pm$0.08 & 11.8$\pm$ 1.0$^{[130]}$ &  3.47$\pm$ 0.09 &  34$^{[32]}$ & 6.335$^{[130]}$ &  4$^{+ 3}_{- 4}$ & 4.3$^{+9.0}_{-0.4}$$^{[130]}$ & $<$-6.36$^{[128]}$ \\ [2.5 pt]
HD186205 &  3.84$\pm$0.25 & 19.6$\pm$ 0.8$^{[107]}$ &  8.3$\pm$ 0.6 &   6$^{[96]}$ &  37.2$^{[96]}$ &  7$^{+ 3}_{- 4}$ & 3.0$^{+1.0}_{-0.5}$$^{[110]}$ & $<$-4.82$^{[128]}$ \\ [2.5 pt]
HD187474 &  1.76$\pm$0.03 &  9.9$\pm$ 0.1$^{[105]}$ &  2.52$\pm$ 0.02 &   0$^{[99]}$ &  6.4 (yr)$^{[106]}$ & 89$^{+ 0}_{- 3}$ & 7.2$^{+2.1}_{-0.2}$$^{[106]}$ & $<$-7.29$^{[128]}$ \\ [2.5 pt]
HD188041$^{mr}$ &  1.40$\pm$0.01 &  8.5$\pm$ 0.1$^{[105]}$ &  2.07$\pm$ 0.04 &   4$^{[99]}$ & 224.0$^{[106]}$ &  0$^{+25}_{- 0}$ & 4.1$^{+16.6}_{-0.4}$$^{[106,130]}$ & $<$-7.46$^{[128]}$ \\ [2.5 pt]
HD189775$^{f}$ &  2.91$\pm$0.11 & 17.5$\pm$ 0.6$^{[107]}$ &  5.6$\pm$ 0.2 &  58$^{[96]}$ & 2.607$^{[96]}$ & 43$^{+11}_{-12}$ & 4.3$^{+0.7}_{-0.7}$$^{[110]}$ & -5.83$\pm$0.04$^{[130]}$ \\ [2.5 pt]
HD192678 &  1.65$\pm$0.27 &  9.0$\pm$ 0.1$^{[43]}$ &  2.5$\pm$ 0.2 &   6$^{[130]}$ &  12.9$^{[23]}$ & 25$^{+11}_{-12}$ & 5.5$^{+4.4}_{-0.1}$$^{[23,130]}$ & $<$-6.79$^{[128]}$ \\ [2.5 pt]
HD196178$^{mr}$ &  2.15$\pm$0.10 & 13.1$\pm$ 0.5$^{[90]}$ &  4.2$\pm$ 0.2 &  50$^{[32]}$ & 1.101$^{[130]}$ & 41$^{+ 7}_{- 9}$ & 3.9$^{+0.4}_{-0.2}$$^{[3,130]}$ & -5.99$\pm$0.04$^{[36]}$ \\ [2.5 pt]
HD196502 &  2.01$\pm$0.39 &  8.9$\pm$ 0.4$^{[43]}$ &  2.55$\pm$ 0.05 &   9$^{[99]}$ &  20.3$^{[25]}$ & 88$^{+ 1}_{-12}$ & 1.7$^{+0.3}_{-0.3}$$^{[1,2,11,116,130]}$ & $<$-6.54$^{[6]}$ \\ [2.5 pt]
HD200775$^{f}$ &  3.95$\pm$0.30 & 18.6$\pm$ 2.0$^{[45]}$ &  9.0$\pm$ 0.8 &  26$^{[45]}$ & 4.381$^{[45]}$ & 89$^{+ 0}_{- 6}$ & 3.9$^{+1.7}_{-0.7}$$^{[45,130]}$ & -6.11$\pm$0.04$^{[130]}$ \\ [2.5 pt]
HD202671 &  2.70$\pm$0.07 & 13.2$\pm$ 0.1$^{[43]}$ &  4.0$\pm$ 0.2 &  20$^{[99]}$ & 1.992$^{[90]}$ &  -- & $<$0.056$^{[53]}$ & $<$-7.04$^{[128]}$ \\ [2.5 pt]
\hline
\hline
\end{tabular}
\end{table*}
 
\begin{table*}
\contcaption{}
\label{partab:continued}
\begin{tabular}{l l l l l l l l l}
Star & $\log{\frac{L_{\rm bol}}{L_\odot}}$ & $\frac{T_{\rm eff}}{{\rm kK}}$ & $\frac{M_*}{{\rm M_\odot}}$ & $\frac{v\sin{i}}{{\rm km~s^{-1}}}$ & $\frac{P_{\rm rot}}{{\rm d}}$ & $\frac{\beta}{{\rm deg}}$ & $\frac{B_{\rm d}}{{\rm kG}}$ & $\log{\frac{L_{\rm rad}}{{\rm L_\odot}}}$ \\
\hline
HD205021$^{f}$ &  4.26$\pm$0.11 & 25.0$\pm$ 1.0$^{[107]}$ & 11.9$\pm$ 1.1 &  34$^{[96]}$ &  12.0$^{[61]}$ & 86$^{+ 2}_{- 4}$ & 0.26$^{+0.03}_{-0.03}$$^{[110]}$ & $<$-7.42$^{[88,130]}$ \\ [2.5 pt]
HD208057$^{f}$ &  3.01$\pm$0.11 & 16.5$\pm$ 1.2$^{[107]}$ &  5.3$\pm$ 0.3 & 105$^{[96]}$ & 1.368$^{[96]}$ & 89$^{+ 0}_{- 0}$ & 0.60$^{+0.20}_{-0.20}$$^{[110]}$ & $<$-7.30$^{[130]}$ \\ [2.5 pt]
HD215441 &  2.31$\pm$0.09 & 14.5$\pm$ 0.4$^{[43]}$ &  3.9$\pm$ 0.1 &   5$^{[99]}$ & 9.490$^{[23]}$ & 37$^{+ 7}_{- 7}$ & 62$^{+6}_{-3}$$^{[23]}$ & -5.13$\pm$0.04$^{[6,14,18,31,75]}$ \\ [2.5 pt]
HD224801 &  2.19$\pm$0.10 & 11.9$\pm$ 0.5$^{[90]}$ &  3.2$\pm$ 0.1 &  30$^{[32]}$ & 3.740$^{[22]}$ &  -- & $>$4.6$^{[1]}$ & $<$-6.61$^{[128]}$ \\ [2.5 pt]
HD260858$^{r}$ &  3.52$\pm$0.27 & 18.0$\pm$ 1.0$^{[43]}$ &  6.7$\pm$ 0.4 &  47$^{[109]}$ &  -- &  -- & $>$1.8$^{[94]}$ & $<$-4.86$^{[6]}$ \\ [2.5 pt]
HD335238 &  1.62$\pm$0.23 &  9.4$\pm$ 0.5$^{[90]}$ &  2.25$\pm$ 0.04 &  -- &  48.7$^{[82]}$ & 85$^{+ 4}_{-44}$ & 5.9$^{+2.0}_{-2.5}$$^{[82,111,130]}$ & $<$-6.18$^{[128]}$ \\ [2.5 pt]
\hline\hline
\end{tabular}
\caption*{
\\
{\bf Reference key}: \\
  1, \protect\cite{1958ApJS....3..141B}; 
  2, \protect\cite{1972ApJ...176..425W}; 
  3, \protect\cite{1980ApJS...42..421B}; 
  4, \protect\cite{1983ApJS...53..151B}; 
  5, \protect\cite{1985AJ.....90.1354B}; 
  6, \protect\cite{1987ApJ...322..902D}; 
  7, \protect\cite{1987ApJ...323..325B}; 
  8, \protect\cite{1987ApJS...64..219T}; 
  9, \protect\cite{1988AN....309..181R}; 
 10, \protect\cite{1990ApJ...365..665S}; 
 11, \protect\cite{1990IzKry..82...69W}; 
 12, \protect\cite{1991A&AS...90..365L}; 
 13, \protect\cite{1991AAS...89..121M}; 
 14, \protect\cite{1992ApJ...393..341L}; 
 15, \protect\cite{1993A&A...269..355B}; 
 16, \protect\cite{1993AA...269..355B}; 
 17, \protect\cite{1994A&A...283..908L}; 
 18, \protect\cite{1996A&A...310..271L}; 
 19, \protect\cite{1997A&AS..124..475M}; 
 20, \protect\cite{1997A&AS..125...65A}; 
 21, \protect\cite{1997AAS..124..475M}; 
 22, \protect\cite{1998A&AS..127..421C}; 
 23, \protect\cite{2000A&A...359..213L}; 
 24, \protect\cite{2001A&A...365..118L}; 
 25, \protect\cite{2001A&A...378..113R}; 
 26, \protect\cite{2001AA...366..121B}; 
 27, \protect\cite{2003A&A...406.1019N}; 
 28, \protect\cite{2003A&A...407..631B}; 
 29, \protect\cite{2003A&A...411..565N}; 
 30, \protect\cite{2004A&A...414..613K}; 
 31, \protect\cite{2004A&A...423.1095L}; 
 32, \protect\cite{2005ESASP.560..571G}; 
 33, \protect\cite{2006A&A...450..763K}; 
 34, \protect\cite{2006A&A...457.1033L}; 
 35, \protect\cite{2006AN....327..289H}; 
 36, \protect\cite{2006ESASP.604...73D}; 
 37, \protect\cite{2006MNRAS.370..629D}; 
 38, \protect\cite{2006MNRAS.372.1804K}; 
 39, \protect\cite{2007A&A...470..685L}; 
 40, \protect\cite{2007AA...475.1053A}; 
 41, \protect\cite{2007AN....328..475H}; 
 42, \protect\cite{2008A&A...481L..99A}; 
 43, \protect\cite{2008AA...491..545N}; 
 44, \protect\cite{2008AstBu..63..128S}; 
 45, \protect\cite{2008MNRAS.385..391A}; 
 46, \protect\cite{2008PASP..120..595A}; 
 47, \protect\cite{2009A&A...499..567K}; 
 48, \protect\cite{2009MNRAS.398.1505S}; 
 49, \protect\cite{2010ApJ...714L.318T}; 
 50, \protect\cite{2010MNRAS.401.2739L}; 
 51, \protect\cite{2010MNRAS.405L..46R}; 
 52, \protect\cite{2010MNRAS.405L..51O}; 
 53, \protect\cite{2011A&A...525A..97M}; 
 54, \protect\cite{2011A&A...536L...6A}; 
 55, \protect\cite{2011AJ....141..169B}; 
 56, \protect\cite{2011ApJ...726...24K}; 
 57, \protect\cite{2011MNRAS.414.2602D}; 
 58, \protect\cite{2012MNRAS.419.1610G}; 
 59, \protect\cite{2012MNRAS.420..757W}; 
 60, \protect\cite{2012MNRAS.426.1003S}; 
 61, \protect\cite{2013A&A...555A..46H}; 
 62, \protect\cite{2013MNRAS.432.1687B}; 
 63, \protect\cite{2014A&A...562A.143F}; 
 64, \protect\cite{2014A&A...565A..83K}; 
 65, \protect\cite{2014A&A...567A..28A}; 
 66, \protect\cite{2014MNRAS.440..182S}; 
 67, \protect\cite{2014MNRAS.444.1442S}; 
 68, \protect\cite{2015A&A...583A.115B}; 
 69, \protect\cite{2015AA...583A.115B}; 
 70, \protect\cite{2015AJ....150....2K}; 
 71, \protect\cite{2015AstBu..70..444R}; 
 72, \protect\cite{2015MNRAS.447.1418Y}; 
 73, \protect\cite{2015MNRAS.449.3945S}; 
 74, \protect\cite{2015MNRAS.451.2015O}; 
 75, \protect\cite{2015MNRAS.452.1245C}; 
 76, \protect\cite{2015MNRAS.453.2163S}; 
 77, \protect\cite{2016A&A...586A..30K}; 
 78, \protect\cite{2016A&A...588A.138R}; 
 79, \protect\cite{2016AstBu..71..302R}; 
 80, \protect\cite{2016AstBu..71..436R}; 
 81, \protect\cite{2016MNRAS.460.1811S}; 
 82, \protect\cite{2017A&A...601A..14M}; 
 83, \protect\cite{2017A&A...605A..13K}; 
 84, \protect\cite{2017ASPC..510..220M}; 
 85, \protect\cite{2017ApJ...834..142K}; 
 86, \protect\cite{2017AstBu..72..165R}; 
 87, \protect\cite{2017AstBu..72..391R}; 
 88, \protect\cite{2017MNRAS.465.2160K}; 
 89, \protect\cite{2017MNRAS.467.2820L}; 
 90, \protect\cite{2017MNRAS.468.2745N}; 
 91, \protect\cite{2017MNRAS.471.2286S}; 
 92, \protect\cite{2018A&A...618L...2J}; 
 93, \protect\cite{2018AJ....155..196R}; 
 94, \protect\cite{2018AstBu..73..178R}; 
 95, \protect\cite{2018MNRAS.473.3367O}; 
 96, \protect\cite{2018MNRAS.475.5144S}; 
 97, \protect\cite{2018MNRAS.476..562L}; 
 98, \protect\cite{2018MNRAS.478L..39S}; 
 99, \protect\cite{2018MNRAS.480.2953G}; 
100, \protect\cite{2019A&A...621A..47K}; 
101, \protect\cite{2019ApJ...877..123D}; 
102, \protect\cite{2019AstBu..74...55R}; 
103, \protect\cite{2019AstBu..74...62M}; 
104, \protect\cite{2019MNRAS.482.3950S}; 
105, \protect\cite{2019MNRAS.483.2300S}; 
106, \protect\cite{2019MNRAS.483.3127S}; 
107, \protect\cite{2019MNRAS.485.1508S}; 
108, \protect\cite{2019MNRAS.486.5558S}; 
109, \protect\cite{2019MNRAS.487.5922G}; 
110, \protect\cite{2019MNRAS.490..274S}; 
111, \protect\cite{2020AstBu..75..294R}; 
112, \protect\cite{2020MNRAS.493.3293B}; 
113, \protect\cite{2020MNRAS.493.4657L}; 
114, \protect\cite{2020MNRAS.499.5379S}; 
115, \protect\cite{2020MNRAS.499L..72L}; 
116, \protect\cite{2021A&A...652A..31B}; 
117, \protect\cite{2021AstBu..76...39R}; 
118, \protect\cite{2021AstBu..76..163R}; 
119, \protect\cite{2021MNRAS.502.5200W}; 
120, \protect\cite{2021MNRAS.502.5438P}; 
121, \protect\cite{2021MNRAS.504.3203S}; 
122, \protect\cite{2021MNRAS.504.4850S}; 
123, \protect\cite{2021MNRAS.506.2296E}; 
124, \protect\cite{2021MNRAS.507.1979L}; 
125, \protect\cite{2021ApJ...921....9D}; 
126, \protect\cite{2021MNRAS.506.5328K}; 
127, \protect\cite{2021arXiv210904043D}; 
128, Drake (priv. comm.); 
129, Shultz et al., in prep.; 
130, This Work
.}
\end{table*}

%% file: fluxtab.tex
\begin{table*}
\caption[]{Measured radio flux density (in mJy) at various wavelengths. Subscripts indicate the wavelength in cm. The superscripts beside each star indicate the works from which the fluxes were obtained according to the following key: $a$, \protect\cite{1987ApJ...322..902D}; $b$, \protect\cite{1992ApJ...393..341L}; $c$, \protect\cite{1994A&A...283..908L}; $d$, \protect\cite{1996A&A...310..271L}; $e$, \protect\cite{2004A&A...423.1095L}; $f$, \protect\cite{2006ESASP.604...73D}; $g$, \protect\cite{2015MNRAS.452.1245C}; $h$, \protect\cite{2017ApJ...834..142K}; $i$, \protect\cite{2017MNRAS.465.2160K}; $j$, \protect\cite{2017MNRAS.467.2820L}; $k$, \protect\cite{2018MNRAS.476..562L}; $l$, \protect\cite{2019ApJ...877..123D}; $m$, \protect\cite{2020MNRAS.493.4657L}; $n$, \protect\cite{2020MNRAS.499L..72L}; $o$, \protect\cite{2021MNRAS.502.5438P}; $p$, \protect\cite{2021MNRAS.507.1979L}; $q$, \protect\cite{2021ApJ...921....9D}; $r$, \protect\cite{2021arXiv210904043D}; $s$, Drake (priv. comm.); $t$, This Work;  stars with an $X$ were removed from the analysis.}
\label{fluxtab}
\begin{tabular}{l | r r r r r r r r}
\hline\hline
Star & $F_{0.3}$ & $F_1$ & $F_2$ & $F_3$ & $F_6$ & $F_{13}$ & $F_{20}$ & $F_{50}$\\
\hline
HD108$^{i}$ &  -- & $<0.04$ &  -- & $<0.02$ &  -- &  -- &  -- &  -- \\ 
HD886$^{s,X}$ &  -- &  -- &  -- & $<0.21$ &  -- &  -- &  -- &  -- \\ 
HD3360$^{it}$ &  -- & $<0.04$ &  -- & $<0.02$ &  -- & $<0.03$ & $<0.10$ &  -- \\ 
HD5737$^{a}$ &  -- &  -- &  -- &  -- & $<0.27$ &  -- &  -- &  -- \\ 
HD11502$^{b,X}$ &  -- &  -- &  -- &  -- & $<0.18$ &  -- &  -- &  -- \\ 
HD11503$^{bt}$ &  -- &  -- &  -- &  -- & $<0.18$ &  -- &  -- & $0.25 \pm 0.03$ \\ 
HD12447$^{f}$ &  -- &  -- &  -- & $0.37 \pm 0.09$ &  -- &  -- &  -- & $1.00 \pm 0.10$ \\ 
HD12767$^{s}$ &  -- &  -- &  -- & $<0.47$ &  -- &  -- &  -- &  -- \\ 
HD19832$^{f}$ &  -- &  -- &  -- & $0.45 \pm 0.12$ &  -- &  -- &  -- &  -- \\ 
HD20629$^{s}$ &  -- &  -- &  -- & $<0.28$ &  -- &  -- &  -- &  -- \\ 
HD21699$^{a}$ &  -- &  -- &  -- &  -- & $<0.42$ &  -- &  -- &  -- \\ 
HD22470$^{a}$ &  -- &  -- &  -- &  -- & $<0.36$ &  -- &  -- &  -- \\ 
HD22920$^{s}$ &  -- &  -- &  -- & $<0.27$ &  -- &  -- &  -- &  -- \\ 
HD23408$^{a,X}$ &  -- &  -- &  -- &  -- & $<0.19$ &  -- &  -- &  -- \\ 
HD25267$^{s}$ &  -- &  -- &  -- & $<0.26$ &  -- &  -- &  -- &  -- \\ 
HD27309$^{f}$ &  -- &  -- & $0.38 \pm 0.02$ & $0.33 \pm 0.02$ &  -- &  -- &  -- &  -- \\ 
HD28843$^{a}$ &  -- &  -- &  -- &  -- & $<0.25$ &  -- &  -- &  -- \\ 
HD32633$^{b}$ &  -- &  -- &  -- &  -- & $<0.18$ &  -- &  -- &  -- \\ 
HD34452$^{ab}$ &  -- &  -- &  -- &  -- & $0.48 \pm 0.10$ &  -- &  -- &  -- \\ 
HD35298$^{bef}$ & $1.61 \pm 0.43$ &  -- &  -- & $0.29 \pm 0.07$ & $0.28 \pm 0.06$ &  -- &  -- &  -- \\ 
HD35456$^{s}$ &  -- &  -- &  -- & $<0.20$ &  -- &  -- &  -- &  -- \\ 
HD35502$^{ft}$ &  -- &  -- &  -- & $2.97 \pm 0.10$ &  -- &  -- &  -- & $0.64 \pm 0.16$ \\ 
HD35575$^{c}$ &  -- &  -- &  -- &  -- & $<0.02$ &  -- &  -- &  -- \\ 
HD36313$^{f}$ &  -- &  -- &  -- & $0.49 \pm 0.06$ &  -- &  -- &  -- &  -- \\ 
HD36429$^{s}$ &  -- &  -- &  -- & $<0.20$ &  -- &  -- &  -- &  -- \\ 
HD36485$^{abeg}$ & $<0.79$ &  -- & $<0.70$ &  -- & $0.95 \pm 0.10$ &  -- & $1.20 \pm 0.10$ & $0.59 \pm 0.32$ \\ 
HD36526$^{bt}$ &  -- &  -- &  -- &  -- & $<0.18$ &  -- &  -- & $<0.10$ \\ 
HD36540$^{s}$ &  -- &  -- &  -- & $<0.21$ &  -- &  -- &  -- &  -- \\ 
HD36629$^{b,X}$ &  -- &  -- &  -- &  -- & $<0.35$ &  -- &  -- &  -- \\ 
HD36668$^{s}$ &  -- &  -- &  -- & $<0.21$ &  -- &  -- &  -- &  -- \\ 
HD36916$^{bc}$ &  -- &  -- &  -- &  -- & $<0.07$ &  -- &  -- &  -- \\ 
HD36960$^{c,X}$ &  -- &  -- &  -- &  -- & $<0.12$ &  -- &  -- &  -- \\ 
HD37017$^{abdeghr}$ & $<0.28$ & $3.10 \pm 0.40$ & $2.10 \pm 0.20$ &  -- & $2.60 \pm 0.06$ &  -- & $2.40 \pm 0.12$ & $0.44 \pm 0.08$ \\ 
HD37022$^{g}$ &  -- &  -- &  -- &  -- &  -- &  -- & $<4.95$ &  -- \\ 
HD37041$^{c,X}$ &  -- &  -- &  -- &  -- & $<0.30$ &  -- &  -- &  -- \\ 
HD37043$^{a,X}$ &  -- &  -- &  -- &  -- & $<0.24$ &  -- &  -- &  -- \\ 
HD37058$^{b}$ &  -- &  -- &  -- &  -- & $<0.31$ &  -- &  -- &  -- \\ 
HD37061$^{ht}$ &  -- &  -- &  -- &  -- & $0.68 \pm 0.07$ &  -- &  -- & $0.54 \pm 0.14$ \\ 
HD37129$^{a}$ &  -- &  -- &  -- &  -- & $<0.20$ &  -- &  -- &  -- \\ 
HD37140$^{s}$ &  -- &  -- &  -- & $<0.20$ &  -- &  -- &  -- &  -- \\ 
HD37150$^{f,X}$ &  -- &  -- &  -- & $0.51 \pm 0.07$ &  -- &  -- &  -- &  -- \\ 
HD37151$^{s}$ &  -- &  -- &  -- & $<0.21$ &  -- &  -- &  -- &  -- \\ 
HD37210$^{s}$ &  -- &  -- &  -- & $<0.26$ &  -- &  -- &  -- &  -- \\ 
HD37321$^{b,X}$ &  -- &  -- &  -- &  -- & $<0.27$ &  -- &  -- &  -- \\ 
HD37479$^{abdeg}$ & $<1.03$ & $4.90 \pm 0.40$ & $3.70 \pm 0.30$ &  -- & $3.90 \pm 0.10$ &  -- & $3.20 \pm 0.11$ & $1.25 \pm 0.28$ \\ 
HD37642$^{f}$ &  -- &  -- &  -- & $0.60 \pm 0.07$ &  -- &  -- &  -- &  -- \\ 
HD37742$^{it}$ &  -- & $2.77 \pm 0.05$ &  -- & $1.10 \pm 0.01$ &  -- & $0.49 \pm 0.03$ & $0.28 \pm 0.05$ & $0.95 \pm 0.10$ \\ 
HD37752$^{c}$ &  -- &  -- &  -- &  -- & $<0.08$ &  -- &  -- &  -- \\ 
HD37776$^{at}$ &  -- &  -- &  -- &  -- & $<0.25$ &  -- &  -- & $<0.10$ \\ 
HD37808$^{ce}$ & $<0.36$ &  -- &  -- &  -- & $0.97 \pm 0.26$ &  -- &  -- &  -- \\ 
HD40312$^{b}$ &  -- &  -- & $<0.46$ &  -- & $0.31 \pm 0.06$ &  -- &  -- &  -- \\ 
\hline\hline
\end{tabular}
\end{table*}
 
\begin{table*}
\contcaption{}
\label{fluxtab:continued}
\begin{tabular}{l | r r r r r r r r}
\hline
\hline
Star & $F_{0.3}$ & $F_1$ & $F_2$ & $F_3$ & $F_6$ & $F_{13}$ & $F_{20}$ & $F_{50}$\\
\hline
HD41269$^{c}$ &  -- &  -- &  -- &  -- & $<0.08$ &  -- &  -- &  -- \\ 
HD42509$^{c}$ &  -- &  -- &  -- &  -- & $<0.06$ &  -- &  -- &  -- \\ 
HD43819$^{c}$ &  -- &  -- &  -- &  -- & $<0.06$ &  -- &  -- &  -- \\ 
HD45105$^{c}$ &  -- &  -- &  -- &  -- & $<0.06$ &  -- &  -- &  -- \\ 
HD45583$^{fr}$ &  -- &  -- &  -- & $0.27 \pm 0.05$ &  -- &  -- &  -- & $0.22 \pm 0.04$ \\ 
HD46328$^{i}$ &  -- & $<0.05$ &  -- & $<0.02$ &  -- & $<45.00$ &  -- &  -- \\ 
HD47129$^{i}$ &  -- & $0.39 \pm 0.04$ &  -- & $0.21 \pm 0.01$ &  -- & $<0.03$ &  -- &  -- \\ 
HD47777$^{i}$ &  -- & $<0.03$ &  -- & $<0.04$ &  -- &  -- &  -- &  -- \\ 
HD49333$^{b}$ &  -- &  -- &  -- &  -- & $<0.22$ &  -- &  -- &  -- \\ 
HD49606$^{s}$ &  -- &  -- &  -- & $<0.20$ &  -- &  -- &  -- &  -- \\ 
HD50204$^{c}$ &  -- &  -- &  -- &  -- & $<0.06$ &  -- &  -- &  -- \\ 
HD51418$^{s}$ &  -- &  -- &  -- & $<0.21$ &  -- &  -- &  -- &  -- \\ 
HD52589$^{s}$ &  -- &  -- &  -- & $<0.17$ &  -- &  -- &  -- &  -- \\ 
HD55522$^{t}$ &  -- &  -- &  -- &  -- &  -- &  -- &  -- & $<0.37$ \\ 
HD57219$^{a,X}$ &  -- &  -- &  -- &  -- & $<0.33$ &  -- &  -- &  -- \\ 
HD57682$^{gi}$ &  -- & $<0.03$ &  -- & $<0.03$ &  -- & $<0.02$ & $<0.16$ &  -- \\ 
HD58260$^{ai}$ &  -- & $<0.08$ &  -- & $<0.04$ & $<0.18$ & $<0.03$ &  -- &  -- \\ 
HD60344$^{a}$ &  -- &  -- &  -- &  -- & $<0.20$ &  -- &  -- &  -- \\ 
HD61556$^{ot}$ &  -- &  -- &  -- &  -- &  -- &  -- & $2.86 \pm 0.81$ & $1.21 \pm 0.18$ \\ 
HD64740$^{t}$ &  -- &  -- &  -- &  -- &  -- &  -- &  -- & $0.06 \pm 0.01$ \\ 
HD65339$^{ab}$ &  -- &  -- &  -- &  -- & $<0.18$ &  -- &  -- &  -- \\ 
HD66665$^{i}$ &  -- & $<0.03$ &  -- & $<0.02$ &  -- & $<42.00$ &  -- &  -- \\ 
HD66765$^{t}$ &  -- &  -- &  -- &  -- &  -- &  -- &  -- & $<0.10$ \\ 
HD77653$^{o}$ &  -- &  -- &  -- &  -- &  -- &  -- & $6.87 \pm 1.60$ & $6.87 \pm 1.60$ \\ 
HD78556A$^{c}$ &  -- &  -- &  -- &  -- & $<0.06$ &  -- &  -- &  -- \\ 
HD79158$^{af}$ &  -- &  -- &  -- & $0.45 \pm 0.05$ & $<0.62$ &  -- &  -- &  -- \\ 
HD89822$^{a,X}$ &  -- &  -- &  -- &  -- & $<0.48$ &  -- &  -- &  -- \\ 
HD89897$^{c}$ &  -- &  -- &  -- &  -- & $<0.06$ &  -- &  -- &  -- \\ 
HD90044$^{c}$ &  -- &  -- &  -- &  -- & $<0.05$ &  -- &  -- &  -- \\ 
HD97441A$^{c,X}$ &  -- &  -- &  -- &  -- & $<0.12$ &  -- &  -- &  -- \\ 
HD98664$^{c}$ &  -- &  -- &  -- &  -- & $<0.10$ &  -- &  -- &  -- \\ 
HD100340$^{a}$ &  -- &  -- &  -- &  -- & $<0.43$ &  -- &  -- &  -- \\ 
HD105382$^{o}$ &  -- &  -- &  -- &  -- &  -- &  -- & $2.20 \pm 0.63$ & $2.20 \pm 0.63$ \\ 
HD109387$^{a,X}$ &  -- &  -- &  -- &  -- & $<0.18$ &  -- &  -- &  -- \\ 
HD112413$^{af}$ &  -- &  -- &  -- & $0.29 \pm 0.03$ & $<0.46$ &  -- &  -- &  -- \\ 
HD115735$^{c}$ &  -- &  -- &  -- &  -- & $<0.07$ &  -- &  -- &  -- \\ 
HD118022$^{p}$ &  -- &  -- &  -- & $0.50 \pm 0.25$ &  -- &  -- &  -- &  -- \\ 
HD120709$^{b,X}$ &  -- &  -- &  -- &  -- & $<0.24$ &  -- &  -- &  -- \\ 
HD120710$^{b,X}$ &  -- &  -- &  -- &  -- & $<0.21$ &  -- &  -- &  -- \\ 
HD122532$^{b}$ &  -- &  -- &  -- &  -- & $<0.18$ &  -- &  -- &  -- \\ 
HD124224$^{cdefoq}$ & $1.32 \pm 0.31$ & $4.30 \pm 0.10$ & $4.60 \pm 0.10$ & $4.07 \pm 0.14$ & $3.80 \pm 0.07$ &  -- &  -- & $2.19 \pm 0.24$ \\ 
HD125248$^{s}$ &  -- &  -- &  -- & $<0.22$ &  -- &  -- &  -- &  -- \\ 
HD125823$^{a}$ &  -- &  -- &  -- &  -- & $<0.55$ &  -- &  -- &  -- \\ 
HD126515$^{b}$ &  -- &  -- &  -- &  -- & $<0.18$ &  -- &  -- &  -- \\ 
HD130557$^{c,X}$ &  -- &  -- &  -- &  -- & $<0.18$ &  -- &  -- &  -- \\ 
HD131120$^{b}$ &  -- &  -- &  -- &  -- & $<0.21$ &  -- &  -- &  -- \\ 
HD133029$^{b}$ &  -- &  -- &  -- &  -- & $<0.22$ &  -- &  -- &  -- \\ 
HD133652$^{f}$ &  -- &  -- &  -- & $0.24 \pm 0.07$ &  -- &  -- &  -- &  -- \\ 
HD133880$^{fg}$ &  -- &  -- &  -- & $4.08 \pm 0.16$ &  -- &  -- & $14.45 \pm 0.11$ & $2.04 \pm 0.17$ \\ 
HD135679$^{c}$ &  -- &  -- &  -- &  -- & $<0.05$ &  -- &  -- &  -- \\ 
\hline\hline
\end{tabular}
\end{table*}
 
\begin{table*}
\contcaption{}
\label{fluxtab:continued}
\begin{tabular}{l | r r r r r r r r}
\hline
\hline
Star & $F_{0.3}$ & $F_1$ & $F_2$ & $F_3$ & $F_6$ & $F_{13}$ & $F_{20}$ & $F_{50}$\\
\hline
HD137193$^{b}$ &  -- &  -- &  -- &  -- & $<0.21$ &  -- &  -- &  -- \\ 
HD137909$^{a}$ &  -- &  -- &  -- &  -- & $<0.44$ &  -- &  -- &  -- \\ 
HD138764$^{c}$ &  -- &  -- &  -- &  -- & $<0.06$ &  -- &  -- &  -- \\ 
HD142184$^{ko}$ &  -- & $160.00 \pm 20.00$ & $143.00 \pm 2.00$ & $121.00 \pm 2.00$ &  -- &  -- & $7.64 \pm 1.75$ & $7.64 \pm 1.75$ \\ 
HD142250$^{s}$ &  -- &  -- &  -- & $<0.21$ &  -- &  -- &  -- &  -- \\ 
HD142301$^{bcde}$ & $<1.28$ & $<0.56$ & $1.50 \pm 0.30$ & $2.76 \pm 2.76$ & $4.90 \pm 0.06$ &  -- & $1.81 \pm 1.81$ &  -- \\ 
HD142884$^{b}$ &  -- &  -- &  -- &  -- & $<0.18$ &  -- &  -- &  -- \\ 
HD142990$^{bcdel}$ & $<0.54$ & $<0.52$ & $<0.42$ & $1.98 \pm 1.98$ & $2.27 \pm 0.07$ &  -- & $<3.30$ & $1.10 \pm 0.20$ \\ 
HD143473$^{f}$ &  -- &  -- &  -- & $0.50 \pm 0.07$ &  -- &  -- &  -- &  -- \\ 
HD143699$^{bs}$ &  -- &  -- &  -- & $<0.21$ & $0.21 \pm 0.06$ &  -- &  -- &  -- \\ 
HD144217$^{b,X}$ &  -- &  -- &  -- &  -- & $<0.21$ &  -- &  -- &  -- \\ 
HD144218$^{b,X}$ &  -- &  -- &  -- &  -- & $<0.21$ &  -- &  -- &  -- \\ 
HD144334$^{bce}$ & $<0.92$ &  -- &  -- &  -- & $0.42 \pm 0.07$ &  -- & $<0.80$ &  -- \\ 
HD144661$^{b,X}$ &  -- &  -- &  -- &  -- & $<0.18$ &  -- &  -- &  -- \\ 
HD144844$^{bs,X}$ &  -- &  -- &  -- & $<0.20$ & $0.19 \pm 0.06$ &  -- & $<0.49$ &  -- \\ 
HD145102$^{s}$ &  -- &  -- &  -- & $<0.15$ &  -- &  -- &  -- &  -- \\ 
HD145482$^{b}$ &  -- &  -- &  -- &  -- & $<0.18$ &  -- &  -- &  -- \\ 
HD145501C$^{bcr}$ &  -- &  -- &  -- & $1.67 \pm 0.06$ & $2.06 \pm 0.07$ &  -- & $1.57 \pm 0.18$ & $1.24 \pm 0.12$ \\ 
HD145501D$^{b,X}$ &  -- &  -- &  -- & $<0.60$ & $<0.42$ &  -- & $<0.54$ &  -- \\ 
HD145502$^{b,X}$ &  -- &  -- &  -- & $<0.38$ & $<0.36$ &  -- &  -- &  -- \\ 
HD146001$^{b}$ &  -- &  -- &  -- & $<0.22$ & $0.32 \pm 0.07$ &  -- & $<0.84$ &  -- \\ 
HD147010$^{b}$ &  -- &  -- &  -- &  -- & $<0.20$ &  -- &  -- &  -- \\ 
HD147084$^{s,X}$ &  -- &  -- &  -- & $<0.17$ &  -- &  -- &  -- &  -- \\ 
HD147550$^{c}$ &  -- &  -- &  -- &  -- & $<0.17$ &  -- &  -- &  -- \\ 
HD147890$^{s}$ &  -- &  -- &  -- & $<0.20$ &  -- &  -- &  -- &  -- \\ 
HD147932$^{n}$ &  -- &  -- & $20.00 \pm 2.00$ & $20.00 \pm 2.00$ & $20.00 \pm 2.00$ & $15.00 \pm 2.00$ & $10.00 \pm 2.00$ &  -- \\ 
HD147933$^{m}$ &  -- & $11.00 \pm 0.10$ & $11.00 \pm 0.10$ & $12.00 \pm 0.10$ & $12.00 \pm 0.10$ &  -- &  -- &  -- \\ 
HD148112$^{a}$ &  -- &  -- &  -- &  -- & $<0.47$ &  -- &  -- &  -- \\ 
HD148199$^{b}$ &  -- &  -- &  -- &  -- & $<0.45$ &  -- &  -- &  -- \\ 
HD149438$^{i}$ &  -- & $<0.04$ &  -- & $<0.02$ &  -- & $<36.00$ &  -- &  -- \\ 
HD149822$^{c}$ &  -- &  -- &  -- &  -- & $<0.07$ &  -- &  -- &  -- \\ 
HD151346$^{b}$ &  -- &  -- &  -- &  -- & $<0.18$ &  -- &  -- &  -- \\ 
HD152107$^{ab}$ &  -- &  -- &  -- & $<0.15$ & $<0.40$ &  -- &  -- &  -- \\ 
HD156424$^{i}$ &  -- & $0.49 \pm 0.03$ &  -- & $0.38 \pm 0.01$ &  -- &  -- &  -- &  -- \\ 
HD157779$^{c}$ &  -- &  -- &  -- &  -- & $<0.07$ &  -- &  -- &  -- \\ 
HD159376$^{c}$ &  -- &  -- &  -- &  -- & $<0.06$ &  -- &  -- &  -- \\ 
HD162374$^{s,X}$ &  -- &  -- &  -- & $<0.33$ &  -- &  -- &  -- &  -- \\ 
HD163472$^{i}$ &  -- & $<0.03$ &  -- & $<0.02$ &  -- & $<0.03$ &  -- &  -- \\ 
HD164429$^{cf}$ &  -- &  -- &  -- & $0.30 \pm 0.05$ & $0.30 \pm 0.06$ &  -- &  -- &  -- \\ 
HD165474$^{s}$ &  -- &  -- &  -- & $<0.14$ &  -- &  -- &  -- &  -- \\ 
HD166182$^{s,X}$ &  -- &  -- &  -- & $<0.21$ &  -- &  -- &  -- &  -- \\ 
HD168785$^{s}$ &  -- &  -- &  -- & $<0.27$ &  -- &  -- &  -- &  -- \\ 
HD168814$^{c,X}$ &  -- &  -- &  -- &  -- & $<0.08$ &  -- &  -- &  -- \\ 
HD168856$^{s}$ &  -- &  -- &  -- & $<0.20$ &  -- &  -- &  -- &  -- \\ 
HD170000$^{f}$ &  -- &  -- &  -- & $0.45 \pm 0.05$ &  -- &  -- &  -- &  -- \\ 
HD170973$^{s}$ &  -- &  -- &  -- & $<0.16$ &  -- &  -- &  -- &  -- \\ 
HD171247$^{ce}$ & $<0.03$ &  -- &  -- &  -- & $3.04 \pm 0.09$ &  -- &  -- &  -- \\ 
HD174638$^{c,X}$ &  -- &  -- &  -- &  -- & $3.55 \pm 0.06$ &  -- &  -- &  -- \\ 
HD175132$^{cs}$ &  -- &  -- &  -- & $<0.14$ & $<0.07$ &  -- &  -- &  -- \\ 
HD175156$^{b,X}$ &  -- &  -- &  -- &  -- & $<0.18$ &  -- &  -- &  -- \\ 
\hline\hline
\end{tabular}
\end{table*}
 
\begin{table*}
\contcaption{}
\label{fluxtab:continued}
\begin{tabular}{l | r r r r r r r r}
\hline
\hline
Star & $F_{0.3}$ & $F_1$ & $F_2$ & $F_3$ & $F_6$ & $F_{13}$ & $F_{20}$ & $F_{50}$\\
\hline
HD175362$^{bcdt}$ &  -- &  -- &  -- & $<0.21$ & $0.38 \pm 0.06$ &  -- & $0.27 \pm 0.04$ & $0.54 \pm 0.08$ \\ 
HD175744$^{cs}$ &  -- &  -- &  -- & $<0.17$ & $<0.02$ &  -- &  -- &  -- \\ 
HD176582$^{f}$ &  -- &  -- &  -- & $0.46 \pm 0.05$ &  -- &  -- &  -- &  -- \\ 
HD177003$^{c}$ &  -- &  -- &  -- &  -- & $<0.07$ &  -- &  -- &  -- \\ 
HD177410$^{c}$ &  -- &  -- &  -- &  -- & $<0.07$ &  -- &  -- &  -- \\ 
HD179527$^{s}$ &  -- &  -- &  -- & $<0.16$ &  -- &  -- &  -- &  -- \\ 
HD182180$^{jt}$ &  -- & $18.00 \pm 0.10$ & $20.00 \pm 0.10$ & $20.00 \pm 0.10$ &  -- &  -- &  -- & $4.17 \pm 0.42$ \\ 
HD183056$^{c}$ &  -- &  -- &  -- &  -- & $<0.06$ &  -- &  -- &  -- \\ 
HD183339$^{s}$ &  -- &  -- &  -- & $<0.17$ &  -- &  -- &  -- &  -- \\ 
HD184927$^{cst}$ &  -- &  -- &  -- & $<0.17$ & $<0.07$ &  -- & $<0.11$ & $<0.19$ \\ 
HD184961$^{s}$ &  -- &  -- &  -- & $<0.22$ &  -- &  -- &  -- &  -- \\ 
HD186205$^{s}$ &  -- &  -- &  -- & $<0.21$ &  -- &  -- &  -- &  -- \\ 
HD187474$^{s}$ &  -- &  -- &  -- & $<0.23$ &  -- &  -- &  -- &  -- \\ 
HD188041$^{s}$ &  -- &  -- &  -- & $<0.21$ &  -- &  -- &  -- &  -- \\ 
HD189775$^{t}$ &  -- &  -- &  -- &  -- &  -- &  -- & $0.41 \pm 0.07$ & $1.09 \pm 0.12$ \\ 
HD191612$^{it}$ &  -- & $<0.03$ &  -- & $<0.02$ &  -- & $<0.04$ & $<0.24$ & $<0.34$ \\ 
HD192678$^{s}$ &  -- &  -- &  -- & $<0.17$ &  -- &  -- &  -- &  -- \\ 
HD196178$^{f}$ &  -- &  -- &  -- & $3.00 \pm 0.07$ &  -- &  -- &  -- &  -- \\ 
HD196502$^{a}$ &  -- &  -- &  -- &  -- & $<0.80$ &  -- &  -- &  -- \\ 
HD200775$^{t}$ &  -- &  -- &  -- &  -- &  -- &  -- & $0.30 \pm 0.06$ & $<0.37$ \\ 
HD202671$^{s}$ &  -- &  -- &  -- & $<0.15$ &  -- &  -- &  -- &  -- \\ 
HD205021$^{it}$ &  -- & $<0.03$ &  -- & $<0.03$ &  -- & $<0.03$ & $<0.15$ & $<0.25$ \\ 
HD207538$^{a,X}$ &  -- &  -- &  -- &  -- & $<0.35$ &  -- &  -- &  -- \\ 
HD207840$^{s,X}$ &  -- &  -- &  -- & $<0.21$ &  -- &  -- &  -- &  -- \\ 
HD208057$^{t}$ &  -- &  -- &  -- &  -- &  -- &  -- &  -- & $<0.05$ \\ 
HD208266$^{s}$ &  -- &  -- &  -- & $<0.36$ &  -- &  -- &  -- &  -- \\ 
HD209339$^{a}$ &  -- &  -- &  -- &  -- & $<0.29$ &  -- &  -- &  -- \\ 
HD214993$^{s,X}$ &  -- &  -- &  -- & $<0.26$ &  -- &  -- &  -- &  -- \\ 
HD215441$^{abdeg}$ & $<0.34$ & $<0.16$ & $0.60 \pm 0.10$ &  -- & $1.10 \pm 0.10$ &  -- & $1.49 \pm 0.10$ & $0.98 \pm 0.10$ \\ 
HD216916$^{s,X}$ &  -- &  -- &  -- & $<0.21$ &  -- &  -- &  -- &  -- \\ 
HD223128$^{s}$ &  -- &  -- &  -- & $<0.24$ &  -- &  -- &  -- &  -- \\ 
HD224801$^{s}$ &  -- &  -- &  -- & $<0.23$ &  -- &  -- &  -- &  -- \\ 
HD224926$^{a,X}$ &  -- &  -- &  -- &  -- & $<0.20$ &  -- &  -- &  -- \\ 
HD260858$^{a}$ &  -- &  -- &  -- &  -- & $<0.18$ &  -- &  -- &  -- \\ 
HD264111$^{a}$ &  -- &  -- &  -- &  -- & $<0.18$ &  -- &  -- &  -- \\ 
HD335238$^{s}$ &  -- &  -- &  -- & $<0.16$ &  -- &  -- &  -- &  -- \\ 
ALS8988$^{i}$ &  -- & $<0.04$ &  -- & $<0.02$ &  -- & $<30.00$ &  -- &  -- \\ 
ALS9522$^{i}$ &  -- & $<0.03$ &  -- & $0.08 \pm 0.01$ &  -- & $<0.03$ &  -- &  -- \\ 
CPD-271791$^{a}$ &  -- &  -- &  -- &  -- & $<0.18$ &  -- &  -- &  -- \\ 
CPD282561$^{i}$ &  -- &  -- &  -- & $<0.02$ &  -- & $<0.03$ &  -- &  -- \\ 
NGC1624-2$^{gi}$ &  -- & $<0.04$ &  -- & $<0.02$ &  -- & $<0.02$ & $<0.21$ &  -- \\ 
\hline\hline
\end{tabular}
\end{table*}